\ifdefined\XeTeXrevision\else
\pdfoutput=1
\fi
\documentclass[11pt,onecolumn,logo]{neomme}

\usepackage[authoryear,round]{natbib}
\usepackage{bm}
\usepackage{nicefrac}
\usepackage{fontawesome5}
\usepackage{array}
\usepackage{multirow}
\usepackage{longtable}
\usepackage{subcaption}
\usepackage{tablefootnote}
\usepackage{etoolbox}
\usepackage{wrapfig}
\usepackage{float}
\usepackage{placeins}
\usepackage[most,breakable,skins]{tcolorbox}
\usepackage{algorithm}
\usepackage{algpseudocode}
\usepackage{xurl}

\graphicspath{{assets/figures/}}

\definecolor{retrievalquery}{HTML}{7964E8}
\definecolor{retrievalrank}{HTML}{F14738}
\definecolor{retrievalhit}{HTML}{16A34A}
\newcommand{\code}[1]{\texttt{#1}}
\newcommand{\neomme}{\textit{NeoMME}}
\newcommand{\lik}{\textit{LIK}}
\newcommand{\paramsmall}{\textbf{\textless{}300M}}
\newcommand{\parammedium}{\textbf{300M to 1B}}
\newcommand{\paramlarge}{\textbf{\textgreater{}1B}}
\newcommand{\githublink}[1]{\href{#1}{\faGithub\ \nolinkurl{#1}}}
\newcommand{\flagemoji}[1]{\raisebox{-0.15em}{\includegraphics[height=1em]{flags/#1.png}}}
\newcommand{\hfemoji}{\raisebox{-0.15em}{\includegraphics[height=1em]{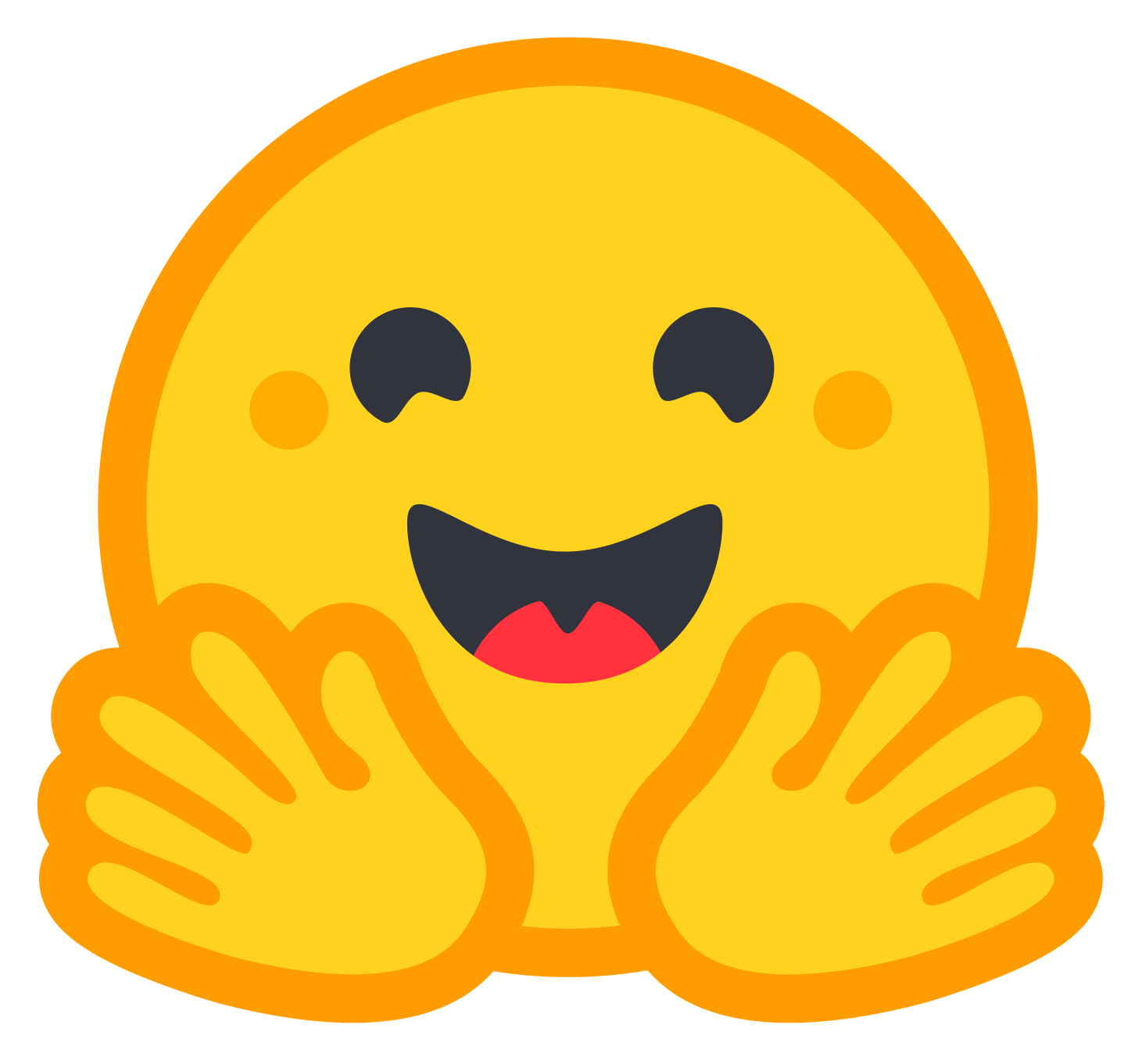}}}
\newcommand{\hflink}[2]{\href{#1}{\hfemoji\ #2}}
\tcbset{
  retrievalshowcase/.style={
    colback=white,
    colframe=black,
    coltitle=white,
    boxrule=0.5mm,
    arc=2mm,
    outer arc=2mm,
    left=1mm,
    right=1mm,
    top=1mm,
    bottom=1mm,
    width=\textwidth,
  },
  retrievalquery/.style={
    fontupper=\scriptsize,
    colback=white,
    colframe=retrievalquery,
    coltitle=white,
    boxrule=0.3mm,
    arc=2mm,
    outer arc=2mm,
    left=1mm,
    right=1mm,
    top=1mm,
    bottom=1mm,
  },
  retrievalresult/.style={
    fontupper=\scriptsize,
    colback=white,
    colframe=retrievalrank,
    coltitle=white,
    boxrule=0.3mm,
    arc=2mm,
    outer arc=2mm,
    left=1mm,
    right=1mm,
    top=1mm,
    bottom=1mm,
  },
  retrievalrelevant/.style={retrievalresult, colframe=retrievalhit},
  retrievalunjudged/.style={retrievalresult, colframe=retrievalrank},
}
\newcommand{\retrievalresultcard}[5]{%
  \begin{tcolorbox}[title=Rank #2, #1]
    \centering\includegraphics[width=\linewidth]{retrieval-cases/#3}\\[3pt]
    \raggedright
    \textbf{MeanMaxSim:} #4\\
    \textbf{Query relevance:} #5
  \end{tcolorbox}%
}
\newcommand{\retrievalcasemeta}[2]{%
  \footnotesize
  \textbf{Task:} #1 \quad
  \textbf{Model:} #2
}
\let\cite\citep
\hypersetup{
  pdftitle={NeoMME: A Single-Tower Multimodal-Native Multilingual Foundation Encoder for Efficient Fine-Tuning and Inference},
  pdfauthor={Aurélien Lac, Tony Wu}
}
\title{\neomme: A Single-Tower Multimodal-Native Multilingual Foundation Encoder
for Efficient Fine-Tuning and Inference}

\author{%
  Aurélien Lac\textsuperscript{*} \qquad Tony Wu\textsuperscript{*}\\
  H Company\\
  \textsuperscript{*}Equal contribution.
}
\correspondingauthor{\texttt{aurelien.lac.pro@gmail.com, tonywu.ai@outlook.com}}
\paperurl{https://hf.co/collections/Hcompany/neomme}
\renewcommand{\today}{}

\begin{abstract}
  Multimodal models often build on architectures designed for generative vision--language modeling, typically combining separately pretrained vision encoders with causal language models. Visual document retrievers such as ColPali repurpose these models as encoders, carrying over the parameter and compute overhead of a VLM for a non-generative task.

We introduce \neomme{}, a family of 260M and 800M-parameter Multimodal and Multilingual bidirectional Encoders that process multilingual text and raw image patches in a single bidirectional Transformer encoder. Both models are pretrained from scratch with a masked discrete-diffusion text objective, conditioned on visible image patches for multimodal examples. Both support a 16,384-token context, enough to encode up to two standard 4K UHD images.

To demonstrate its downstream capabilities, we fine-tune \neomme{} with jointly trained dense and late-interaction heads. On the ViDoRe v3 benchmark, the resulting \neomme{}-Retriever 260M outperforms all evaluated models strictly below 800M parameters with 0.523 nDCG@10, while \neomme{}-Retriever 800M reaches 0.556. At a matched $2048\times2048$ image input size on an NVIDIA L40S, \neomme-260M encodes pages with about $2\times$ the throughput of ColModernVBERT. Hierarchical token pooling and asymmetric quantization compress late-interaction multimodal document embeddings by $255\times$ while preserving over 95\% of baseline nDCG@10. We contribute \neomme{} to Hugging Face Transformers and release the pretrained backbone and retrieval-compatible checkpoints under Apache 2.0 at \url{https://hf.co/collections/Hcompany/neomme}.

\end{abstract}

\begin{document}

\maketitle
\begin{center}
\vspace{-4pt}
\begin{minipage}{\textwidth}
\centering
\includegraphics[width=\linewidth]{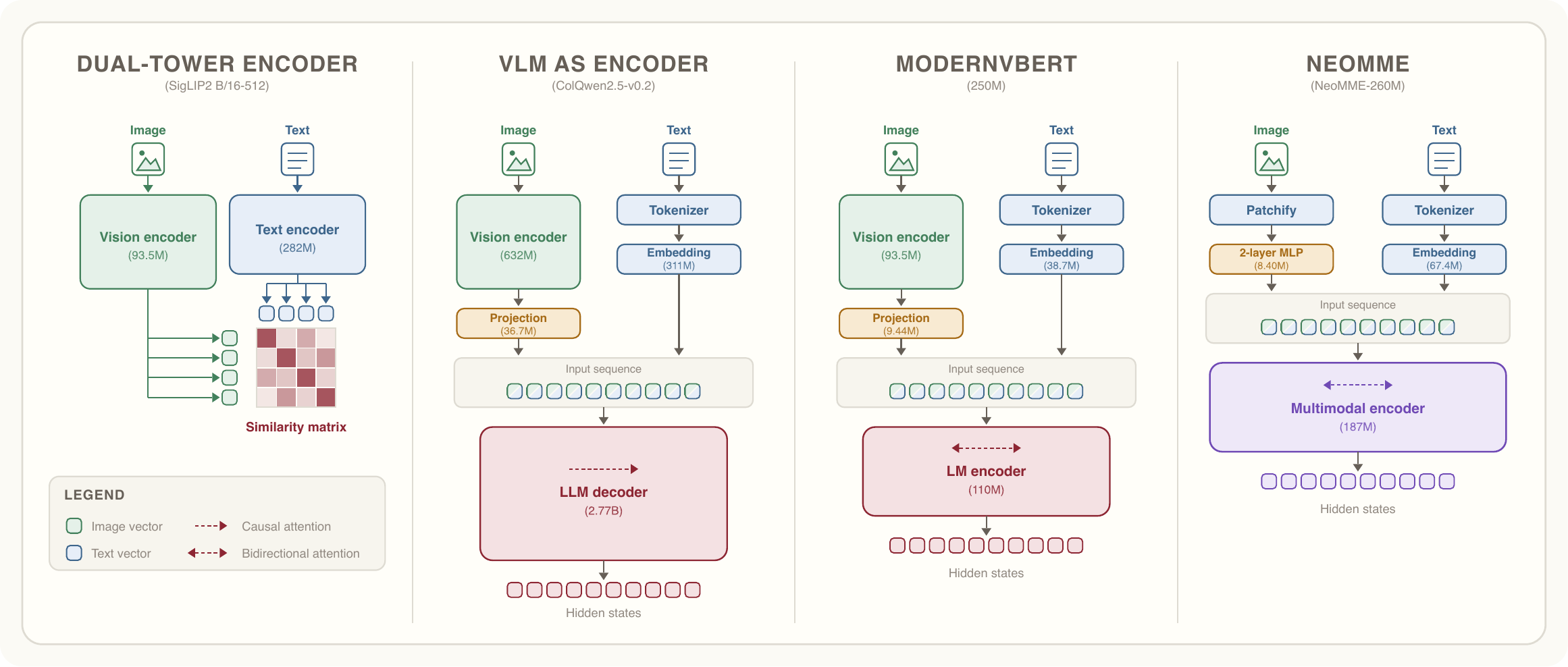}
\captionof{figure}{Unlike dual-tower and VLM encoders, \neomme{} processes image patches and text tokens in one bidirectional Transformer, without a pretrained vision tower or causal decoder.}
\label{fig:input-paths}
\end{minipage}
\vspace{-4pt}
\end{center}
\clearpage

\section{Introduction}
\label{sec:introduction}
\vspace{-0.5em}

Bidirectional encoders are strong models for learning text representations that transfer across tasks. BERT established masked bidirectional pretraining \citep{devlin2019bert}, while ModernBERT brought long-context and efficiency improvements to encoder architectures \citep{warner2024modernbert}. Large Language Models (LLMs) can also be converted into encoders similarly to LLM2Vec and LFM2.5-Encoder \citep{behnamghader2024llm2vec,liquidai2026lfm25encoders}. In matched experiments, Ettin \citep{weller2026seqvsseq} finds that native masked encoders remain stronger than causal decoders and decoder-to-encoder adaptations on classification and retrieval tasks, while native causal decoders remain stronger on generation.

The contrast between encoders and decoders also extends to multimodal architectures. CLIP and SigLIP align independent image and text towers \citep{radford2021clip,zhai2023siglip}, while generative Visual Language Models (VLMs) often project the output of a pretrained visual encoder to a causal language model \citep{alayrac2022flamingo,li2023blip2,beyer2024paligemma}. Visual document retrieval systems reuse both architectural patterns. DSE produces dense embeddings from PDF page screenshots \citep{ma2024dse}, while ColPali keeps the image patch granularity for finer representations using late-interaction \citep{faysse2025colpali}. ModernVBERT is a 250M-parameter model that replaces the causal decoder with a bidirectional encoder while retaining a pretrained SigLIP2 tower \citep{teiletche2025modernvbert,tschannen2025siglip2}. Its retrieval results show that a model of this size can perform visual document retrieval.

An alternative is to process image and text tokens with a single shared Transformer. ViLT, OneR, and M3AE instantiate this design in bidirectional architectures \citep{kim2021vilt,jang2022oner,geng2022m3ae}, while recent tower-free VLMs feed projected image patches directly to generative backbones \citep{chen2024single,diao2024eve}. Sharing the Transformer gives both modalities the same computational path, rather than separate towers with potentially asymmetric architectures and forward passes. It also unifies the model lifecycle: the same backbone can be pretrained, fine-tuned, parallelized, and served across modalities.

With these motivations in view, we introduce \neomme{}, a family of bidirectional multimodal encoders trained entirely from scratch. Text embeddings and raw $32\times32$ pixel patches enter through modality-specific projections and then share every Transformer layer. Dynamic-resolution images retain their aspect ratio, while the 32-pixel patches and long-context architecture keep high-resolution inputs tractable. For text-only examples, pretraining uses a masked discrete-diffusion objective \citep{sahoo2024mdlm,shi2024md4,nie2025llada}. For image--text pairs, the same text-denoising objective is conditioned on visible patches from the corresponding natural or document image.

As a downstream evaluation of the backbone, we fine-tune \neomme{} for visual document retrieval, yielding \neomme{}-Retriever. A single backbone forward pass produces both a dense pooled representation and a late-interaction multi-vector representation. We also release Late-Interaction Kernels (\lik{}), a suite of fused MaxSim kernels that reduces peak VRAM and runtime for late-interaction inference and training \citep{lac2026lik}. On ViDoRe v3 \citep{loison2026vidorev3}, the 260M and 800M models respectively achieve 0.523 and 0.556 nDCG@10, as shown in \autoref{tab:visual-results}. \neomme-260M encodes $2048\times2048$ pages at 51.3 pages per second on an NVIDIA L40S, with $1.97\times$ the throughput of ColModernVBERT at the same input size, as shown in \autoref{fig:indexing-l40s}. Hierarchical token pooling at factor 8, combined with asymmetric quantization to int8 queries and binary documents, makes high-resolution visual document retrieval tractable for large corpora. In our experiments, the two together shrink the \neomme-260M embedding from $\sim$1.5 MB per ViDoRe v3 document to 6 kB, a 255$\times$ compression that retains more than 95\% of the original retrieval quality, as shown in \autoref{fig:compression-frontier}.

\noindent\textbf{Contribution 1: \neomme{}, an efficient Multilingual and Multimodal-native foundational Encoder.} We train a tokenizer and 260M and 800M bidirectional single-tower Transformer backbones from scratch that take raw image patches as input. The evaluated recipe combines multilingual text, code, mathematics, natural images, and document images with long-context dynamic-resolution encoding, and image-conditioned masked-diffusion pretraining. We also contribute \neomme{} to the Hugging Face Transformers library \citep{wolf2020transformers}. The \href{https://huggingface.co/docs/transformers/en/model_doc/neomme}{\neomme{} model documentation} describes the public architecture and API.

\noindent\textbf{Contribution 2: \neomme{}-Retriever, an efficient visual document retrieval embedder.}\footnote{\hflink{https://huggingface.co/spaces/tonywu71/neomme-retriever-demo}{spaces/tonywu71/neomme-retriever-demo}} The model can output both late-interaction and dense representations for deployment flexibility. It is compatible with Sentence Transformers for dense and multi-vector retrieval \citep{reimers2019sentencebert}. We evaluate both 260M and 800M models on visual-document and text retrieval and show that \neomme{}-Retriever is Pareto-optimal on ViDoRe v1, v2, and v3 \citep{mace2025vidorev2}. We also study input resolution, representation storage and compression, indexing throughput, and query-encoding latency.

\section{Related work}
\label{sec:related-work}

\subsection{Text representation models}

BERT established masked bidirectional pretraining \citep{devlin2019bert}, and Sentence-BERT, DPR, and ColBERT adapted encoder representations to dense and late-interaction retrieval \citep{reimers2019sentencebert,karpukhin2020dpr,khattab2020colbert}. ModernBERT, EuroBERT, and mmBERT update this family of models with longer contexts, newer Transformer components, and broader language coverage \citep{warner2024modernbert,boizard2025eurobert,marone2025mmbert}. Ettin compares masked encoders and causal decoders with matched model shapes, data order, and training recipes \citep{weller2026seqvsseq}. Its results are task-specific: the tested encoders are stronger on classification and retrieval, while the decoders are stronger on generation.

Causal decoder models can also be converted into encoders. LLM2Vec enables bidirectional attention before masked and contrastive adaptation \citep{behnamghader2024llm2vec}; LFM2.5-Encoder also converts its causal convolutions \citep{liquidai2026lfm25encoders}; and BidirLM extends this strategy across model scales and modalities \citep{boizard2026bidirlm}.

\subsection{Masked diffusion for representations}

Masked diffusion extends masked-language modeling from one corruption rate to a sampled noise trajectory. DiffusionBERT connected discrete diffusion to BERT-style denoising \citep{he2023diffusionbert}; MDLM and MD4 simplified the objective \citep{sahoo2024mdlm,shi2024md4}; and LLaDA demonstrated its large-scale generative use \citep{nie2025llada}. Diffusion-pretrained hidden states also support retrieval: DiffEmbed learns text embeddings \citep{zhang2025diffembed}, PPLX-Embed converts causal models into multilingual bidirectional encoders \citep{eslami2026diffusion}, and DiffRetriever reads multiple retrieval representations from masked positions \citep{wang2026diffretriever}.

Multimodal work spans several related settings. Masked Diffusion Captioning and LaViDa reconstruct text conditioned on separate image encoders \citep{feng2025masked,li2025lavida}. UniDisc and MMaDA move diffusion into a shared Transformer, but operate on externally tokenized images and primarily target generation \citep{swerdlow2025unidisc,yang2025mmada}. LaViDa and MMaDA have also been contrastively adapted into dense multimodal embedding models \citep{wang2026diffusionvlm}. Among the systems reviewed, none combines continuous raw patches, a vision-tower-free shared bidirectional Transformer trained from random initialization, image-conditioned masked-text diffusion, and subsequent dense and late-interaction visual-document retrieval.

\subsection{Vision--language architectures}

CLIP and SigLIP scale image and text retrieval through separate towers \citep{radford2021clip,zhai2023siglip}. This allows offline image indexing, but with limited interaction between modalities. Generative VLMs introduce deeper fusion while usually retaining a separate visual model. Flamingo inserts gated cross-attention between vision and language streams \citep{alayrac2022flamingo}; BLIP-2 uses a lightweight Querying Transformer to connect a frozen image encoder and language model \citep{li2023blip2}; and LLaVA, PaliGemma and Qwen2-VL project or merge vision-tower features into causal language models \citep{liu2023llava,beyer2024paligemma,wang2024qwen2vl}.

Other studies also developed architectures that share more of the model backbone among modalities. ViLT processes raw projected patches and text with the same bidirectional layers, but initializes them from a pretrained ViT \citep{kim2021vilt,dosovitskiy2021vit}. UFO and Uni-Perceiver reuse one Transformer across unimodal and multimodal tasks \citep{wang2021ufo,zhu2022uniperceiver}, while OneR and M3AE train shared representation backbones from scratch \citep{jang2022oner,geng2022m3ae}. VLMo, BEiT-3, and the masked-prediction EVE share attention while retaining modality-specific experts or visual targets \citep{bao2022vlmo,wang2023beit3,chen2024eve}. That EVE uses fixed-rate masked reconstruction and pretrained initialization, and BEiT-3 relies on a separate visual tokenizer.

Recent vision tower-free VLMs fed patches directly to decoder-oriented backbones. This line of work includes Fuyu, the encoder-free EVE (an unrelated model that shares the EVE name), SOLO, EVEv2, and NEO \citep{bavishi2023fuyu,diao2024eve,chen2024single,diao2025evev2,diao2025neo}. Chameleon instead uses discrete image codes \citep{chameleon2024}, while Gemma~4 Unified includes a from-scratch raw-patch decoder \citep{gemmateam2026gemma4}. These works establish the individual components of shared multimodal processing.

\subsection{Visual document representation and retrieval}
\label{sec:visual-document-retrieval}

Document encoders traditionally combine extracted words, layout coordinates, and image pixels. LayoutLMv3, for example, requires OCR tokens and boxes and uses a pretrained visual tokenizer for masked-image targets \citep{huang2022layoutlmv3}. Visual document retrieval can instead avoid OCR by treating each page as an image and embedding it directly. DSE produces dense page embeddings \citep{ma2024dse}, while ColPali and ColQwen2 preserve page-token representations for MaxSim retrieval \citep{faysse2025colpali}. In practice, visual document retrieval can be used for visual retrieval-augmented generation (Visual RAG): the top-$k$ retrieved pages are fed to a VLM along with a user query so it can answer from them. This keeps layout, tables, figures, and other evidence that text extraction may lose \citep{yu2024visrag,cho2024m3docrag,sun2025visrag2}.

Recent systems improve this recipe without removing inherited visual components. ModernVBERT combines a compact bidirectional encoder with a pretrained SigLIP2 tower \citep{teiletche2025modernvbert,tschannen2025siglip2}; Jina Embeddings v4 derives dense and multi-vector representations from Qwen2.5-VL \citep{gunther2025jina}; and Nemotron ColEmbed V2 scales the model size and embedding space dimension \citep{moreira2026nemotron}.

\subsection{Late-interaction capacity and efficiency}

Single-vector embeddings have a fixed representational capacity. Their dimension limits which top-$k$ document sets can be separated by a fixed score margin, and current embedding models exhibit related failures on the LIMIT benchmark \citep{weller2025limitations}. Multi-vector embeddings can have strictly greater capacity. Some relevance matrices require exponentially large single-vector embeddings but admit polynomial-size multi-vector embeddings, and models using multi-vector representations retain an advantage on the associated ANDOR benchmark after task-specific fine-tuning \citep{agarwal2026multivectors}. These mathematical constructions and synthetic text benchmarks do not establish that late-interaction outperforms dense retrieval on every task or that the same mechanism explains visual document retrieval.

Late-interaction can provide greater expressive capacity than dense retrieval, but it also increases storage and scoring costs: storage scales with the number of input tokens, and MaxSim scoring is more costly than cosine similarity. PLAID prunes candidates through centroids \citep{santhanam2022plaid}, while hierarchical token pooling and MUVERA reduce or transform stored multi-vector representations \citep{clavie2024tokenpooling,jayaram2025muvera}. FLASH-MAXSIM \citep{pony2026flashmaxsim} and MaxSim\footnote{\url{https://github.com/erikkaum/maxsim}} are kernels that fuse MaxSim without materializing the full token-similarity tensor.

\section{Architecture}
\label{sec:architecture}

\neomme{} is a multimodal encoder built around a single bidirectional Transformer optimized for long-context. Modality-specific input layers map text tokens and RGB image patches into a shared hidden space, where the encoder processes them jointly. \autoref{fig:input-paths} compares how dual-tower encoders, decoder-based visual language models, ModernVBERT, and \neomme{} route image and text through their architectures.

\subsection{Tokenizer}

\neomme{}'s tokenizer was trained from scratch for efficiency and multilingual coverage.

\noindent\textbf{Data}. The training mixture used to train \neomme{}'s tokenizer comprises multilingual web text, code, math, and machine-produced image transcripts. It draws English from \href{https://huggingface.co/datasets/HuggingFaceFW/fineweb-edu}{FineWeb-Edu}, 20 additional languages from \href{https://huggingface.co/datasets/epfml/FineWeb2-HQ}{FineWeb2-HQ}, mathematical text from \href{https://huggingface.co/datasets/HuggingFaceTB/finemath}{FineMath}, and 10 programming languages from \href{https://huggingface.co/datasets/bigcode/starcoderdata}{StarCoderData}. Its 131,072-entry vocabulary uses byte fallback, splits every digit, and reserves 64 identifiers for fixed special tokens.

\noindent\textbf{Methodology}. Motivated by the compression results reported for SuperBPE \citep{liu2025superbpe}, we trained a whitespace-unconstrained byte-level byte-pair encoding (BPE) tokenizer. Allowing merges to cross word boundaries enables the vocabulary to capture subwords, common multiword expressions, and formatting patterns such as code indentation. Tokens are limited to 48 bytes.

\noindent\textbf{Performance} Aggregated by total token count over 14 target languages from FLORES-200 devtest \citep{nllb2022}, \neomme{} emits 44.4\% fewer tokens than ModernBERT \citep{warner2024modernbert}, 39.4\% fewer than LFM2.5-Encoder-230M \citep{liquidai2026lfm25encoders}, 6.3\% fewer than mmBERT-base \citep{marone2025mmbert}, and 16.9\% fewer than EuroBERT-210m \citep{boizard2025eurobert}. However, the analysis over the 204 languages in FLORES-200 exposes a weaker coverage outside the original target set. \autoref{tab:tokenizer-metrics} and~\autoref{tab:tokenizer-metrics-b} in \autoref{app:tokenizer} report the per-language results, evaluation protocol, and comparison limitations.

\subsection{Shared multimodal inputs}

\subsubsection{Text tokens}

To minimize \neomme{}'s model size, text tokens use an ALBERT-style factorized embedding \citep{lan2020albert}: a 256-dimensional lookup followed by a linear projection to the model width. Let $E\in\mathbb{R}^{V\times d_e}$ be the token table and $P\in\mathbb{R}^{d\times d_e}$ the projection, where $V$ is the vocabulary size, $d_e$ the embedding rank, and $d$ the model width. For token $x_i$, the input path is
\begin{equation}
  \bm{h}^{(0)}_i=P E_{x_i,:}^{\top}.
  \label{eq:factorized-token-embedding}
\end{equation}

For final hidden state $\bm{h}_i$, the masked-token output path is
\begin{equation}
  \bm{\ell}_i=E P^{\top}\bm{h}_i.
  \label{eq:factorized-token-decoder}
\end{equation}
Because the tied, factorized masked-token decoder reuses both factors, it adds no output-specific parameters.

\subsubsection{Image patches}

Each input image is converted to RGB and partitioned into non-overlapping $32\times32$ patches. An image patch contains $3\times32\times32=3{,}072$ values. These values are then projected to the model input space dimension using layer normalization followed by a 2-layer MLP. The MLP is trained jointly from scratch; no patch-merging module or pretrained vision encoder is used. Structural tokens delimit inputs, image grids, and patch rows, while the segment offsets prevent attention across packed inputs.

At fixed resolution, 32-pixel patches produce approximately one quarter as many image tokens as 16-pixel patches. An earlier Gemma 4-inspired variant combined direct 48-pixel patches, a linear stem, and learned coordinate embeddings \citep{gemmateam2026gemma4}. In the 260M-scale experiments, this configuration appeared to weaken text-reading performance, possibly because each token had to compress a larger image region.

\begin{figure}[H]
  \centering
  \includegraphics[width=\linewidth]{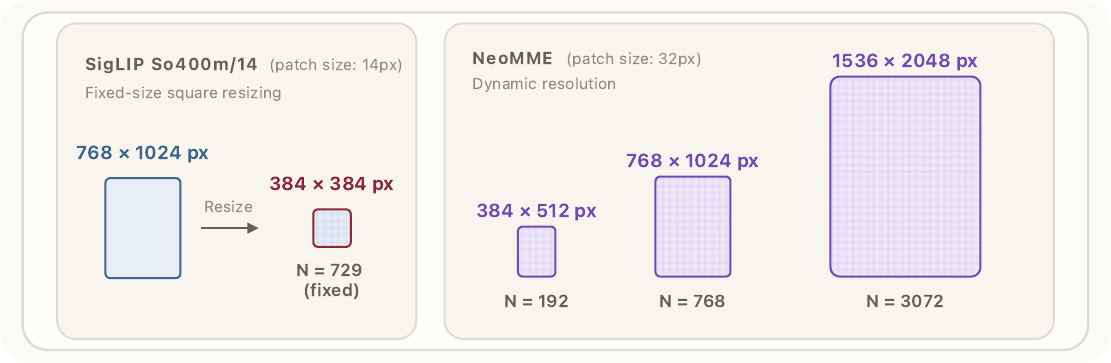}
  \caption{Dynamic-resolution image processing with a variable side-length cap.}
  \label{fig:dynamic-resolution}
\end{figure}

During pretraining and fine-tuning, the image pipeline randomly samples a longest-side cap between 1,024 and 2,048 pixels for each example. Varying the cap exposes the model to different image sizes and reduces overfitting to a fixed resolution. The pipeline preserves aspect ratio and downsamples only when the image exceeds the sampled cap. The resulting patch count varies with image dimensions and resolution, following dynamic-resolution vision--language models such as Qwen2-VL \citep{wang2024qwen2vl}. \autoref{fig:dynamic-resolution} illustrates this trade-off with dimensions that are illustrative rather than benchmark averages.

\subsubsection{2D rotary position embeddings}

\neomme{} extends rotary position embeddings (RoPE) \citep{su2024roformer} to two coordinate axes, following multimodal RoPE designs such as Qwen2-VL. Consecutive rotary frequency pairs alternate between the two axes. This construction preserves the usual one-dimensional ordering of text by assigning token $i$ the coordinate $(i,i)$, whereas image inputs use the axes to represent rows and columns.

For an image whose coordinate base is $b$, the input and image markers receive $(b,b)$ and $(b+1,b+1)$, respectively. A patch at row $r$ and column $c$ then receives $(b+2+r,b+2+c)$, while the row marker appended to each patch row occupies one additional grid column. Text following the image grid returns to diagonal coordinates, starting beyond both grid axes.

Global-attention layers use partial RoPE \citep{khan2026partialrope}, rotating 25\% of each query and key head as in Qwen3-Next \citep{qwen2025qwen3next}. Following the sliding-window--global RoPE split used in Gemma 4, global layers use base $10^6$, whereas sliding-window layers apply full RoPE with base $10^4$.

\subsection{Modern bidirectional backbone}

\begin{figure}[H]
  \centering
  \includegraphics[width=\linewidth]{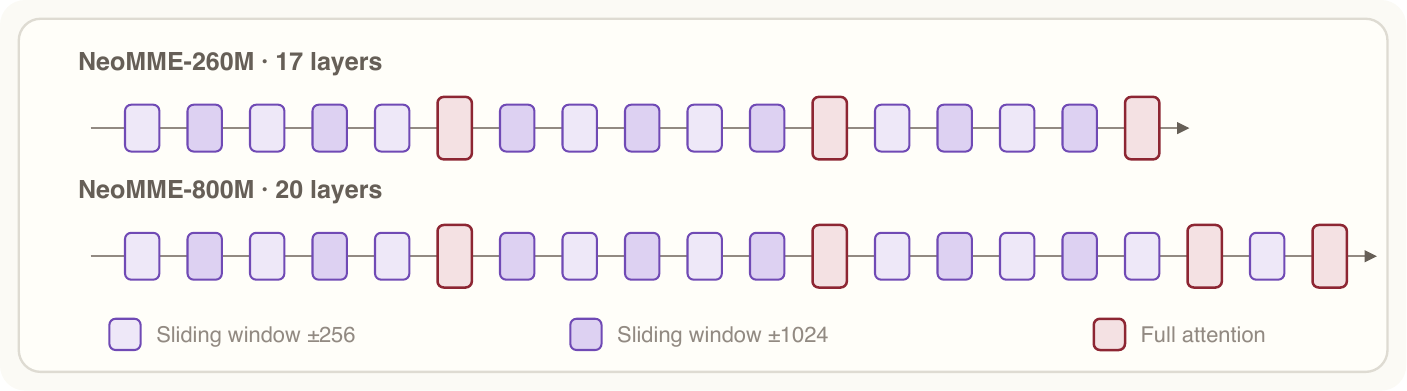}
  \caption{Alternating sliding-window and global-attention layers in the \neomme{} encoder stack.}
  \label{fig:encoder-stack}
\end{figure}

\noindent\textbf{Long-context attention}. The encoder uses bidirectional attention. Both 260M and 800M \neomme{} models support a maximum context length of 16,384 tokens, chosen to accommodate up to two standard $3{,}840\times2{,}160$ 4K UHD images after 32-pixel patching. To reduce attention computation at this limit, most layers use symmetric sliding-window attention, while every sixth layer and the final layer use global attention. Among the sliding-window layers, the half-window alternates between 256 and 1,024 tokens. This design combines Longformer's symmetric sliding-window attention \citep{beltagy2020longformer} with the interleaved sliding-window--global layouts of ModernBERT and Gemma 2 \citep{gemmateam2024gemma2}; the short--long schedule follows modded-nanogpt \citep{jordan2026moddednanogpt}, although the specific 256- and 1,024-token half-windows are \neomme{} design choices. \autoref{fig:encoder-stack} shows the layer sequence for both model sizes.

\noindent\textbf{Attention heads and query--key normalization}. Both sliding-window and global layers use grouped-query attention (GQA) \citep{ainslie2023gqa}. The 260M model has 16 query heads and 4 key-value heads, whereas the 800M model has 28 query heads and 7 key-value heads. Queries and keys are independently root-mean-square normalized before RoPE and, consequently, before the attention dot product, implementing query--key (QK) normalization \citep{dehghani2023vit22b}.

\Needspace*{0.56\textheight}
\begin{wrapfigure}{r}{0.38\textwidth}
  \vspace{-0.6em}
  \centering
  \includegraphics[width=\linewidth]{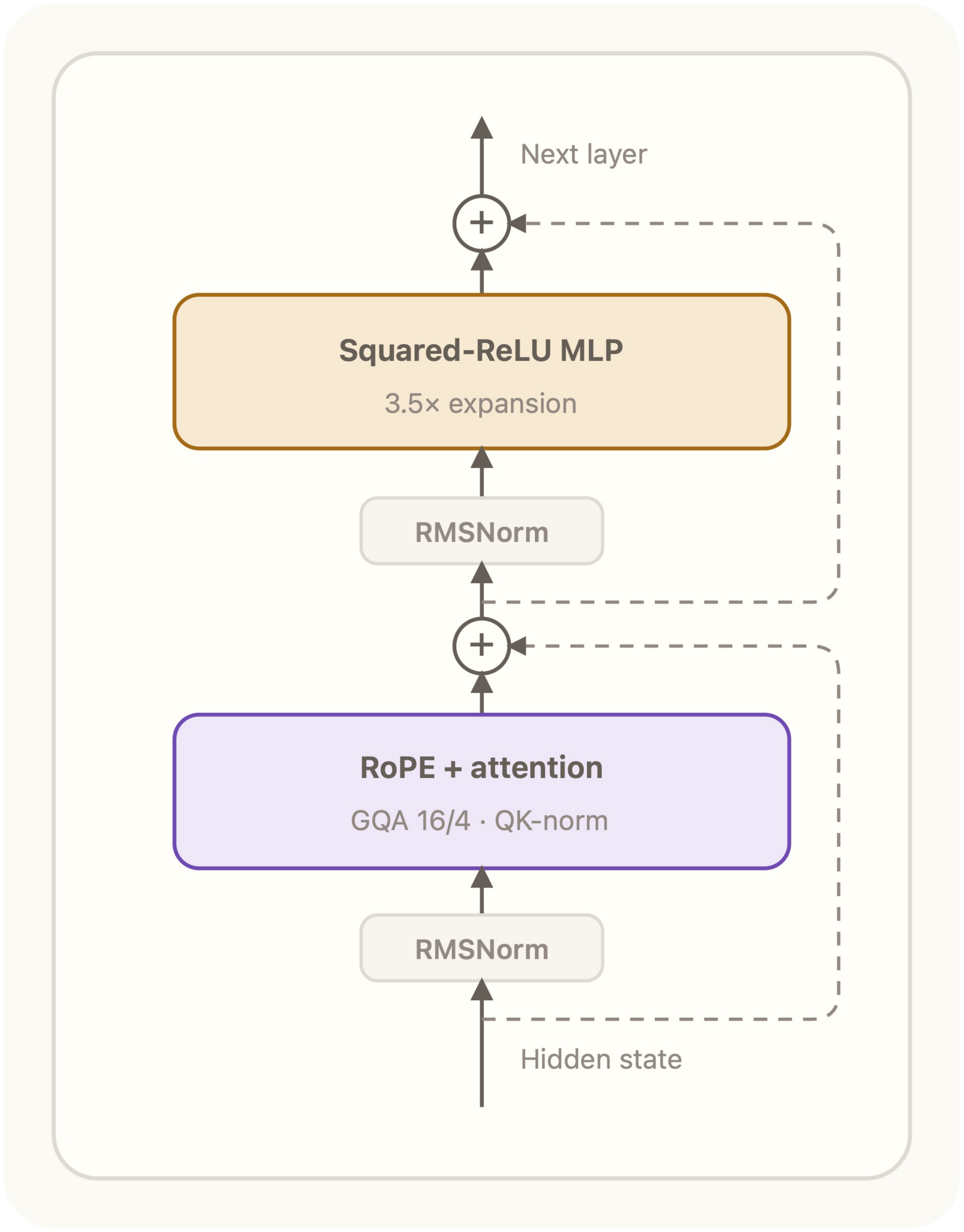}
  \caption{Pre-normalized attention and MLP paths in one \neomme{} encoder layer.}
  \label{fig:encoder-layer}
  \vspace{-0.5em}
\end{wrapfigure}

\noindent\textbf{Block structure}. Each block uses parameter-free root-mean-square pre-normalization \citep{zhang2019rmsnorm} and a squared-ReLU MLP \citep{so2021primer}. Its attention output is further modulated by a query-dependent elementwise sigmoid gate \citep{qiu2025gatedattention}. Meanwhile, each attention and MLP residual branch is multiplied by $(2L)^{-1/2}$, counting the two branches in each of the $L$ layers separately. This is the branch-scaling component of Depth-$\mu$P \citep{yang2024depthmup,bordelon2023depthwise}, although \neomme{} does not otherwise claim to implement the complete Depth-$\mu$P parameterization. The same parameter-free root-mean-square normalization closes the encoder stack. \autoref{fig:encoder-layer} summarizes the normalization, attention, MLP, and residual paths.

\noindent\textbf{Token-indexed value embeddings}. At the first and last global-attention layers, a shared embedding table indexed by the input token identifiers is added to the value vectors. This design is inspired by ResFormer's value-residual learning \citep{zhou2025valueresidual} and follows the token-indexed value-embedding implementation introduced in modded-nanogpt; unlike ResFormer, however, it learns a dedicated table rather than reusing values projected in an earlier layer.

\noindent\textbf{Residual mixing}. Following the embedding shortcut used in modded-nanogpt, each block forms a learned scalar mixture of the current residual stream and the initial normalized multimodal input before its attention mixer. \neomme{} additionally uses a learnable variant of exclusive self-attention \citep{zhai2026xsa}, scaling each head's value-aligned subtraction by a separate coefficient.

\noindent\textbf{Model sizes}. To keep \neomme{} as efficient as possible, we chose model sizes of 260M and 800M parameters. ModernVBERT showed that visual document retrieval was possible with 250M parameters, close to the size of our smaller model \citep{teiletche2025modernvbert}. \neomme{} is released as \neomme-260M\textsuperscript{\ref{fn:neomme-260m-hf}} and \neomme-800M\textsuperscript{\ref{fn:neomme-800m-hf}}. The 800M configuration scales mainly through width, keeping the backbone to 20 layers and 4 global-attention layers at the 16,384-token context length. \autoref{tab:architecture-settings} lists the architecture settings, and \autoref{fig:parameter-allocation} shows the parameter allocation by module group.

\begin{table}[t]
  \centering
  \begin{minipage}[t]{0.43\textwidth}
    \vspace{0pt}
    \centering
    \captionsetup{justification=raggedright,singlelinecheck=false}
    \captionof{table}{\neomme{} architecture settings.}
    \label{tab:architecture-settings}
    \scriptsize
    \setlength{\tabcolsep}{3pt}
    \begin{tabular}{lrr}
      \toprule
      Configuration &
      \neomme-260M\tablefootnote{\label{fn:neomme-260m-hf}\hflink{https://huggingface.co/Hcompany/NeoMME-260M}{Hcompany/NeoMME-260M}.} &
      \neomme-800M\tablefootnote{\label{fn:neomme-800m-hf}\hflink{https://huggingface.co/Hcompany/NeoMME-800M}{Hcompany/NeoMME-800M}.} \\
      \midrule
      Exact parameters & 262,937,906 & 793,715,032 \\
      Embedding rank & 256 & 256 \\
      Hidden width & 1,024 & 1,792 \\
      Encoder layers & 17 & 20 \\
      Global layers & 3 & 4 \\
      Query heads & 16 & 28 \\
      Key-value heads & 4 & 7 \\
      Head dimension & 64 & 64 \\
      MLP width & 3,584 & 6,400 \\
      Context length & 16,384 & 16,384 \\
      Vocabulary & 131,072 & 131,072 \\
      Patch size & 32 & 32 \\
      Patch stem & MLP & MLP \\
      \bottomrule
    \end{tabular}
  \end{minipage}
  \hfill
  \begin{minipage}[t]{0.55\textwidth}
    \vspace{0pt}
    \centering
    \captionsetup{type=figure,justification=raggedright,singlelinecheck=false}
    \includegraphics[width=\linewidth]{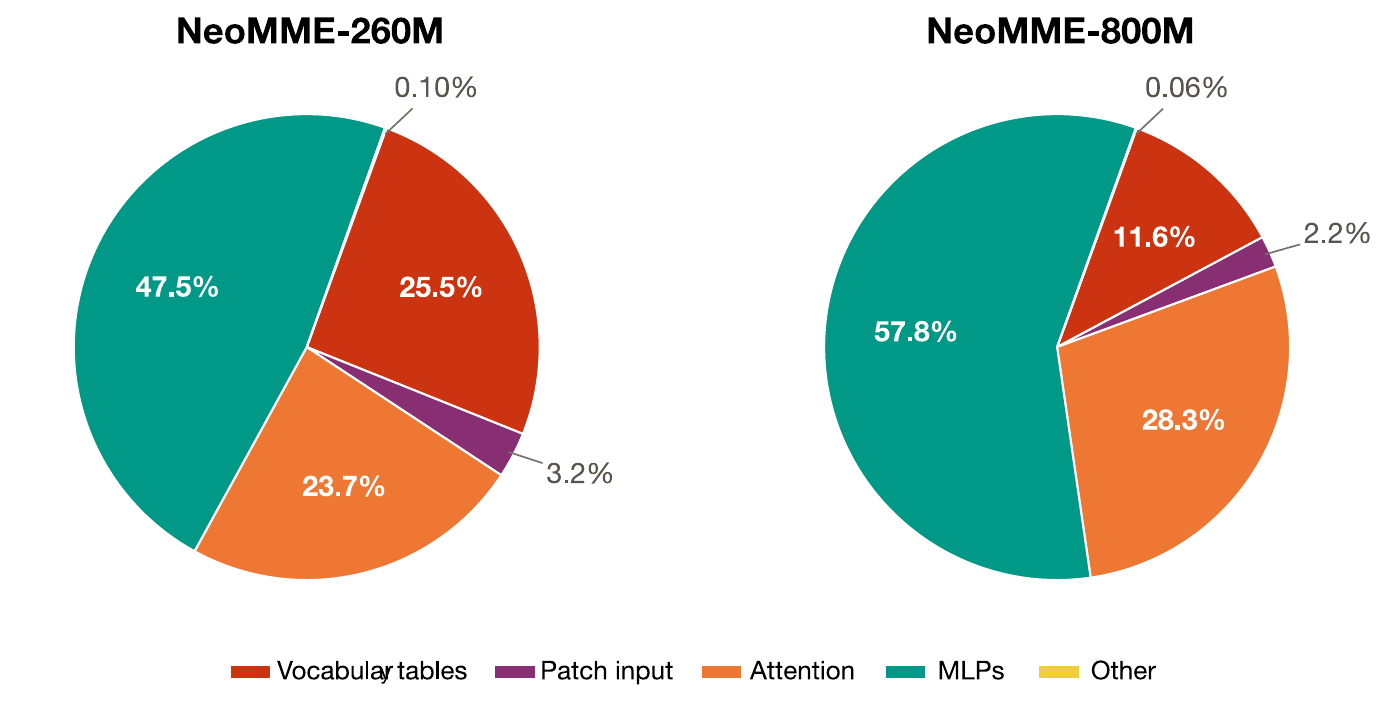}
    \captionof{figure}{Parameter allocation by module group for \neomme-260M and \neomme-800M. Counts include the backbone and input paths but exclude the contrastive-retrieval heads.}
    \label{fig:parameter-allocation}
  \end{minipage}
\end{table}

\subsection{Weight initialization}

\neomme{} initializes the attention output and MLP down projections to zero. This applies the broader zero-last-layer principle introduced by Fixup \citep{zhang2019fixup}, making both residual branches exact no-ops and each block an identity at initialization, without otherwise adopting the complete Fixup parameterization.

The factorized word table is drawn from a zero-mean normal distribution with standard deviation $d_e^{-1/2}$, where $d_e=256$. The same dimension-dependent embedding scale appears in T-Fixup \citep{huang2020tfixup}; in \neomme{}, it keeps the logits produced by the tied factorized decoder in \autoref{eq:factorized-token-decoder} at order-one scale. The token-indexed value table and per-head exclusive-self-attention coefficients are initialized to zero, while the input-shortcut mixture begins with coefficients $(1,0)$ on the current and initial residual streams, respectively. All remaining backbone linear layers use the default \href{https://docs.pytorch.org/docs/2.8/generated/torch.nn.Linear.html}{PyTorch 2.8 \code{torch.nn.Linear} initialization}.

\section{Pretraining}
\label{sec:pretraining}

The pretraining phase trains the shared backbone from random initialization on text-only and multimodal examples, including document images and natural images paired with text. Its data pipeline and masking objective reduce text-only shortcuts and encourage the model to use image evidence.

\subsection{Dataset mixture}

Pretraining samples come from separate text-only and multimodal streams. The text stream uses 14 datasets across 58 configurations, spanning web and PDF text, synthetic educational material, encyclopedic text, mathematics, question answering, and code. The multimodal stream combines mixed visual--text collections, document content, OCR data, and natural images. Sources are sampled according to the weights defined in \autoref{tab:p1-mixture}, with 55\% of packed input tokens coming from the text-only stream and 45\% from the multimodal stream. The planned total is about 524B packed input tokens. \autoref{app:p1-data-mixture} reports every configuration, its sampling weight, and the source descriptions.

\begin{table*}[t]
  \centering
  \caption{Configured pretraining mixture by logical source family.}
  \label{tab:p1-mixture}
  \small
  \setlength{\tabcolsep}{5pt}
  \begin{tabular}{p{0.17\textwidth}p{0.38\textwidth}>{\raggedleft\arraybackslash}p{0.17\textwidth}>{\raggedleft\arraybackslash}p{0.17\textwidth}}
    \toprule
    Group & Dataset & Within-stream weight & Expected tokens \\
    \midrule
    \multicolumn{4}{l}{\textbf{\faFont\ Text only: 55\% of packed input tokens}} \\
    \multirow{4}{0.17\textwidth}{Web and PDF}
      & \hflink{https://huggingface.co/datasets/HuggingFaceFW/fineweb-edu}{FineWeb-Edu} & 0.167660 & 48.346B \\
      & \hflink{https://huggingface.co/datasets/nvidia/Nemotron-CC-v2.1}{Nemotron-CC-v2.1} & 0.072896 & 21.020B \\
      & \hflink{https://huggingface.co/datasets/HuggingFaceFW/finepdfs}{FinePDFs} & 0.131214 & 37.837B \\
      & \hflink{https://huggingface.co/datasets/epfml/FineWeb2-HQ}{FineWeb2-HQ} & 0.349900 & 100.897B \\
    \addlinespace
    \multirow{3}{0.17\textwidth}{Synthetic, reference, and math}
      & \hflink{https://huggingface.co/datasets/HuggingFaceTB/smollm-corpus}{Cosmopedia-v2} & 0.102054 & 29.428B \\
      & \hflink{https://huggingface.co/datasets/wikimedia/wikipedia}{Wikipedia} & 0.087479 & 25.225B \\
      & \hflink{https://huggingface.co/datasets/HuggingFaceTB/finemath}{FineMath} & 0.051027 & 14.714B \\
    \addlinespace
    \multirow{6}{0.17\textwidth}{Question answering}
      & \hflink{https://huggingface.co/datasets/cais/mmlu}{MMLU} & 0.006196 & 1.787B \\
      & \hflink{https://huggingface.co/datasets/openlifescienceai/medmcqa}{MedMCQA} & 0.004009 & 1.156B \\
      & \hflink{https://huggingface.co/datasets/tau/commonsense_qa}{CommonsenseQA} & 0.000219 & 0.063B \\
      & \hflink{https://huggingface.co/datasets/allenai/qasc}{QASC} & 0.000182 & 0.052B \\
      & \hflink{https://huggingface.co/datasets/allenai/openbookqa}{OpenBookQA} & 0.000109 & 0.031B \\
      & \hflink{https://huggingface.co/datasets/allenai/ai2_arc}{AI2 ARC} & 0.000087 & 0.025B \\
    \addlinespace
    Code
      & \hflink{https://huggingface.co/datasets/bigcode/starcoderdata}{StarCoderData} & 0.026972 & 7.778B \\
    \cmidrule(lr){2-4}
      & \textbf{Text-only total} & \textbf{1.000000} & \textbf{288.358B} \\
    \midrule
    \multicolumn{4}{l}{\textbf{\faImage\ Multimodal: 45\% of packed input tokens}} \\
    Mixed collection
      & \hflink{https://huggingface.co/datasets/HuggingFaceM4/FineVision}{FineVision} & 0.500000 & 117.965B \\
    \multirow{3}{0.17\textwidth}{Document and OCR}
      & \hflink{https://huggingface.co/datasets/pixparse/pdfa-eng-wds}{PDFA} + \hflink{https://huggingface.co/datasets/lightonai/LightOnOCR-mix-0126}{LightOnOCR} & 0.300000 & 70.779B \\
      & \hflink{https://huggingface.co/datasets/ahmedheakl/docatlas_instruct}{DocAtlas} & 0.070000 & 16.515B \\
      & \hflink{https://huggingface.co/datasets/nvidia/OCR-Synthetic-Multilingual-v1}{Synthetic multilingual OCR} & 0.030000 & 7.078B \\
    Natural images
      & \hflink{https://huggingface.co/datasets/tomg-group-umd/pixelprose}{PixelProse} & 0.100000 & 23.593B \\
    \cmidrule(lr){2-4}
      & \textbf{Multimodal total} & \textbf{1.000000} & \textbf{235.930B} \\
    \bottomrule
  \end{tabular}
\end{table*}

\subsection{Data pipeline and systems optimization}
At this model scale, input processing can take longer than the forward and backward passes, making it the main bottleneck. We therefore pack documents into 16,384-position streams while keeping them isolated through variable-length attention boundaries. Images are decoded and patchified in background workers, and packed batches are prefetched during the preceding training step. A compute-aware schedule balances multimodal work across data-parallel ranks, while shape-stable batch construction supports compiled training with dynamic image resolution. With FlashAttention-3 \citep{shah2024flashattention3}, Liger Kernel's fused linear cross-entropy \citep{hsu2025ligerkernel}, and \code{torch.compile}, these optimizations keep GPU utilization high during steady-state training.

\subsection{Pretraining objective}

\neomme{} was pretrained as a discrete masked-diffusion denoiser over text, optionally conditioned on visible image patches. Its objective follows the general absorbing-mask training paradigm exemplified by MDLM, MD4, and LLaDA \citep{sahoo2024mdlm,shi2024md4,nie2025llada}.

Let $s$ index a real document segment in a packed sequence. Text-only segments draw a corruption rate $\rho_s\sim\mathcal{U}(0,1)$, whereas multimodal segments draw $\rho_s\sim\mathcal{U}(0.30,1)$. The higher minimum corruption rate reduces reliance on visible textual context and encourages predictions to use the image patches. Given $\rho_s$, each eligible text position is independently replaced by the mask token with probability $\rho_s$. Eligible positions exclude padding, image patches, and structural markers. Unlike BERT, which uses a fixed 15\% selection rate and an 80--10--10 mixture of mask tokens, random tokens, and unchanged tokens \citep{devlin2019bert}, \neomme{} always replaces selected positions with the mask token and computes the loss only at those positions.

Image patches remain visible and condition the prediction of masked text. Neither masked image modeling nor a pixel-prediction head is used, so no pixel-reconstruction, latent-image, or generative-image target contributes to the loss. The same text denoiser can be used at inference time by starting from masked text and iteratively predicting and revealing tokens, optionally conditioned on the visible image patches.

For data-parallel rank $j$, let $\mathcal{M}_j$ contain the masked eligible positions in the packed batch assigned to that rank. The loss for rank $j$ is
\begin{equation}
  \mathcal{L}_j =
  \frac{
    \sum_{i\in\mathcal{M}_j}
    w_i\,\operatorname{CE}
    \left(f_\theta(\tilde{\bm{x}}_j,\bm{p}_j)_i,x_i\right)
  }{
    \sum_{i\in\mathcal{M}_j} w_i
  },
  \qquad
  w_i=\frac{1}{\max(r_{s(i)},r_0)},
  \qquad
  r_0=0.05,
  \label{eq:p1-rank-loss}
\end{equation}
where $s(i)$ is the segment containing position $i$, $\tilde{\bm{x}}_j$ is the corrupted token sequence, $\bm{p}_j$ denotes the image patches when present, and $x_i$ is the original token. We set $\mathcal{L}_j=0$ when $\mathcal{M}_j$ is empty. With $R$ data-parallel ranks, gradient averaging optimizes
\begin{equation}
  \mathcal{L}_{\mathrm{pretrain}}
  = \frac{1}{R}\sum_{j=1}^{R}\mathcal{L}_j .
  \label{eq:pretrain-loss}
\end{equation}

The inverse-rate weighting follows the standard linear absorbing-mask formulation. In practice, \neomme{} caps the reciprocal weight at 20 and normalizes by the realized weight mass on each rank. This limits the influence of rare low-corruption samples and maintains a stable loss scale during packed distributed training.

\subsection{Training settings}

\noindent\textbf{Pretraining configuration}. \autoref{tab:p1-training} summarizes the pretraining hardware, batch, and optimization settings for both 260M and 800M models. The two pretraining runs use AWS \code{p5.48xlarge}\footnote{\label{fn:aws-p5}\url{https://aws.amazon.com/ec2/instance-types/p5/}} instances with Elastic Fabric Adapter (EFA)\footnote{\label{fn:aws-efa}\url{https://aws.amazon.com/hpc/efa/}} networking. We used SkyPilot \citep{yang2023skypilot}\footnote{\url{https://skypilot.ai/}} to launch and manage these AWS training jobs. Each run processes 1,048,576 packed input tokens per global step for 500,000 steps, for a planned total of 524.288 billion packed input tokens.

\begin{table}[t]
  \centering
  \caption{Pretraining hardware, batch, and optimization settings.}
  \label{tab:p1-training}
  \small
  \begin{tabular}{p{0.45\linewidth}rr}
    \toprule
    Setting & \neomme-260M & \neomme-800M \\
    \midrule
    \multicolumn{3}{l}{\textbf{Hardware and batch}} \\
    Nodes & 2 & 4 \\
    H100 accelerators & 16 & 32 \\
    Packed sequences per GPU & 4 & 2 \\
    Multimodal sequence share & 45\% & 45\% \\
    Compute precision & bfloat16 & bfloat16 \\
    Packed input tokens per global step & 1,048,576 & 1,048,576 \\
    Patch cap per rank & 57,344 & 32,768 \\
    \addlinespace
    \multicolumn{3}{l}{\textbf{Optimization}} \\
    NorMuon peak learning rate & 0.012 & 0.010 \\
    AdamW peak learning rate & 0.0013 & 0.00075 \\
    Warmup steps & 300 & 300 \\
    Final decay fraction & 0.10 & 0.10 \\
    Final learning-rate floor & 1\% of peak & 1\% of peak \\
    Gradient clipping norm & 1.0 & 1.0 \\
    Cautious-decay coefficient & 0.01 & 0.01 \\
    Embedding weight decay & 0.01 & 0.01 \\
    One-dimensional weight decay & 0 & 0 \\
    \bottomrule
  \end{tabular}
\end{table}

\noindent\textbf{Optimizers}. \neomme{} routes parameters according to module type and tensor rank. Embedding tables use a custom MasterAdamW implementation, while all other matrix-valued parameters use NorMuon \citep{li2026normuon}, which augments Muon's orthogonalized matrix updates \citep{jordan2024muon} with per-neuron second-moment normalization. The latter group includes the factorized embedding projection, attention projections, and MLP matrices. One-dimensional parameters use MasterAdamW without weight decay. Both optimizers retain full-precision master weights and moments while the model parameters remain in bfloat16. NorMuon applies cautious weight decay, restricting decay to coordinates where the parameter and optimizer-update signs agree \citep{chen2025cautiousweightdecay}. MasterAdamW applies decoupled weight decay to embedding tables \citep{loshchilov2019adamw}.

To reduce optimizer overhead, same-shaped NorMuon matrices are processed in batches and their updates are sharded across GPUs within each node. Since gradients are synchronized globally, nodes can repeat this computation independently without an additional inter-node optimizer collective.

\noindent\textbf{Learning-rate scheduler}. We use a warmup--stable--decay (WSD) learning-rate schedule \citep{hu2024wsd}. Both learning rates increase linearly over 300 updates, remain at their peaks, and decay linearly over the final 10\% of training to 1\% of their peak values. We release the predecay checkpoints of \neomme-260M\footnote{\hflink{https://huggingface.co/Hcompany/NeoMME-260M-Pretrain-predecay-s450000}{Hcompany/NeoMME-260M-Pretrain-predecay-s450000}.} and \neomme-800M\footnote{\hflink{https://huggingface.co/Hcompany/NeoMME-800M-Pretrain-predecay-s450000}{Hcompany/NeoMME-800M-Pretrain-predecay-s450000}.} at the stable-to-decay boundary, following Pythia and ModernBERT \citep{biderman2023pythia,warner2024modernbert}. Researchers can restart training from these checkpoints and anneal on domain-appropriate data for their intended use. The 260M and 800M contrastive runs, however, initialize from the step-500,000 checkpoints rather than from the predecay ones.

\subsection{Text masking and image sensitivity}

Low corruption rates can leave many masked targets predictable from text alone, a known limitation of multimodal masked-language modeling \citep{bitton2021dataefficient}. To weaken this shortcut, multimodal segments draw their corruption rates from $\mathcal{U}(0.30,1)$. Prior work similarly finds that vision--language pretraining benefits from substantially higher text masking rates \citep{verma2022uniform}. With higher corruption rates, the model receives less textual context, so image evidence can contribute more.

We measure this behavior using a cross-modal input-ablation probe \citep{frank2021crossmodal}. At each probe event $t$, we mask 30\%, 60\%, or 90\% of eligible text positions and denote the masked positions by $\mathcal{M}_t$. We run the model twice on the same masked text, first with the page patches $I_t$ and then with a zero tensor $\bm{0}$ of the same shape.

For either patch input $Z$, let $p_\theta(v\mid\widetilde{\bm{x}}_t,Z)_i$ be the predicted probability of vocabulary token $v$ at masked position $i$, where $\widetilde{\bm{x}}_t$ is the masked text and $x_{t,i}$ is the original token. We define masked-token accuracy $A_t(Z)$, event-level image gain $G_t$, and mean image gain $\overline{G}$ as
\begin{equation}
  A_t(Z)
  =
  \frac{1}{|\mathcal{M}_t|}
  \sum_{i\in\mathcal{M}_t}
  \mathbb{1}\!\left[
    \arg\max_{v\in\mathcal{V}}
    p_\theta(v\mid\widetilde{\bm{x}}_t,Z)_i
    =
    x_{t,i}
  \right],
  \qquad
  G_t=A_t(I_t)-A_t(\bm{0}),
  \qquad
  \overline{G}=\frac{1}{T}\sum_{t=1}^{T}G_t.
  \label{eq:image-gain}
\end{equation}
The logged value $G_t$ compares exact token recovery for one probe event, and $\overline{G}$ averages those event-level differences over $T$ probe events. A positive $G_t$ means that visible page patches improve recovery of the masked text.

Over the final 50 probe events, both models have positive mean image gain at every tested corruption rate. Image gain increases as textual context is removed, reaching 38.4 percentage points for \neomme-260M and 40.5 percentage points for \neomme-800M at 90\% masking. Positive image gain shows that the models use page patches to improve masked-token recovery. We examine image-conditioned generation separately in \autoref{sec:p1-generation}.

\begin{figure*}[t]
  \centering
  \includegraphics[width=\textwidth]{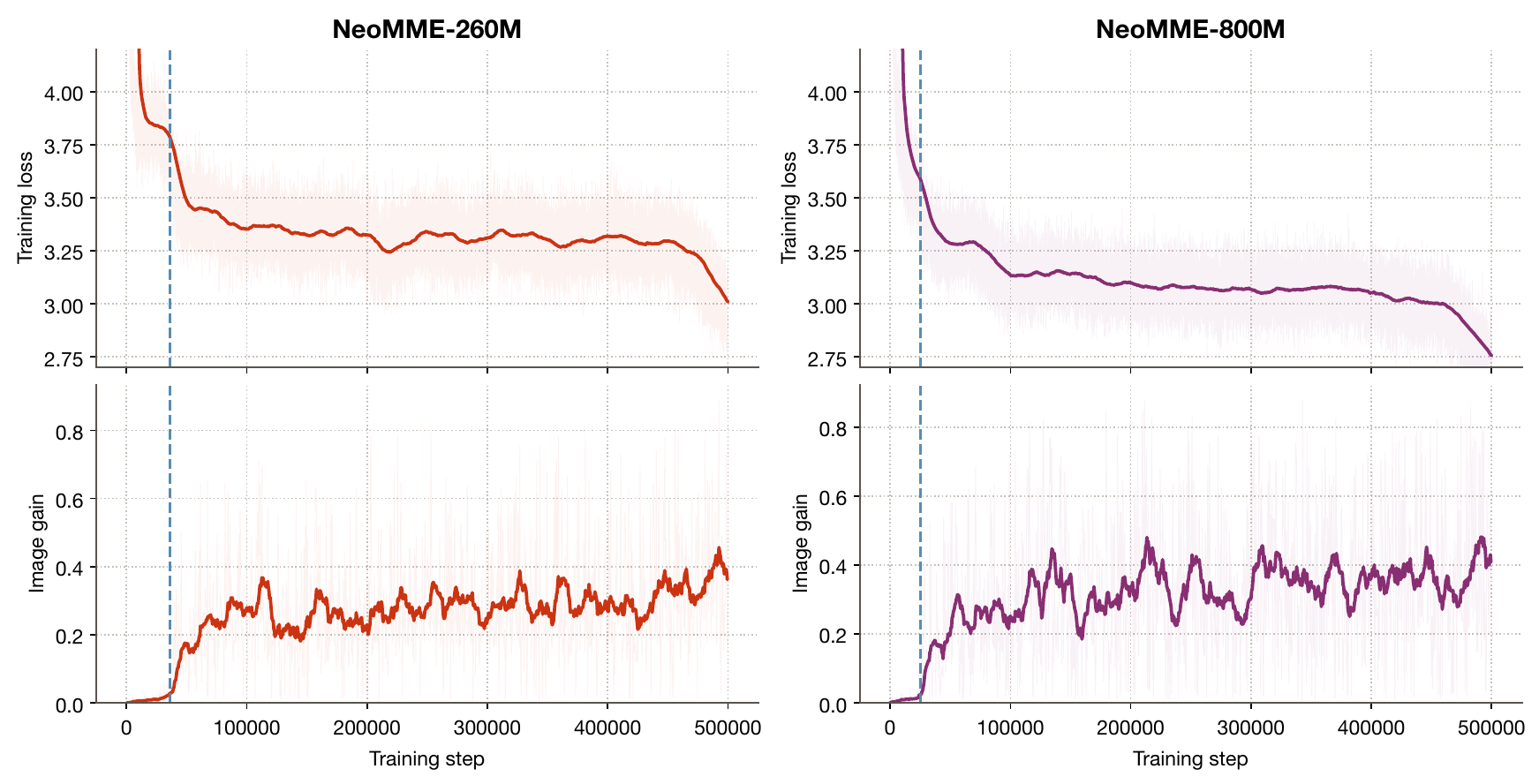}
  \caption{Pretraining loss and image gain at 90\% masking for both model sizes. Lines show raw and smoothed measurements, with dashed lines marking observed loss transitions.}
  \label{fig:p1-transitions}
\end{figure*}

\autoref{fig:p1-transitions} shows the training loss and 90\% image gain for both models. The \neomme-260M loss enters a lower regime around step 36,000, while the \neomme-800M loss has a visible change around step 25,000. Image gain rises from near zero during early training for both models. The shared training-step axis shows when the observed loss transitions occur relative to when image gain appears. The final loss decline coincides with the scheduled WSD learning-rate decay to 1\% of its peak. Such a decline is expected during WSD annealing.

\Needspace*{0.35\textheight}
\subsection{Image-conditioned generation}
\label{sec:p1-generation}

\begin{wrapfigure}{r}{0.42\textwidth}
  \vspace{-8pt}
  \centering
  \includegraphics[width=\linewidth]{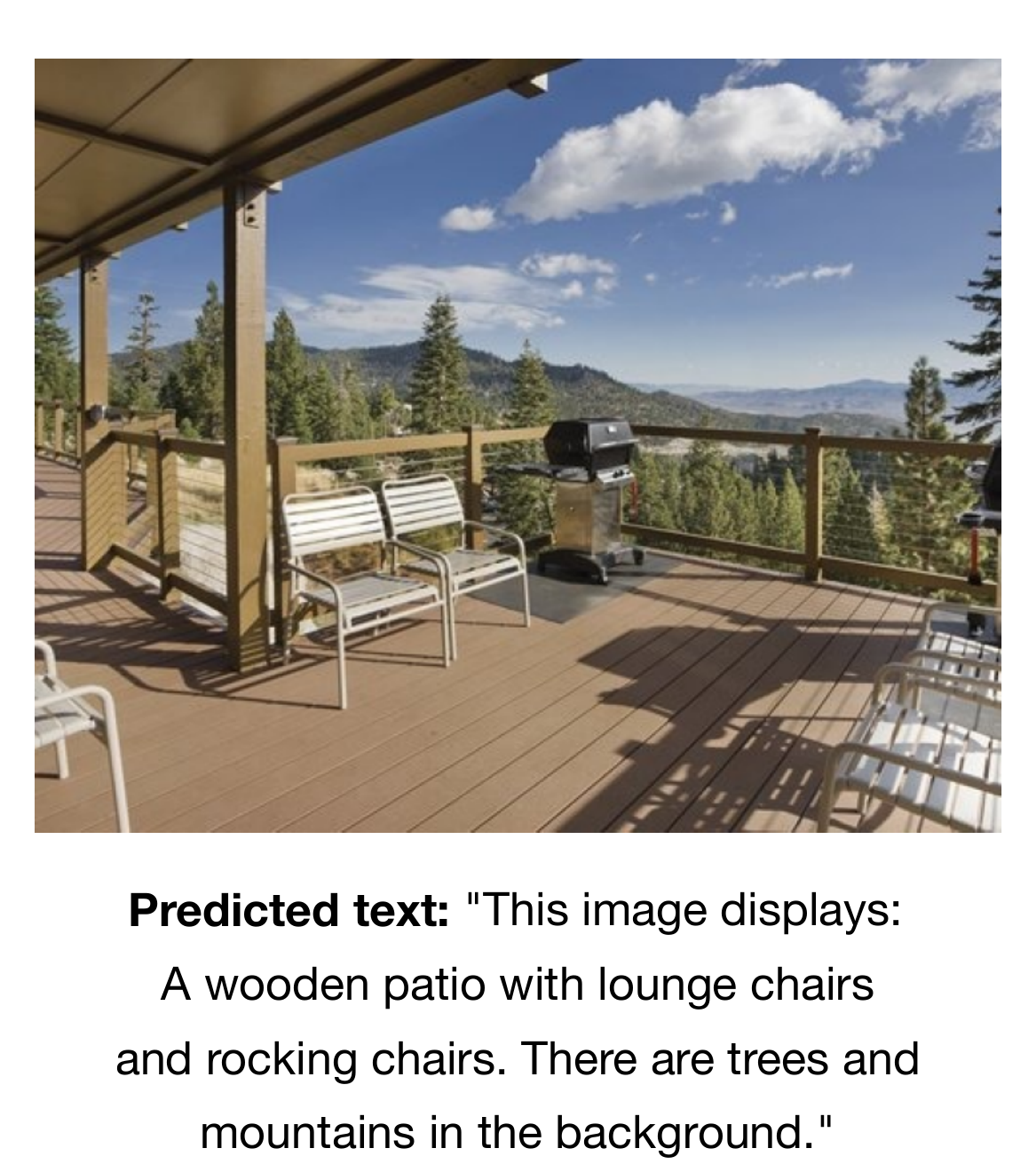}
  \caption{Natural-image input and generated text from the step-500,000 pretrained \neomme-260M model.}
  \label{fig:p1-generation}
  \vspace{-8pt}
\end{wrapfigure}

We tested the step-500,000 pretrained \neomme-260M\textsuperscript{\ref{fn:neomme-260m-hf}}\allowbreak{} model on unprompted image-conditioned text generation without captioning or optical character recognition fine-tuning. Starting from visible image patches and a fully masked text canvas, the model iteratively predicts and reveals tokens through the tied vocabulary head used during pretraining.

We tested uniform and blockwise LLaDA-style decoding. We paired cosine remaining-mask schedules with MaskGIT confidence ranking \citep{chang2022maskgit}, using either greedy or nucleus sampling \citep{holtzman2020nucleus}. The examples shown here and in the appendix use fully masked, unprompted runs. The canvas matches the reference length up to 48 tokens for captions and 64 tokens for document text, so the model receives the output length but no reference tokens. \autoref{fig:p1-generation} shows one natural-image caption, with additional natural-image and document examples in \autoref{tab:generation-examples} in \autoref{app:generation-examples}. Together with the reading probe, the selected outputs show that the model uses image content, but they do not measure average generation quality.

\section{Retrieval}
\label{sec:retrieval}

To evaluate the representations learned from the pretrained \neomme{} encoder backbone, we fine-tune \neomme{} for document retrieval with joint late-interaction and dense objectives. We then evaluate the resulting \neomme{}-Retriever 260M and 800M models on both visual document and text retrieval benchmarks, and compare document-indexing throughput and query-encoding latency against other retrievers.

\subsection{Problem formulation}

\noindent\textbf{Retrieval.} A retrieval system scores how relevant a document $d$ from a corpus $\mathcal{C}=\{d_1,\ldots,d_{N_{\mathcal C}}\}$ is to a query $q$ from a query space $\mathcal{Q}$. Computing the similarity score $s(q,d)\in\mathbb{R}$ for every $d\in\mathcal{C}$ produces a ranking from which the system returns the most relevant documents. A document is an atomic item that the system can index and return. In text information retrieval, text is usually split into chunks \citep{karpukhin2020dpr}. On the other hand, visual document retrieval use screenshots of a PDF page as the retrieval unit \citep{ma2024dse,faysse2025colpali}.

\noindent\textbf{Indexing and querying.} Retrieval systems generally have an offline indexing phase and an online query phase. The system encodes and indexes corpus documents offline. At query time, it encodes $q$, scores the query against the indexed documents, and returns a ranking. The offline representation must therefore preserve evidence that a query can use without another document-encoder forward pass.

\begin{figure*}[t]
  \centering
  \includegraphics[width=\textwidth]{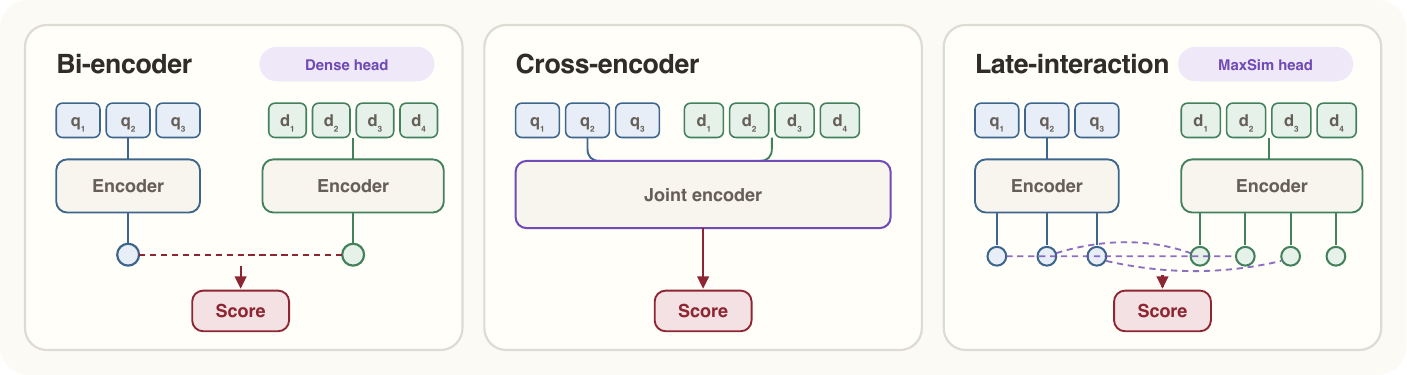}
  \caption{Dense, cross-encoder, and late-interaction retrieval scoring patterns.}
  \label{fig:retrieval-scoring}
\end{figure*}

\noindent\textbf{Retrieval architectures.} We compare three scoring architectures that differ in when query and document representations interact and whether document representations can be indexed offline, as shown in \autoref{fig:retrieval-scoring}.

\noindent\textbf{Dense bi-encoder.} A dense bi-encoder maps $q$ and $d$ independently to normalized vectors $\bm{e}_q,\bm{e}_d\in\mathbb{R}^m$ and scores their dot product \citep{reimers2019sentencebert}:
\begin{equation}
  s_{\mathrm{dense}}(q,d)=\langle\bm{e}_q,\bm{e}_d\rangle.
  \label{eq:dense-score}
\end{equation}
The document vector can be computed offline and reused. Its one-vector form also allows approximate nearest-neighbor (ANN) indexes to search large corpora (1B+ documents) at high speeds (down to $13.3 \mu s/\text{query}$ with Faiss) \citep{indyk1998approximate,jegou2011product,johnson2019billion}.

\noindent\textbf{Cross-encoder.} A cross-encoder processes every query--document pair jointly, allowing full token interaction and often strong relevance scoring, but requiring a separate forward pass for each pair \citep{nogueira2019passage}. Cross-encoders therefore cannot produce reusable document representations for offline indexing and are generally used to rerank a small document set returned by a faster first-stage retriever. The same multistage design includes encoder--decoder rerankers such as monoT5 \citep{nogueira2020document}. Neural reranking now also covers multimodal retrieval with models that rank both text documents and page images \citep{ananya2026lightonrerank}.

\noindent\textbf{Late-interaction.} Late-interaction encodes $q$ and $d$ independently as normalized token-vector matrices $\bm{Q}=[\bm{q}_1,\ldots,\bm{q}_{L_q}]^\top \in\mathbb{R}^{L_q\times r}$ and $\bm{D}=[\bm{d}_1,\ldots,\bm{d}_{L_d}]^\top \in\mathbb{R}^{L_d\times r}$, where $r=128$ is the representation width. Retaining one vector per token preserves token-level matches that dense pooling removes. Standard ANN indexes directly support the one-vector dense score, while the set-to-set late-interaction score requires direct scoring or a specialized multi-vector retrieval method. Each query vector $\bm{q}_s\in\mathbb{R}^r$ for $s\in\{1,\ldots,L_q\}$ and each document vector $\bm{d}_t\in\mathbb{R}^r$ for $t\in\{1,\ldots,L_d\}$ has unit length. ColBERT and ColPali use MaxSim \citep{khattab2020colbert}. mLateOn uses MeanMaxSim, which normalizes MaxSim by the query length, because its authors observed slightly better performance in their experiments \citep{sourty2026denseon}:
\begin{equation}
  s_{\mathrm{late}}(q,d)=
  \frac{1}{L_q}\sum_{s=1}^{L_q}
  \max_{1\leq t\leq L_d}\langle\bm{q}_s,\bm{d}_t\rangle.
  \label{eq:late-score}
\end{equation}

\subsection{Retrieval architecture}

\noindent\begin{minipage}[t]{0.51\textwidth}
\vspace{0pt}
\noindent\textbf{Two retrieval heads.} The final hidden states from the shared backbone feed a jointly trained late-interaction head and a dense head as shown in \autoref{fig:retrieval-heads}. An ablation shows that joint training improves late-interaction retrieval scores on both evaluated benchmark suites. The ablation details are shown in \autoref{app:ablations}.

\noindent\textbf{Late-interaction head.} A learned linear layer projects every final hidden state to 128 dimensions and normalizes the resulting vector. Similarly to ColBERT and ColPali, text inputs generate one vector per token, and images have one token per patch.

\noindent\textbf{Dense head.} The dense head uses mean pooling to average the final hidden state vectors and then applies L2 normalization. The dense head is trained using Matryoshka \citep{kusupati2022matryoshka}: \neomme{}-Retriever 260M uses widths of 128, 256, 512, and 1,024 dimensions and the 800M model uses the same four widths and a fifth width of 1,792 dimensions.
\end{minipage}\hfill
\begin{minipage}[t]{0.46\textwidth}
  \vspace{0pt}
  \centering
  \includegraphics[width=\linewidth]{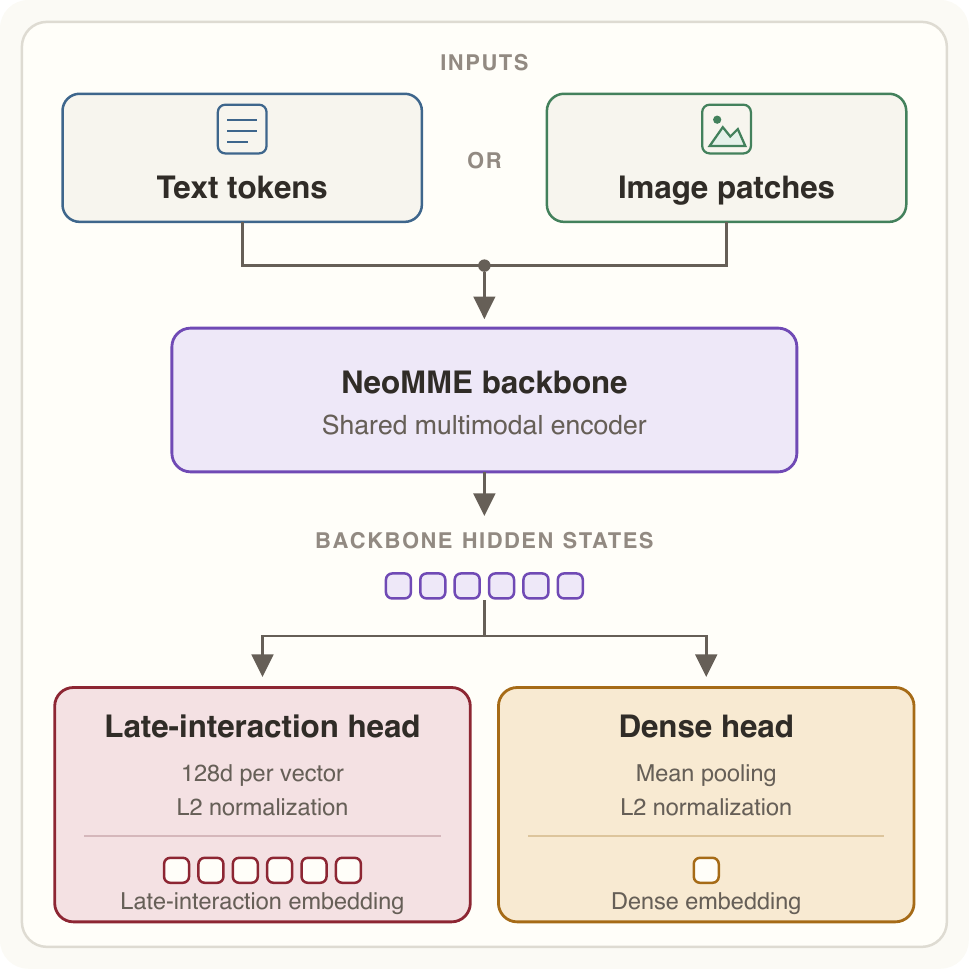}
  \captionof{figure}{Late-interaction and dense retrieval heads for both \neomme{} model sizes.}
  \label{fig:retrieval-heads}
\end{minipage}

\noindent\textbf{Deployment flexibility.} A single backbone pass produces both representations for a query or document, and the parameter-free dense head adds only mean pooling and normalization to the computation already required by late-interaction. A system can use late-interaction directly on a smaller corpus. On a very large corpus, it can use the dense representation with an ANN index to retrieve a shorter list and then rerank that list with late-interaction, following a common multistage retrieval design \citep{hofstatter2022colberter,formal2024splate}.

\subsection{Dataset mixture}

Retrieval training combines separate text-only and multimodal streams of query--document pairs. Training on both streams lets \neomme{}-Retriever retrieve over mixed-modality corpora, where documents are either text chunks or page images, though we do not evaluate that setting here. The text stream uses five source pools spanning English retrieval, multilingual retrieval, and code search. The multimodal stream uses four page-image datasets, including VisRAG \citep{yu2024visrag}. Sources are sampled according to configured weights that define their intended shares within each stream. The weights in \autoref{tab:retrieval-mixture} are normalized within each stream, and the expected batch shares use one text batch for every two multimodal batches. We decontaminated the training mixture against every retrieval benchmark we evaluate by masking matched queries and documents. \autoref{app:retrieval-details} reports source counts and the query-generation procedure.

\noindent\textbf{Query augmentation.} Before hard-negative mining, we augmented the multimodal stream with queries generated from page images. Qwen3.5-9B\footnote{\label{fn:qwen35-9b-hf}\hflink{https://huggingface.co/Qwen/Qwen3.5-9B}{Qwen/Qwen3.5-9B}.} generated two queries for each of 100,000 sampled pages, and 191,915 queries survived parsability and self-containment filtering, raising the multimodal stream from 760,826 to 952,741 queries without adding pages. Qwen3-VL-Reranker-8B\footnote{\label{fn:qwen3-vl-reranker-hf}\hflink{https://huggingface.co/Qwen/Qwen3-VL-Reranker-8B}{Qwen/Qwen3-VL-Reranker-8B}.} then filters the generated pairs during hard-negative mining. The generated queries add six query types, three output formats, and six languages. \autoref{app:retrieval-details} gives the sampling and filtering procedure. This dataset augmentation is separate from the 10 learned query-expansion tokens described in \autoref{app:retrieval-processor}.

\noindent\textbf{Hard negatives.} Text examples use the hard negatives provided with the mLateOn multilingual dataset referenced in \autoref{tab:retrieval-mixture}. Visual hard negatives are self-mined in two stages: a first fine-tuning run without explicit hard negatives mines a 32-candidate window for each visual query, and a second run repeats the protocol with seven negatives sampled from that window. \autoref{app:contrastive-training} details the mining and filtering steps.

\begin{table*}[t]
  \centering
  \caption{Configured retrieval training mixture by logical source family.}
  \label{tab:retrieval-mixture}
  \small
  \setlength{\tabcolsep}{5pt}
  \begin{tabular}{p{0.17\textwidth}p{0.38\textwidth}>{\raggedleft\arraybackslash}p{0.17\textwidth}>{\raggedleft\arraybackslash}p{0.17\textwidth}}
    \toprule
    Group & Dataset & Within-stream weight & Expected batch share \\
    \midrule
    \multicolumn{4}{l}{\textbf{\faFont\ Text only}} \\
    \multirow{3}{0.17\textwidth}{LightOn retrieval}
      & \hflink{https://huggingface.co/datasets/lightonai/embeddings-fine-tuning-filtered-en}{Filtered English retrieval} & 0.09 & 3.00\% \\
      & \hflink{https://huggingface.co/collections/lightonai/mdenseon-and-mlateon}{mLateOn multilingual} & 0.73 & 24.33\% \\
      & \hflink{https://huggingface.co/datasets/lightonai/embeddings-fine-tuning-filtered-code}{Code retrieval} & 0.13 & 4.33\% \\
    \addlinespace
    \multirow{2}{0.17\textwidth}{Other retrieval}
      & Additional multilingual retrieval & 0.03 & 1.00\% \\
      & \hflink{https://huggingface.co/datasets/lightonai/embeddings-fine-tuning-filtered-en}{Organic retrieval (en)} and \hflink{https://huggingface.co/datasets/lightonai/embeddings-fine-tuning-filtered-fr}{Organic retrieval (fr)} & 0.02 & 0.67\% \\
    \cmidrule(lr){2-4}
      & \textbf{Text-only total} & \textbf{1.00} & \textbf{$1/3$} \\
    \midrule
    \multicolumn{4}{l}{\textbf{\faImage\ Multimodal}} \\
    \multirow{4}{0.17\textwidth}{Document images}
      & \hflink{https://huggingface.co/datasets/vidore/colpali_train_set}{ColPali} & 0.25 & 16.67\% \\
      & Multilingual document images & 0.25 & 16.67\% \\
      & \hflink{https://huggingface.co/datasets/openbmb/VisRAG-Ret-Train-In-domain-data}{VisRAG} & 0.25 & 16.67\% \\
      & \hflink{https://huggingface.co/datasets/openbmb/VisRAG-Ret-Train-Synthetic-data}{VisRAG synthetic} & 0.25 & 16.67\% \\
    \cmidrule(lr){2-4}
      & \textbf{Multimodal total} & \textbf{1.00} & \textbf{$2/3$} \\
    \bottomrule
  \end{tabular}
\end{table*}

\subsection{Contrastive objective}

For each training query $q$, let $\mathcal{C}_{\mathcal{B}}(q)\subseteq\mathcal{C}$ be the in-batch corpus scored for $q$, including its positive document $d^+$. The late-interaction loss is
\begin{equation}
  \mathcal{L}_{\mathrm{late}} =
  -\log
  \frac{\exp(s_{\mathrm{late}}(q,d^+)/\tau)}
  {\sum_{d\in\mathcal{C}_{\mathcal{B}}(q)}\exp(s_{\mathrm{late}}(q,d)/\tau)},
  \qquad \tau=0.02.
  \label{eq:late-loss}
\end{equation}

Let $\mathcal{K}$ contain the trained Matryoshka widths, and let $s_{\mathrm{dense}}^{(k)}$ score vectors truncated to width $k$ and renormalized. The dense head uses the same in-batch corpus and temperature at every width:
\begin{equation}
  \mathcal{L}_{\mathrm{dense}} =
  -\frac{1}{|\mathcal{K}|}\sum_{k\in\mathcal{K}}
  \log
  \frac{\exp(s_{\mathrm{dense}}^{(k)}(q,d^+)/\tau)}
  {\sum_{d\in\mathcal{C}_{\mathcal{B}}(q)}
  \exp(s_{\mathrm{dense}}^{(k)}(q,d)/\tau)}.
  \label{eq:dense-loss}
\end{equation}

The retrieval objective is
\begin{equation}
  \mathcal{L}_{\mathrm{retrieval}} =
  \mathcal{L}_{\mathrm{late}}+\mathcal{L}_{\mathrm{dense}}.
  \label{eq:retrieval-loss}
\end{equation}
Both terms have weight one. The recipe uses no teacher logits and no knowledge distillation.

\subsection{Optimization}

\noindent\textbf{Large-batch contrastive training.} The retrieval objective scores each query against the full in-batch corpus, which makes a direct backward pass memory intensive. Ordinary gradient accumulation over independent micro-batches would change the InfoNCE objective because each query would see fewer in-batch negatives \citep{oord2018cpc,gao2021gradcache}. GradCache preserves the full in-batch corpus while recomputing encoder activations in smaller chunks. \autoref{app:retrieval-details} gives the micro-batch sizes and the rest of the training settings.

\noindent\textbf{Late-Interaction Kernels (\lik{}).} A direct implementation of MaxSim materializes the full query-token by document-token similarity tensor in high-bandwidth memory. That tensor dominates scoring memory in this recipe, because a $2048\times2048$ page produces a 4,162-long multi-vector embedding and every query is scored against the entire in-batch corpus. \lik{} instead computes the score in tiles and keeps only a running maximum per query token, storing the winning document-token index so that the backward pass can route gradients to it \citep{lac2026lik}. The computation is exact, so scores and the loss match the direct implementation. At 4,096 document tokens, \lik{} lowers the MaxSim memory peak from 672 MB to 193 MB and the backward pass from 1.82 ms to 0.54 ms, while the direct implementation runs out of memory at 8,192 tokens. At the \neomme-260M training shape, the throughput difference is within run-to-run variation, so the benefit here is memory headroom rather than speed. \autoref{app:lik} gives the kernel details and the full measurements.

\subsection{Evaluation protocol}
\label{sec:evaluation-protocol}

We evaluate each retrieval setting on its standard benchmarks: ViDoRe v3, v2, and v1 for visual document retrieval \citep{mace2025vidorev2,loison2026vidorev3}, and BEIR-15 for text retrieval \citep{thakur2021beir}. We use only the eight public tasks from ViDoRe v3. Each benchmark task contains queries, a document corpus, and relevance judgments. The evaluator first averages each metric over the judged queries within a task, then averages the task scores so each task has equal weight. Each model uses its official processor and the scoring settings recommended by its authors.

\noindent\textbf{\neomme{} evaluation settings.} ViDoRe uses a downscale-only longest-side cap of 2,048 pixels and exact all-pairs MeanMaxSim scoring. BEIR-15 uses 8,192-token query and document limits, with each document represented by its title followed by its body. The late-interaction head uses a compressed FastPLAID\footnote{\url{https://github.com/lightonai/fast-plaid}} index for approximate retrieval followed by exact rescoring of retrieved documents \citep{santhanam2022plaid}, while the dense head uses the same pipeline with one normalized vector per input. Both \neomme{} models use the output at step 20,000. \autoref{app:retrieval-evaluation-details} gives the omitted BEIR datasets and further evaluation details.

\noindent\textbf{Reported comparison results.} External benchmark results come from the \href{https://mteb-leaderboard.hf.space/}{MTEB Leaderboard} or the cited model reports. The ViDoRe v3 comparisons use per-task MTEB scores aggregated over the same eight tasks as our results, and we report each benchmark's default metric \citep{muennighoff2023mteb}.

\subsection{Main results}

\begin{figure}[h!]
  \centering
  \includegraphics[width=\textwidth]{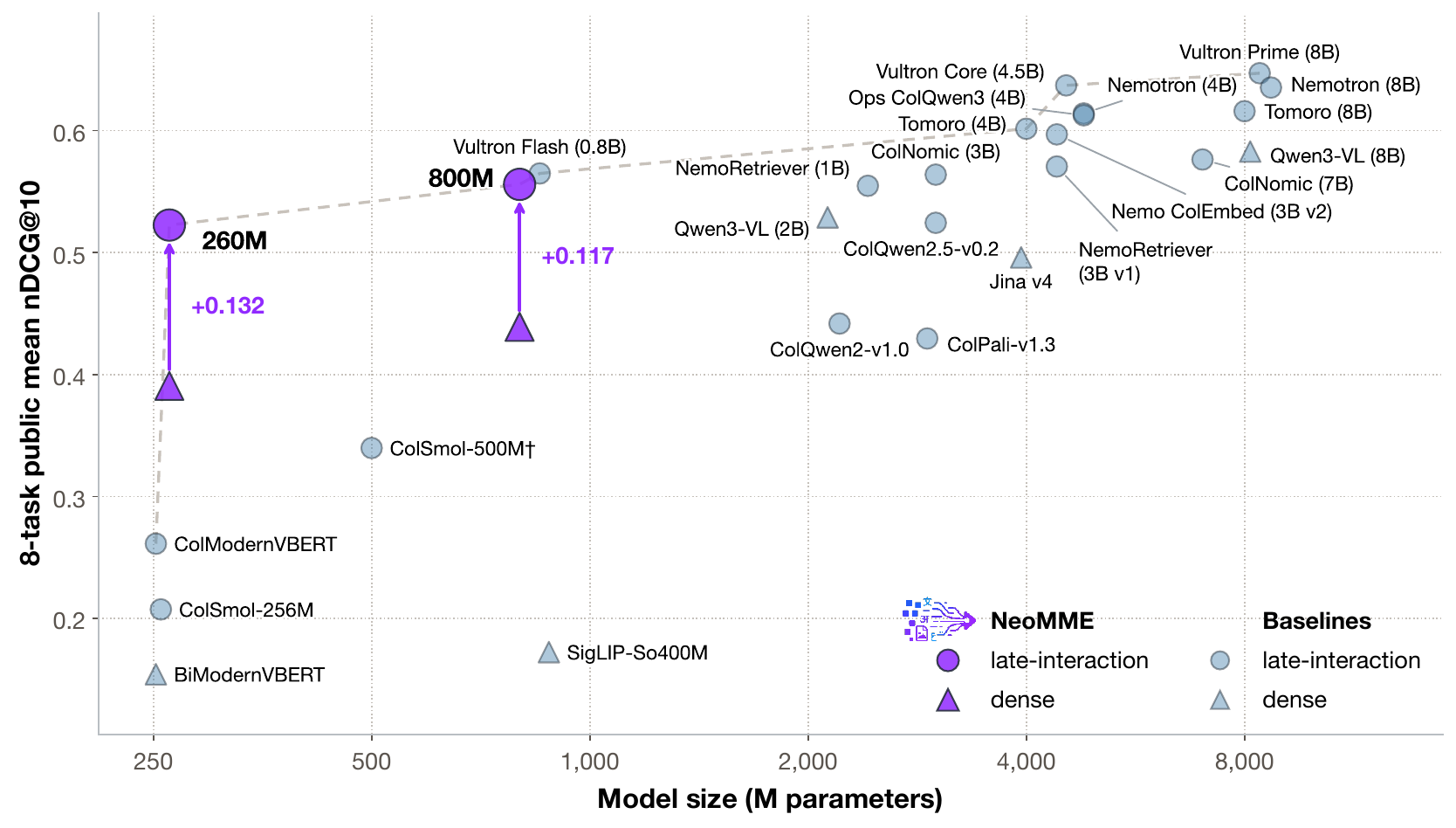}
  \caption{ViDoRe v3 nDCG@10 versus model size. \neomme{}-Retriever 260M and 800M sit on the Pareto frontier. Arrows mark the gain of each \neomme{} late-interaction head over its dense head. \autoref{tab:vidore-v3-pareto-models} gives the full results.}
  \label{fig:vidore-v3-model-size}
\end{figure}

On visual document retrieval, \neomme{} is competitive for its size. The 260M model leads all models below 300M parameters, and the 800M model scores within 0.9 nDCG@10 points of the similarly sized Vultron Flash on ViDoRe v3. \autoref{tab:visual-results} compares \neomme{} with selected visual retrievers across three parameter ranges, while \autoref{tab:text-results} shows how its dense and late-interaction heads transfer to text retrieval. \autoref{tab:vidore-v3-pareto-models} reports every model in \autoref{fig:vidore-v3-model-size}, including ViDoRe v1 and v2 scores where reported and the ViDoRe v3 aggregate. We use ViDoRe v3 as the primary visual benchmark because it was built with a more rigorous methodology and is less saturated than ViDoRe v1 and v2.

\noindent\textbf{Small visual retrievers.} \neomme-260M is the strongest model below 300M parameters and strictly below 800M across all ViDoRe benchmarks as shown in \autoref{tab:visual-results}. Compared with the best other model below 300M parameters on each benchmark, it improves nDCG by 26.1 points on ViDoRe v3, 11.5 points on v2, and 5.4 points on v1. Its ViDoRe v3 score of 0.523 is within 0.2 points of the 3.75B-parameter ColQwen2.5 score while being $14.4\times$ smaller by parameter count.

\noindent\textbf{Scaling visual retrievers to 800M.} Increasing the model size improves every \neomme{} aggregate and both retrieval heads. The ViDoRe gains are 3.3 points on v3, 3.7 points on v2, and 1.5 points on v1. On ViDoRe v3, \neomme-800M reaches 0.556, within 0.9 point of the slightly larger Vultron Flash ($\sim$850M) and 3.2 points above the 3.75B-parameter ColQwen2.5. We report document-encoding throughput separately in \autoref{sec:indexing-efficiency}. On BEIR-15, scaling adds 2.5 points to late-interaction and 6.3 points to dense retrieval. \autoref{app:retrieval-results} gives the full metric and domain breakdowns.

\noindent\textbf{Text retrieval.} Late-interaction exceeds dense retrieval by 18.3 points for \neomme-260M and 14.4 points for \neomme-800M. The 800M model reaches 0.513 with late-interaction and 0.369 with dense retrieval on BEIR-15.

These results show that the same encoder backbone can
retrieve both page images and text, with late-interaction the stronger of the two representations.

\begin{table}[h!]
  \centering
  \begin{minipage}[t]{0.61\textwidth}
    \vspace{0pt}
    \centering
    \captionsetup{justification=raggedright,singlelinecheck=false}
    \captionof{table}{Visual document retrieval performance on the ViDoRe benchmarks.}
    \label{tab:visual-results}
    \scriptsize
    \setlength{\tabcolsep}{3pt}
    \begin{tabular}{lrrrr}
      \toprule
      \multicolumn{2}{c}{Model details} &
      \multicolumn{3}{c}{ViDoRe (nDCG@$k$)} \\
      \cmidrule(lr){1-2}\cmidrule(lr){3-5}
      Model & Params. & v3 (@10) & v2 (@5) & v1 (@5) \\
      \midrule
      \multicolumn{5}{l}{\paramsmall} \\
      ColModernVBERT\tablefootnote{\label{fn:results-colmodernvbert}\hflink{https://huggingface.co/ModernVBERT/colmodernvbert}{ModernVBERT/colmodernvbert}.}
      & 250M & 0.261\textsuperscript{\textdagger} & 0.407\textsuperscript{\ensuremath{\ddagger}} & 0.806\textsuperscript{\ensuremath{\ddagger}} \\
      ColSmol-256M\tablefootnote{\label{fn:results-colsmol}\hflink{https://huggingface.co/vidore/colSmol-256M}{vidore/colSmol-256M}.}\textsuperscript{\textdagger}
      & 256M & 0.207 & 0.348 & 0.797 \\
      \neomme-260M\tablefootnote{\label{fn:results-neomme-260m}\hflink{https://huggingface.co/Hcompany/NeoMME-260M-Retriever}{Hcompany/NeoMME-260M-Retriever}.}\textsuperscript{\ensuremath{\ddagger}}
      & 260M & \textbf{0.523} & \textbf{0.522} & \textbf{0.860} \\
      \midrule
      \multicolumn{5}{l}{\parammedium} \\
      ColSmol-500M\tablefootnote{\label{fn:results-colsmol-500m}\hflink{https://huggingface.co/vidore/colSmol-500M}{vidore/colSmol-500M}.}
      & 500M & 0.340\textsuperscript{\ensuremath{\ddagger}} & 0.455\textsuperscript{\textdagger} & 0.825\textsuperscript{\textdagger} \\
      Vultron Flash\tablefootnote{\label{fn:results-vultron}\hflink{https://huggingface.co/vultr/VultronRetrieverFlash-Qwen3.5-0.8B}{vultr/VultronRetrieverFlash-Qwen3.5-0.8B} \citep{georgiou2026vultronflash}.}\textsuperscript{\textdagger}
      & 850M & \textbf{0.565} & \textbf{0.604} & \textbf{0.882} \\
      \neomme-800M\tablefootnote{\label{fn:results-neomme-800m}\hflink{https://huggingface.co/Hcompany/NeoMME-800M-Retriever}{Hcompany/NeoMME-800M-Retriever}.}\textsuperscript{\ensuremath{\ddagger}}
      & 800M & 0.556 & 0.559 & 0.874 \\
      \midrule
      \multicolumn{5}{l}{\paramlarge} \\
      ColQwen2.5-v0.2\tablefootnote{\label{fn:results-colqwen25}\hflink{https://huggingface.co/vidore/colqwen2.5-v0.2}{vidore/colqwen2.5-v0.2}.}\textsuperscript{\textdagger}
      & 3.75B & \textbf{0.524} & \textbf{0.601} & \textbf{0.895} \\
      ColPali v1.3\tablefootnote{\label{fn:results-colpali}\hflink{https://huggingface.co/vidore/colpali-v1.3}{vidore/colpali-v1.3}.}\textsuperscript{\textdagger}
      & 2.92B & 0.430 & 0.547 & 0.848 \\
      \bottomrule
    \end{tabular}
  \end{minipage}
  \hfill
  \begin{minipage}[t]{0.37\textwidth}
    \vspace{0pt}
    \centering
    \captionsetup{justification=raggedright,singlelinecheck=false}
    \captionof{table}{BEIR-15 text retrieval results.}
    \label{tab:text-results}
    \scriptsize
    \setlength{\tabcolsep}{3pt}
    \begin{tabular}{lrr}
      \toprule
      Model & Params. & nDCG@10 \\
      \midrule
      \multicolumn{3}{l}{\textbf{Dense}} \\
      GTE-ModernBERT\tablefootnote{\hflink{https://huggingface.co/Alibaba-NLP/gte-modernbert-base}{Alibaba-NLP/gte-modernbert-base}.}\textsuperscript{\textdagger}
      & 149M & 0.5519 \\
      DenseOn\tablefootnote{\hflink{https://huggingface.co/lightonai/DenseOn}{lightonai/DenseOn}.}\textsuperscript{\S}
      & 149M & 0.5620 \\
      \neomme-260M\textsuperscript{\ref{fn:results-neomme-260m}}\textsuperscript{\ensuremath{\ddagger}} & 260M & 0.3055 \\
      \neomme-800M\textsuperscript{\ref{fn:results-neomme-800m}}\textsuperscript{\ensuremath{\ddagger}} & 800M & 0.3686 \\
      \midrule
      \multicolumn{3}{l}{\textbf{Late-interaction}} \\
      ColBERTv2\tablefootnote{\hflink{https://huggingface.co/colbert-ir/colbertv2.0}{colbert-ir/colbertv2.0} \citep{santhanam2022colbertv2}.}\textsuperscript{\P} & 110M & 0.4863 \\
      LateOn\tablefootnote{\hflink{https://huggingface.co/lightonai/LateOn}{lightonai/LateOn}.}\textsuperscript{\S} & 149M & 0.5722 \\
      \neomme-260M\textsuperscript{\ref{fn:results-neomme-260m}}\textsuperscript{\ensuremath{\ddagger}} & 260M & 0.4881 \\
      \neomme-800M\textsuperscript{\ref{fn:results-neomme-800m}}\textsuperscript{\ensuremath{\ddagger}} & 800M & 0.5126 \\
      \bottomrule
    \end{tabular}
  \end{minipage}
  \par\vspace{2pt}
  \scriptsize
  \textsuperscript{\textdagger} Scores from MTEB. \textsuperscript{\S} Scores from \citet{sourty2026denseon}. \textsuperscript{\P} Scores from \href{https://huggingface.co/blog/lightonai/denseon-lateon}{LightOn's blog post}. \textsuperscript{\ensuremath{\ddagger}} Results from our evaluation.
\end{table}

\subsection{Resolution impact}
\label{sec:resolution-impact}

Image resolution trades retrieval quality against indexing cost. Higher longest-side caps produce more image tokens, which require more encoder FLOPs and more late-interaction storage. Across ViDoRe v1, v2, and v3, reducing the cap from 2,048 to 1,536 pixels lowers MeanMaxSim nDCG@10 by at most 1.8\% for either model, while a square page uses about 57\% as many vectors and raw float32 bytes. At 1,024 pixels, the relative quality loss ranges from 3.0 to 12.7\%, and at 768 pixels it ranges from 11.7 to 36.1\%. \autoref{tab:resolution-impact} reports the full quality grid and square-page representation sizes.

We keep a downscale-only 2,048-pixel cap as the default because it preserves the best measured retrieval quality without adding tokens to smaller native pages. Forcing every page to a 2,048-pixel longest side through upsampling also does not improve either head. A 1,536-pixel cap is a lower-cost option when a small quality loss is acceptable.

\subsection{Compression and storage}

While a dense embedding stores one fixed-width vector per document, late-interaction storage scales linearly with the number of document vectors. A larger processed image has more patches, so the late-interaction embedding has more vectors, and the index uses more storage. \autoref{sec:evaluation-protocol} uses a 2,048-pixel longest-side cap for retrieval performance, but compression is required to make this setup tractable for large document corpora, because high-resolution images are encoded into large late-interaction embeddings. Hierarchical token pooling reduces the number of stored vectors, and asymmetric quantization reduces the number of bytes per vector. We evaluate each method separately, then combine them for the 260M model to measure their joint quality--storage trade-off.

\begin{figure*}[t]
  \centering
  \begin{subfigure}[t]{0.68\textwidth}
    \centering
    \includegraphics[width=\linewidth]{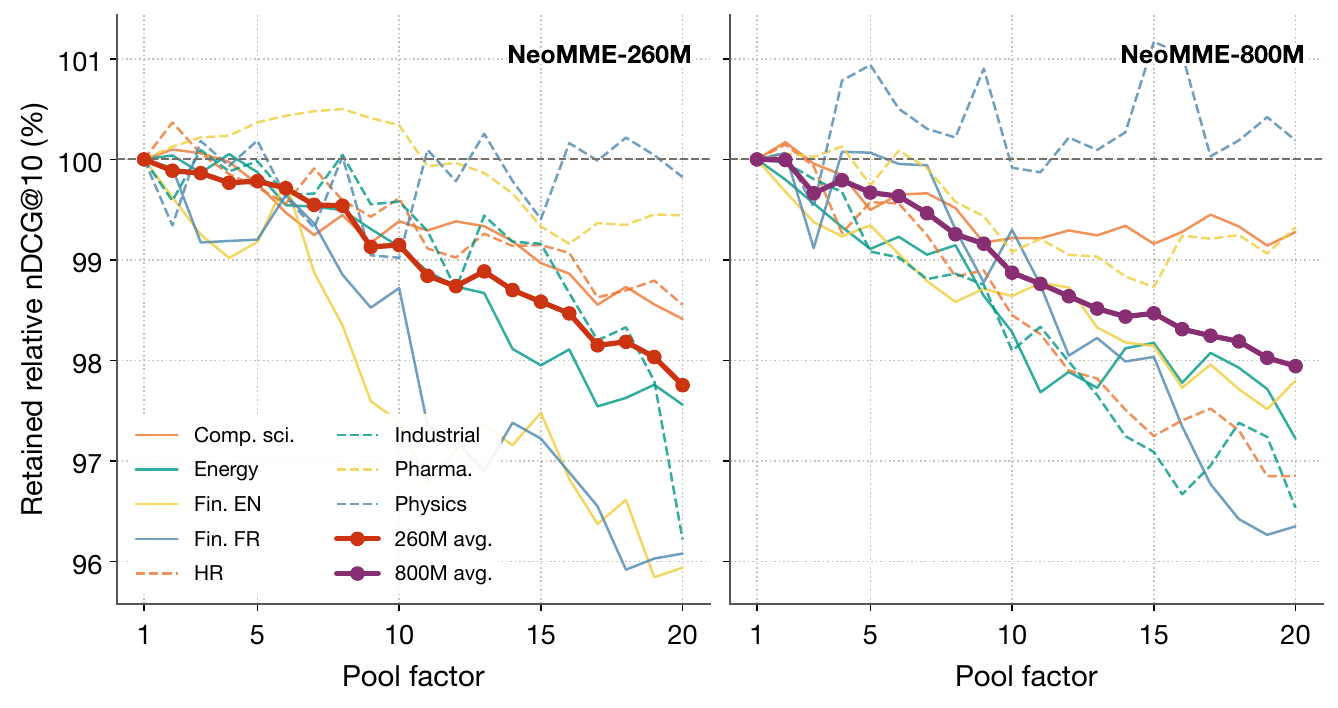}
    \caption{Quality by pool factor for both model sizes.}
    \label{fig:token-pooling}
  \end{subfigure}
  \hfill
  \begin{subfigure}[t]{0.30\textwidth}
    \centering
    \includegraphics[width=\linewidth]{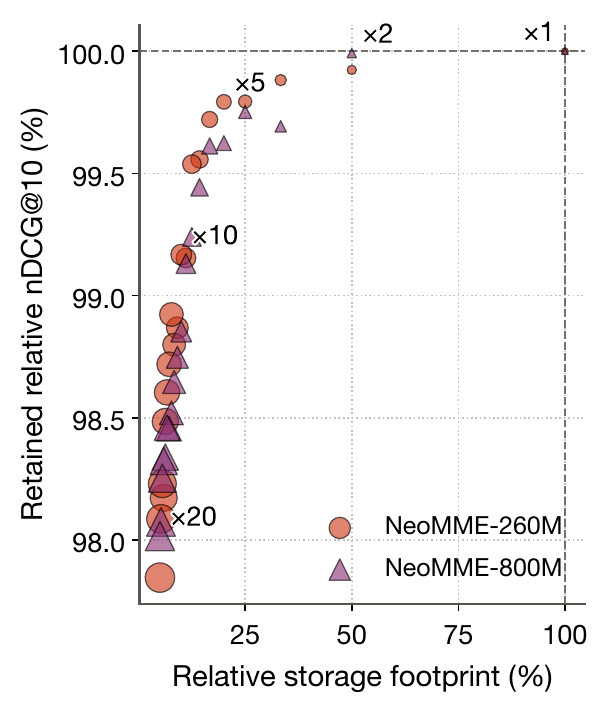}
    \caption{Quality versus storage for both model sizes.}
    \label{fig:token-pooling-storage}
  \end{subfigure}
  \caption{Hierarchical token pooling on ViDoRe v3: impact of the pool factor on retained retrieval quality for both \neomme{}-Retriever models.}
\end{figure*}

\noindent\textbf{Hierarchical token pooling.} Hierarchical token pooling clusters similar document vectors and replaces each cluster with its mean \citep{clavie2024tokenpooling}. At pool factor 7, both \neomme{} models reduce the storage used by their document vectors by about $7\times$ while retaining more than 99\% of factor-one ViDoRe v3 nDCG@10. In the paper's unquantized text-retrieval experiments, factor three is the practical near-lossless boundary: it reduces average quality by less than 1\%, while factors four and six reduce it by about 3\% and 10\%. In our visual document experiments, factor 10 retains 99.2\% of \neomme-260M quality and 98.9\% of \neomme-800M quality, while factor 20 still retains 97.8\% and 98.0\%, respectively. The models, datasets, and evaluation protocols differ, so this is not a controlled modality comparison. However, the wider range of low-loss pooling factors suggests that token pooling is especially effective for visual document retrieval, possibly because page images contain more redundant patch representations than text contains redundant token representations. \autoref{fig:token-pooling} shows all measured factors for both sizes, and \autoref{fig:token-pooling-storage} shows the corresponding storage trade-off. The experiment does not measure index overhead or serving latency.

\begin{center}
  \begin{minipage}{\textwidth}
  \begin{minipage}[t]{0.52\textwidth}
    \vspace{0pt}
    \centering
    \captionsetup{justification=raggedright,singlelinecheck=false}
    \captionof{table}{Int8 document representations reduce packed storage by $3.9\times$ while retaining at least 99.96\% of float32 nDCG@10, and binary documents reduce storage by $32\times$. Arrows give the change from the float32 baseline in nDCG@10 points (1 point = 0.01). All results use the evaluation protocol in \autoref{sec:evaluation-protocol}.}
    \label{tab:asymmetric-quant}
    \scriptsize
    \setlength{\tabcolsep}{2.5pt}
    \begin{tabular}{@{}lllrr@{}}
      \toprule
      Model & Query & Document & nDCG@10 & \shortstack{Storage\\(kB)} \\
      \midrule
      \multirow{4}{*}{\neomme-260M\textsuperscript{\ref{fn:results-neomme-260m}}}
        & float32 & float32 & \textbf{0.5226} & $1536.7 \pm 201.1$ \\
        & int8 & int8 & \begin{tabular}[t]{@{}r@{}}0.5224\\[-1pt]\tiny\textcolor{red!70!black}{\ensuremath{\downarrow}\,$-0.02$}\end{tabular} & $390.2 \pm 51.1$ \\
        & int8 & binary & \begin{tabular}[t]{@{}r@{}}0.5068\\[-1pt]\tiny\textcolor{red!70!black}{\ensuremath{\downarrow}\,$-1.58$}\end{tabular} & $48.0 \pm 6.3$ \\
        & binary & binary & \begin{tabular}[t]{@{}r@{}}0.4960\\[-1pt]\tiny\textcolor{red!70!black}{\ensuremath{\downarrow}\,$-2.66$}\end{tabular} & $48.0 \pm 6.3$ \\
      \midrule
      \multirow{4}{*}{\neomme-800M\textsuperscript{\ref{fn:results-neomme-800m}}}
        & float32 & float32 & \textbf{0.5560} & $1536.7 \pm 201.1$ \\
        & int8 & int8 & \begin{tabular}[t]{@{}r@{}}0.5559\\[-1pt]\tiny\textcolor{red!70!black}{\ensuremath{\downarrow}\,$-0.01$}\end{tabular} & $390.2 \pm 51.1$ \\
        & int8 & binary & \begin{tabular}[t]{@{}r@{}}0.5369\\[-1pt]\tiny\textcolor{red!70!black}{\ensuremath{\downarrow}\,$-1.91$}\end{tabular} & $48.0 \pm 6.3$ \\
        & binary & binary & \begin{tabular}[t]{@{}r@{}}0.5253\\[-1pt]\tiny\textcolor{red!70!black}{\ensuremath{\downarrow}\,$-3.07$}\end{tabular} & $48.0 \pm 6.3$ \\
      \bottomrule
    \end{tabular}
  \end{minipage}
  \hfill
  \begin{minipage}[t]{0.46\textwidth}
    \vspace{0pt}
    \centering
    \captionsetup{justification=raggedright,singlelinecheck=false}
    \includegraphics[width=\linewidth]{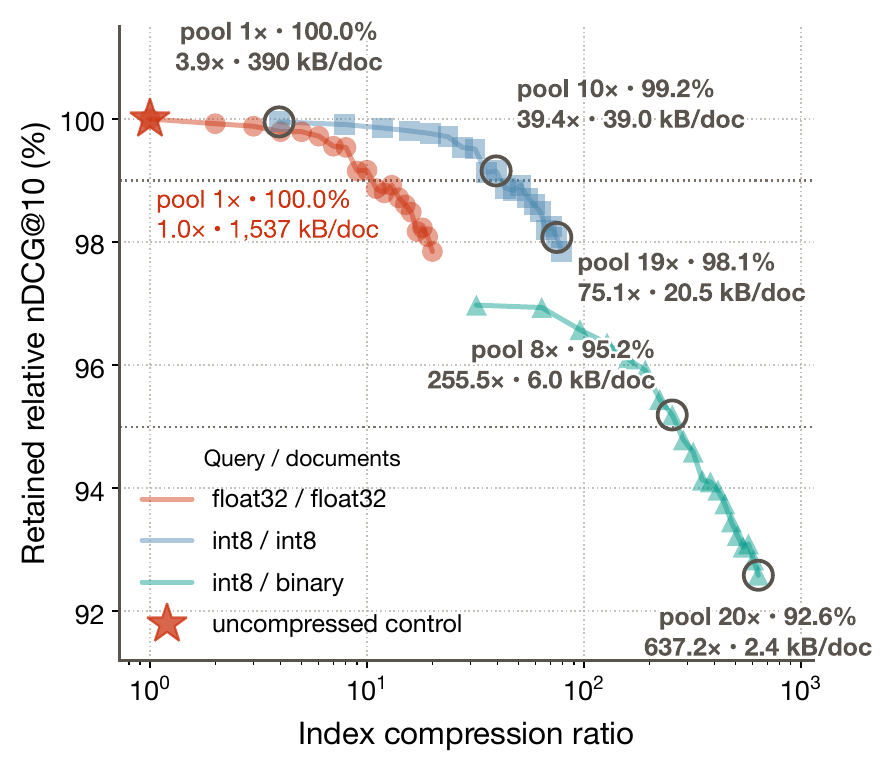}
    \captionof{figure}{Quality--storage frontier for the \neomme-260M late-interaction index on ViDoRe v3. Labels show pool factor, retained quality, compression, and storage.}
    \label{fig:compression-frontier}
  \end{minipage}
  \end{minipage}
\end{center}

\noindent\textbf{Asymmetric quantization.} Asymmetric quantization stores document vectors at low precision because stored documents dominate index size while queries are encoded once per search \citep{mixedbread2026asymmetric}. We test float32 query--document scoring, matched int8 scoring, int8 queries with binary documents, and matched binary scoring. The packed indexes use per-token symmetric int8 codes with float16 scales or sign-bit binary codes and are built from released late-interaction embeddings. We verified that packed scores match the float reference within $5\times10^{-5}$ on every query, with an observed maximum difference of $1.8\times10^{-7}$. The float32 baselines match the main retrieval scores. Storage counts vector codes, int8 scales, document offsets, and identifiers, while excluding query bytes, approximate nearest-neighbor routing, and filesystem overhead. \autoref{tab:asymmetric-quant} shows that quantizing queries to int8 costs at most 0.0002 nDCG@10, while int8 and binary document indexes are $3.9\times$ and $32\times$ smaller than float32. Matched binary query--document scoring can also reduce retrieval latency because packed binary dot products can use bitwise XOR and population-count instructions. Realizing that speedup requires specialized kernels that operate directly on packed bits; dequantizing before scoring removes the advantage. Developing and benchmarking these kernels is outside the scope of this study, so we report quality and storage rather than optimized retrieval latency.

\noindent\textbf{Combined compression.} We also measure every combination of token pooling and query and document quantization for \neomme-260M. The study reports task-macro nDCG@10 relative to the float32 baseline and computes compression from deployed bytes. Quantizing queries to int8 alongside the documents costs at most 0.0002 nDCG@10 at any document setting. Pool factor 10 with int8 queries and int8 documents reduces storage from 1536.7 kB to 39.0 kB ($39.4\times$) and retains 99.16\% of the baseline nDCG@10. Pool factor 8 with int8 queries and binary documents reduces storage from 1536.7 kB to 6.0 kB ($255.5\times$) and retains 95.19\%. \autoref{fig:compression-frontier} shows the three selected precision pairs across all measured pool factors.

\subsection{Indexing efficiency}
\label{sec:indexing-efficiency}

\begin{center}
  \centering
  \includegraphics[width=\linewidth]{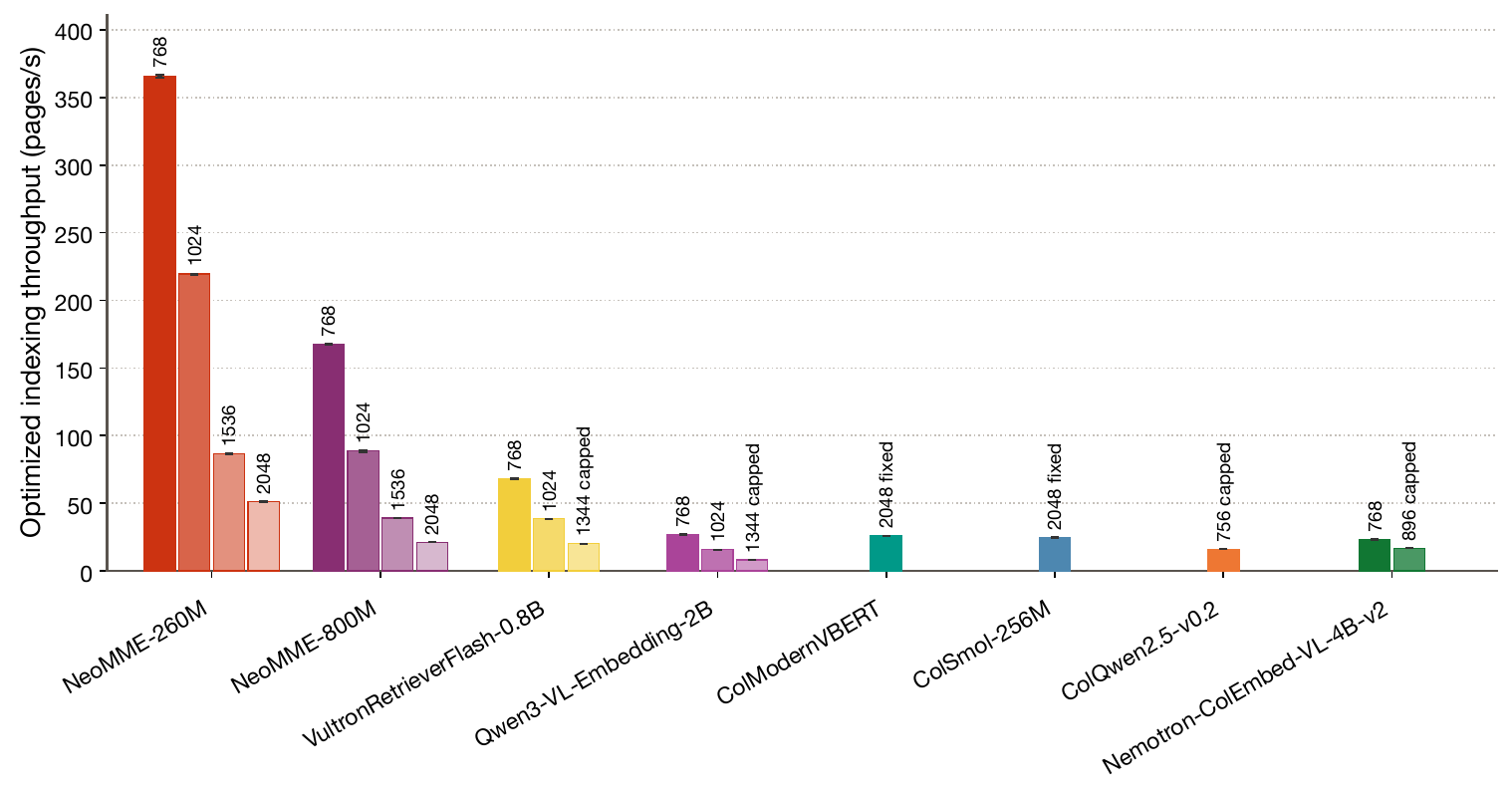}
  \captionof{figure}{Compiled document-encoding throughput over the resolution grid on one NVIDIA L40S.}
  \label{fig:indexing-l40s}
\end{center}

\autoref{fig:indexing-l40s} compares model-side document-encoding throughput. We run each model's official image processor before timing to separate processor runtime from model and accelerator execution. The processor still determines each model's input shape and token count. The timer covers device transfer, the model forward pass, output unpadding, and contiguous float32 serialization. Each row uses a separately calibrated batch, and the figure includes only the image sizes that each processor actually encodes. \autoref{app:indexing-details} gives the complete protocol and cross-hardware results.

At $2048\times2048$, \neomme-260M reaches 51.3 pages per second, which is $1.97\times$ ColModernVBERT's throughput of 26.0 pages per second. \neomme-800M reaches 39.2 pages per second at $1536\times1536$, 94\% faster than Vultron Flash's 20.2 pages per second at its documented $1344\times1344$ square. At the $2048\times2048$ resolution used for its retrieval-quality results, \neomme-800M still indexes faster, at 21.2 pages per second. The unmatched comparison favors Vultron because \neomme-800M processes 31\% more image pixels. \autoref{tab:indexing-results} reports the largest effective square for every model.

Document encoding happens offline, while query encoding runs for every request. In interactive applications such as Visual RAG, described in \autoref{sec:visual-document-retrieval}, query-encoding latency adds to the time between a user query and the generated answer. The measured \neomme{} latency is small on the tested hardware. \autoref{app:query-encoding-details} reports the mean and 95th-percentile latency.

\newpage
\subsection{Interpretability}

\begin{wrapfigure}{r}{0.42\textwidth}
  \vspace{-0.8em}
  \centering
  \includegraphics[width=\linewidth,trim={0 25pt 0 24pt},clip]{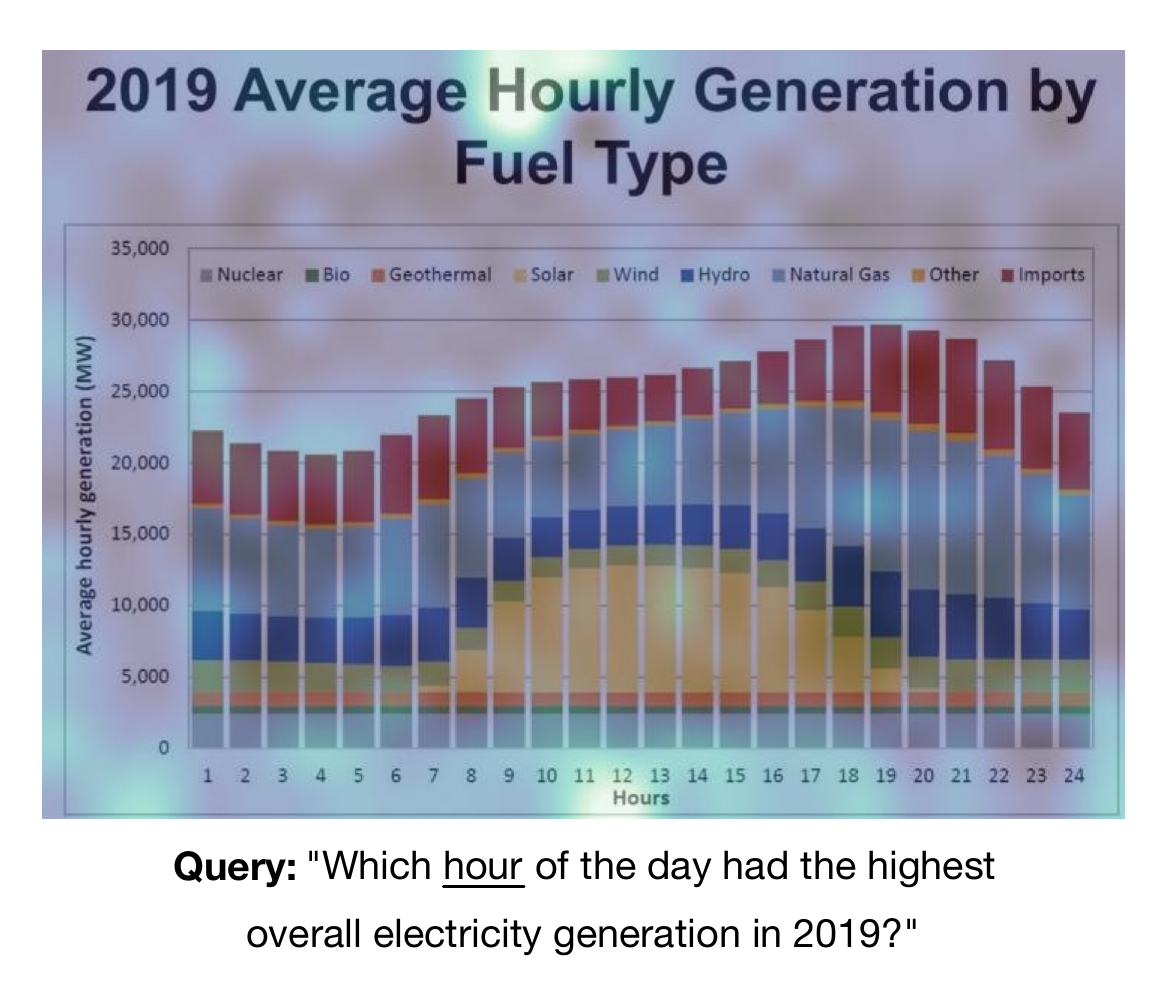}
  \caption{Similarity map for the query token \code{hour}, overlaid on a document from the ViDoRe syntheticDocQA energy test set.\protect\footnotemark}
  \label{fig:similarity-energy}
  \vspace{-0.6em}
\end{wrapfigure}
\footnotetext{\hflink{https://huggingface.co/datasets/vidore/syntheticDocQA_energy_test}{vidore/syntheticDocQA\_energy\_test}.}

We reuse ColPali's similarity-map methodology to visualize late-interaction scoring. For each query token, \neomme{} identifies the most similar document positions, and MeanMaxSim averages these per-token maxima into a query-to-page score. The map colors each document patch by its cosine similarity to the token representation. \autoref{fig:similarity-energy} asks ``Which hour of the day had the highest overall electricity generation in 2019?'' and shows the map for the underlined token \code{hour}. The strong responses around the chart title, the horizontal-axis label, and the numbered hours suggest that \neomme-260M has learned emergent OCR-like behavior without a separate OCR pipeline.

The visualization does not capture the full scoring process because MeanMaxSim also considers nonspatial structural positions that cannot be placed on the page. Some responses on background patches may also reflect the high-norm register-like tokens observed in Vision Transformers \citep{darcet2024registers}, rather than local visual evidence. We therefore treat the map as a qualitative diagnostic, not a complete explanation of the retrieval score.

\section{Other downstream tasks}
\label{sec:downstream}

Retrieval is the principal downstream application of \neomme{}. Nevertheless, we test whether the same compact backbone can serve as an initialization for language classification, token labeling, document classification, and natural-image tasks. The following experiments use the \neomme{}-260M and measure adaptability rather than superiority over models specialized for a single modality, and assess whether one shared backbone can support task-specific text, document-image, and multimodal fine-tuning.

\subsection{Language transfer}
We evaluate the 17-task suite used for LFM2.5-Encoder, comprising eight GLUE tasks, four SuperGLUE tasks, and five multilingual tasks \citep{wang2018glue,wang2019superglue,liquidai2026lfm25encoders}. We follow its task definitions and disjoint selection and reporting seeds, while retaining \neomme{}'s matched NorMuon--AdamW optimization \citep{li2026normuon,loshchilov2019adamw}. The comparison is therefore closely aligned, but not a same-harness reproduction of Liquid's training recipe.

\neomme{} reaches a 17-task mean of 75.3, within 4 points of the 79.3 reported by the dedicated LFM2.5-Encoder-230M text model. The gap is small for a similarly sized backbone whose capacity and pretraining are shared between multilingual text and raw image patches rather than optimized exclusively for language understanding. \neomme{} exceeds that peer on PAWS-X, MASSIVE Intent, SeaHorse, MRPC, and WSC, although its 46.4 Matthews correlation on CoLA shows a clear weakness in acceptability judgments. The same backbone reaches $85.5 \pm 0.2$ accuracy on the 8K-context LEDGAR task and 89.3 entity F1 in a one-seed CoNLL-2003 experiment \citep{sang2003conll}. This is additional evidence of transfer to long-document classification and token labeling.

\subsection{Visual transfer}
Pretraining does not optimize a direct target over image tokens: patches remain visible and contribute through masked transcript recovery, without pixel reconstruction, image classification, or an image-level contrastive objective. Unlike vision encoders trained explicitly to organize a global image representation \citep{radford2021clip,zhai2023siglip}, \neomme{} is therefore not expected to expose a strong frozen natural-image embedding space without adaptation. Frozen 16-shot probes across 10 natural-image classification tasks average 13.2 accuracy \citep{xiao2025mieb}, while task-specific fine-tuning reaches $77.1 \pm 0.3$ on Food101, $63.9 \pm 0.2$ on Oxford Pets, and $46.8 \pm 6.3$ on Stanford Cars. These natural-image experiments use an exploratory evaluation protocol and are reported as adaptation diagnostics rather than leaderboard results. Transfer is stronger for document images, which are closer to the pretraining data and objective. On RVL-CDIP \citep{harley2015rvlcdip}, a frozen first-token probe reaches $51.6 \pm 0.3$ accuracy, while fine-tuning on only 6,000 examples, less than 2\% of the available training set, raises accuracy to $81.5 \pm 0.6$. RVL-CDIP contains known annotation noise and train--test overlap \citep{larson2026revisingrvlcdip}, so this result is best interpreted as evidence of rapid document adaptation rather than clean out-of-distribution generalization.

\section{Limitations}
\label{sec:limitations}

This work was completed under limited time and compute. While we believe the \neomme{} results are interesting, the study at hand has several limitations.

\noindent\textbf{Pretraining scale.} \neomme{} uses much less pretraining data than ModernBERT. Each model processes about 524B packed input tokens, including about 290B from the text-only stream, while ModernBERT trains on about 2T text tokens \citep{warner2024modernbert}. Although the accounting methods differ, ModernBERT sees roughly seven times more text-only data. Longer runs could test how \neomme{} responds to more text, more images, and more training steps.

\noindent\textbf{Visual objective.} \neomme{} has no objective that predicts or reconstructs image content. Pretraining predicts masked text while image patches remain visible, so the model receives no direct image-token or global image-representation target. The weak frozen natural-image results in \autoref{sec:downstream} may follow from this choice. Matched studies of image-token prediction, pixel reconstruction, and image-level contrastive objectives could separate the effect of the objective from the data mixture and architecture.

\noindent\textbf{Retrieval supervision.} \neomme{} uses much less text retrieval supervision than SOTA retrieval system such as mLateOn. During retrieval training, each model processes about 430K sampled pure-text query examples and 850K sampled image-query examples. In comparison, mLateOn processes about 660M contrastive query--document examples followed by about 16M multilingual hard-negative examples \citep{sourty2026denseon}. Larger matched training sets could show how much of the BEIR difference comes from data scale rather than model architecture, objectives, or evaluation code.

\noindent\textbf{Mixed-modality corpora.} Training \neomme{}-Retriever on both text and page-image retrieval data places text chunks and page images in one embedding space, so we can rank both from a single candidate pool. However, we did not run this evaluation, though two benchmarks in the literature cover it: UniDoc-Bench scores joint retrieval over a unified corpus of 70,000 PDF pages \citep{peng2025unidocbench}, and MixBench targets mixed-modality search over collections of images, texts, and multimodal documents \citep{li2025mixbench}. Future work could measure \neomme{}-Retriever on both, then extend them to the multilingual page-image corpora that the model targets.

\noindent\textbf{Visual document data.} Visual retrieval lacks a large multilingual dataset with broad coverage of document types and layouts. Real-world PDFs are difficult to translate while preserving their layout and visual content. Future work could generate documents from code, translate their source content, and render matched pages in many languages. Releasing the resulting dataset would extend the translation augmentation used for mLateOn from text retrieval to visual document retrieval.

\noindent\textbf{Language coverage for retrieval.} Visual retrieval data has much narrower language coverage than text retrieval data. The text stream draws heavily on mLateOn's translate-train corpus, which covers English plus Arabic, French, German, Italian, Norwegian, Portuguese, Spanish, and Swedish, while another multilingual pool covers 13 languages. The visual stream contains pages in English, French, German, Italian, and Spanish, while generated queries add Portuguese. Synthetic document generation could expand page and query coverage across more scripts and regions, especially for low-resource languages.

\noindent\textbf{Ablations.} The current results do not isolate the effects of major architecture, model-scaling, data, and training choices. Future work could compare these factors under matched compute and evaluation settings to identify which choices drive pretraining, retrieval, and transfer performance.

\noindent\textbf{Distillation.} Neither pretraining nor retrieval fine-tuning uses teacher supervision. Future work could distill the 800M model into the 260M model at both stages and compare against matched training without distillation.

\noindent\textbf{Data bias.} Training data from the web and machine-generated sources can reproduce social, geographic, topical, and language biases \citep{singh2025globalmmlu}. A dedicated evaluation could measure how often the biases affect representations, retrieval rankings, and generated text, then compare data filtering and balancing methods.

\noindent\textbf{Safety.} Iterative masked-token sampling can produce harmful, private, or biased text even though \neomme{} is not a conversational model. Future work could measure this behavior across text and image inputs and different decoding methods.

\section{Conclusion}
\label{sec:conclusion}

\neomme{} addresses common limitations of multimodal encoders built by repurposing separate image and text encoders. The model maps multilingual text tokens and raw $32\times32$ image patches into one long-context bidirectional Transformer encoder trained from scratch. Dynamic resolution preserves image aspect ratios without patch merging, while the same masked-diffusion objective supports text-only and image-conditioned denoising. The 260M and 800M models share this architecture and are trained on multilingual text, code, mathematics, natural images, and document images.

\neomme{}-Retriever validates the shared backbone on visual document and text retrieval. One forward pass produces both dense and late-interaction representations for flexibility at inference time. The 260M model outperforms all evaluated models strictly below 800M parameters and, at a matched $2048\times2048$ input size on an NVIDIA L40S, encodes pages at about $2\times$ ColModernVBERT's throughput. The 800M model is competitive with models several times its size, and it encodes pages at much higher resolution for comparable throughput. Moreover, we show that hierarchical token pooling and asymmetric quantization reduce the late-interaction multimodal embeddings from roughly 1.5 MB to 6 kB per page, a $255\times$ compression, while retaining more than 95\% of the original retrieval quality.

We release the \neomme{} model checkpoints and a day-zero Hugging Face Transformers implementation to give practitioners a foundation to build efficient multimodal and multilingual representation models and advance research on the topic.

\section*{Acknowledgements}
We thank H Company for supporting our work and providing the compute to train \neomme{}.

\bibliographystyle{plainnat} \bibliography{references}

\appendix
\clearpage

\section{Training and implementation details}
\label{app:training-data}
This section describes the data mixtures, optimization settings, hardware, input formats, and sampling procedures used for pretraining and retrieval training.
\subsection{Pretraining data mixture}
\label{app:p1-data-mixture}

\subsubsection{Text-only stream}
\label{app:p1-text-mixture}

\autoref{tab:p1-text-mixture} and~\autoref{tab:p1-text-mixture-b} report the 58 datasets used in the pretraining text mixture, and the dataset names link to their Hugging Face dataset cards. The listed weights are configured source-sampling weights before document packing, not observed shares of tokens, documents, or examples. \autoref{tab:p1-text-descriptions} groups related language and configuration rows into dataset families while the two weight tables keep those rows separate.

\begin{table*}[htbp]
  \centering
  \caption{Pretraining text-source weights for English-dominant and multilingual web and PDF text.}
  \label{tab:p1-text-mixture}
  \small
  \begin{minipage}[t]{0.46\textwidth}
    \centering
    \textbf{English-dominant web and multilingual PDF text}\\[3pt]
    \begin{tabular}{lr}
      \toprule
      Dataset and config & Weight \\
      \midrule
      \multicolumn{2}{l}{\hflink{https://huggingface.co/datasets/HuggingFaceFW/fineweb-edu}{HuggingFaceFW/fineweb-edu}} \\
      \quad \code{default} & 0.167660 \\
      \multicolumn{2}{l}{\hflink{https://huggingface.co/datasets/nvidia/Nemotron-CC-v2.1}{nvidia/Nemotron-CC-v2.1}} \\
      \quad \code{High-Quality} & 0.072896 \\
      \addlinespace
      \multicolumn{2}{l}{\hflink{https://huggingface.co/datasets/HuggingFaceFW/finepdfs}{HuggingFaceFW/finepdfs}} \\
      \quad \code{eng\_Latn} & 0.072896 \\
      \quad \code{fra\_Latn} & 0.014579 \\
      \quad \code{deu\_Latn} & 0.014579 \\
      \quad \code{spa\_Latn} & 0.007290 \\
      \quad \code{rus\_Cyrl} & 0.007290 \\
      \quad \code{arb\_Arab} & 0.007290 \\
      \quad \code{cmn\_Hani} & 0.007290 \\
      \bottomrule
    \end{tabular}
  \end{minipage}
  \hfill
  \begin{minipage}[t]{0.49\textwidth}
    \centering
    \textbf{Multilingual web text}\\[3pt]
    \begin{tabular}{lr}
      \toprule
      \multicolumn{2}{l}{\hflink{https://huggingface.co/datasets/epfml/FineWeb2-HQ}{epfml/FineWeb2-HQ}} \\
      Config & Weight \\
      \midrule
      \code{deu\_Latn} & 0.021869 \\
      \code{fra\_Latn} & 0.021869 \\
      \code{spa\_Latn} & 0.021869 \\
      \code{rus\_Cyrl} & 0.021869 \\
      \code{cmn\_Hani} & 0.021869 \\
      \code{jpn\_Jpan} & 0.021869 \\
      \code{ita\_Latn} & 0.021869 \\
      \code{por\_Latn} & 0.021869 \\
      \code{pol\_Latn} & 0.014579 \\
      \code{nld\_Latn} & 0.014579 \\
      \code{ind\_Latn} & 0.014579 \\
      \code{tur\_Latn} & 0.014579 \\
      \code{ces\_Latn} & 0.014579 \\
      \code{vie\_Latn} & 0.014579 \\
      \code{swe\_Latn} & 0.014579 \\
      \code{fas\_Arab} & 0.014579 \\
      \code{arb\_Arab} & 0.014579 \\
      \code{ell\_Grek} & 0.014579 \\
      \code{dan\_Latn} & 0.014579 \\
      \code{hun\_Latn} & 0.014579 \\
      \bottomrule
    \end{tabular}
  \end{minipage}
\end{table*}

\begin{table*}[htbp]
  \centering
  \caption{Pretraining text-source weights for synthetic, reference, mathematics, question answering, and code.}
  \label{tab:p1-text-mixture-b}
  \small
  \begin{minipage}[t]{0.49\textwidth}
    \centering
    \textbf{Synthetic, reference, and mathematics}\\[3pt]
    \begin{tabular}{lr}
      \toprule
      Dataset and config & Weight \\
      \midrule
      \multicolumn{2}{l}{\hflink{https://huggingface.co/datasets/HuggingFaceTB/smollm-corpus}{HuggingFaceTB/smollm-corpus}} \\
      \quad \code{cosmopedia-v2} & 0.102054 \\
      \multicolumn{2}{l}{\hflink{https://huggingface.co/datasets/wikimedia/wikipedia}{wikimedia/wikipedia}} \\
      \quad \code{20231101.en} & 0.021869 \\
      \quad \code{20231101.fr} & 0.007290 \\
      \quad \code{20231101.de} & 0.007290 \\
      \quad \code{20231101.es} & 0.007290 \\
      \quad \code{20231101.ru} & 0.007290 \\
      \quad \code{20231101.zh} & 0.007290 \\
      \quad \code{20231101.ja} & 0.007290 \\
      \quad \code{20231101.it} & 0.007290 \\
      \quad \code{20231101.pt} & 0.007290 \\
      \quad \code{20231101.ar} & 0.007290 \\
      \multicolumn{2}{l}{\hflink{https://huggingface.co/datasets/HuggingFaceTB/finemath}{HuggingFaceTB/finemath}} \\
      \quad \code{finemath-4plus} & 0.051027 \\
      \bottomrule
    \end{tabular}
  \end{minipage}
  \hfill
  \begin{minipage}[t]{0.46\textwidth}
    \centering
    \textbf{Question answering and code}\\[3pt]
    \begin{tabular}{lr}
      \toprule
      Dataset, config & Weight \\
      \midrule
      \hflink{https://huggingface.co/datasets/cais/mmlu}{MMLU},
        \code{auxiliary\_train} & 0.006196 \\
      \hflink{https://huggingface.co/datasets/openlifescienceai/medmcqa}{MedMCQA},
        \code{default} & 0.004009 \\
      \hflink{https://huggingface.co/datasets/tau/commonsense_qa}{CommonsenseQA},
        \code{default} & 0.000219 \\
      \hflink{https://huggingface.co/datasets/allenai/qasc}{QASC},
        \code{default} & 0.000182 \\
      \hflink{https://huggingface.co/datasets/allenai/openbookqa}{OpenBookQA},
        \code{main} & 0.000109 \\
      \hflink{https://huggingface.co/datasets/allenai/ai2_arc}{AI2 ARC},
        \code{ARC-Easy} & 0.000058 \\
      \hflink{https://huggingface.co/datasets/allenai/ai2_arc}{AI2 ARC},
        \code{ARC-Challenge} & 0.000029 \\
      \addlinespace
      \multicolumn{2}{l}{\hflink{https://huggingface.co/datasets/bigcode/starcoderdata}{bigcode/starcoderdata}} \\
      \quad \code{python} & 0.005103 \\
      \quad \code{javascript} & 0.003645 \\
      \quad \code{typescript} & 0.002916 \\
      \quad \code{java} & 0.002916 \\
      \quad \code{c} & 0.002187 \\
      \quad \code{cpp} & 0.002187 \\
      \quad \code{go} & 0.002187 \\
      \quad \code{rust} & 0.002187 \\
      \quad \code{sql} & 0.001822 \\
      \quad \code{shell} & 0.001822 \\
      \bottomrule
    \end{tabular}
  \end{minipage}
\end{table*}

\begin{table}[H]
  \centering
  \caption{Pretraining text source descriptions.}
  \label{tab:p1-text-descriptions}
  \small
  \begin{tabular}{p{0.19\textwidth}p{0.24\textwidth}p{0.49\textwidth}}
    \toprule
    Group & Dataset family & Description \\
    \midrule
    \multirow{4}{0.19\textwidth}{Web and PDF}
      & \hflink{https://huggingface.co/datasets/HuggingFaceFW/fineweb-edu}{FineWeb-Edu} & English web text filtered for educational quality. \\
      & \hflink{https://huggingface.co/datasets/nvidia/Nemotron-CC-v2.1}{Nemotron-CC-v2.1} & High-quality English Common Crawl text curated by NVIDIA. \\
      & \hflink{https://huggingface.co/datasets/HuggingFaceFW/finepdfs}{FinePDFs} & Extracted PDF text in English, French, German, Spanish, Russian, Arabic, and Chinese. \\
      & \hflink{https://huggingface.co/datasets/epfml/FineWeb2-HQ}{FineWeb2-HQ} & High-quality web text in 20 languages. \\
    \addlinespace
    \multirow{3}{0.19\textwidth}{Synthetic, reference, and math}
      & \hflink{https://huggingface.co/datasets/HuggingFaceTB/smollm-corpus}{Cosmopedia-v2} & Synthetic educational text generated from web and reference material. \\
      & \hflink{https://huggingface.co/datasets/wikimedia/wikipedia}{Wikipedia} & Encyclopedic articles in 10 languages. \\
      & \hflink{https://huggingface.co/datasets/HuggingFaceTB/finemath}{FineMath} & Mathematical web text from the \code{finemath-4plus} configuration. \\
    \addlinespace
    \multirow{6}{0.19\textwidth}{Question answering}
      & \hflink{https://huggingface.co/datasets/cais/mmlu}{MMLU} & Multiple-choice questions spanning academic and professional subjects. \\
      & \hflink{https://huggingface.co/datasets/openlifescienceai/medmcqa}{MedMCQA} & Multiple-choice medical entrance-exam questions. \\
      & \hflink{https://huggingface.co/datasets/tau/commonsense_qa}{CommonsenseQA} & Multiple-choice questions that test everyday commonsense knowledge. \\
      & \hflink{https://huggingface.co/datasets/allenai/qasc}{QASC} & Multiple-choice science questions paired with supporting facts. \\
      & \hflink{https://huggingface.co/datasets/allenai/openbookqa}{OpenBookQA} & Elementary science questions based on an open book of scientific facts. \\
      & \href{https://huggingface.co/datasets/allenai/ai2_arc}{\hfemoji\ AI2 ARC} & Grade-school science questions from the Easy and Challenge configurations. \\
    \addlinespace
    Code
      & \hflink{https://huggingface.co/datasets/bigcode/starcoderdata}{StarCoderData} & Source code in Python, JavaScript, TypeScript, Java, C, C++, Go, Rust, SQL, and shell. \\
    \bottomrule
  \end{tabular}
\end{table}
\subsubsection{Visual--text stream}
\label{app:p1-vision-mixture}

The visual--text stream combines five source families covering natural images, rendered documents, PDF pages, charts, diagrams, and synthetic images paired with captions, optical character recognition text, recaptions, Markdown transcriptions, or synthetic text. \autoref{tab:p1-vision-mixture} reports each family's configured sampling weight and role. These weights define the distribution within the visual--text stream, not proportions of the complete pretraining mixture or observed shares of images, pixels, or tokens.

\begin{table}[H]
  \centering
  \caption{Pretraining visual--text source weights and descriptions.}
  \label{tab:p1-vision-mixture}
  \scriptsize
  \begin{tabular}{p{0.28\textwidth}p{0.08\textwidth}p{0.54\textwidth}}
    \toprule
    Dataset family & Weight & Description \\
    \midrule
    \hflink{https://huggingface.co/datasets/HuggingFaceM4/FineVision}{FineVision}
      & 0.50 & A selected mixture of 45 vision--language and document datasets, predominantly in English. \\
    \hflink{https://huggingface.co/datasets/pixparse/pdfa-eng-wds}{PDFA} and \hflink{https://huggingface.co/datasets/lightonai/LightOnOCR-mix-0126}{LightOnOCR}
      & 0.30 & PDF page images paired with extracted text and model-generated recaptions where available. \\
    \hflink{https://huggingface.co/datasets/tomg-group-umd/pixelprose}{PixelProse}
      & 0.10 & URL-fetched natural images paired with Gemini-generated captions. \\
    \hflink{https://huggingface.co/datasets/ahmedheakl/docatlas_instruct}{DocAtlas}
      & 0.07 & Rendered document images paired with Markdown targets in 82 languages. \\
    \href{https://huggingface.co/datasets/nvidia/OCR-Synthetic-Multilingual-v1}{\hfemoji\ Synthetic multilingual OCR}
      & 0.03 & Synthetic document images and text in Japanese, Korean, simplified Chinese, and traditional Chinese. \\
    \midrule
    \textbf{Total} & \textbf{1.00} & \\
    \bottomrule
  \end{tabular}
\end{table}

\FloatBarrier
\subsection{Retrieval training settings}
\label{app:retrieval-details}

\neomme{}-Retriever was fine-tuned using one \code{p5.48xlarge}\textsuperscript{\ref{fn:aws-p5}} node with eight H100 accelerators. \autoref{tab:p2-training-settings} summarizes the retrieval-training settings for both model sizes. \autoref{tab:p1-training} reports the more detailed pretraining settings in the main pretraining section.

\begin{table*}[htbp]
  \centering
  \caption{Retrieval-training settings.}
  \label{tab:p2-training-settings}
  \small
  \begin{tabular}{p{0.30\linewidth}rr}
    \toprule
    Setting & \neomme-260M & \neomme-800M \\
    \midrule
    \multicolumn{3}{l}{\textbf{Initialization and schedule}} \\
    Pretraining initialization &
    Step 500,000\textsuperscript{\ref{fn:neomme-260m-hf}} &
    Step 500,000\textsuperscript{\ref{fn:neomme-800m-hf}} \\
    Nodes & 1 & 1 \\
    H100 accelerators & 8 & 8 \\
    Steps & 20,000 & 20,000 \\
    Sequences per rank & 8 & 8 \\
    \addlinespace
    \multicolumn{3}{l}{\textbf{Optimization}} \\
    NorMuon learning rate & 0.0015 & 0.0010 \\
    AdamW learning rate & 0.00025 & 0.0001667 \\
    Warmup & 100 steps & 100 steps \\
    Schedule & Cosine to 1\% of peak & Cosine to 1\% of peak \\
    \addlinespace
    \multicolumn{3}{l}{\textbf{Representations}} \\
    Dense pooling & Mean pooling & Mean pooling \\
    Late-interaction width & 128 & 128 \\
    Dense widths & 128, 256, 512, 1,024 &
    128, 256, 512, 1,024, 1,792 \\
    \addlinespace
    \multicolumn{3}{l}{\textbf{Batch construction}} \\
    Hard negatives & 7 from a 32-candidate window & 7 from a 32-candidate window \\
    Query expansion tokens & 10 & 10 \\
    Temperature & 0.02 & 0.02 \\
    GradCache micro-batch & 24 & 12 \\
    Text:image batch ratio & 1:2 & 1:2 \\
    \bottomrule
  \end{tabular}
\end{table*}

\subsection{Processor}
\label{app:retrieval-processor}

\definecolor{fmtquery}{RGB}{126,52,153}
\definecolor{fmtdoc}{RGB}{27,112,170}
\definecolor{fmtimage}{RGB}{213,94,0}
\definecolor{fmtrow}{RGB}{0,145,120}
\definecolor{fmtmask}{RGB}{194,37,37}
\definecolor{fmtpatch}{RGB}{92,92,92}
\newcommand{\queryfmt}[1]{\textcolor{fmtquery}{\texttt{#1}}}
\newcommand{\docfmt}[1]{\textcolor{fmtdoc}{\texttt{#1}}}
\newcommand{\imagefmt}[1]{\textcolor{fmtimage}{\texttt{#1}}}
\newcommand{\rowfmt}[1]{\textcolor{fmtrow}{\texttt{#1}}}
\newcommand{\maskfmt}[1]{\textcolor{fmtmask}{\texttt{#1}}}
\newcommand{\patchfmt}[1]{\textcolor{fmtpatch}{\texttt{#1}}}

The processor notation uses $q_s$ for a query token, $d_t$ for a document token, $L_q$ and $L_d$ for the query and document lengths, and $p_{r,c}$ for the patch at grid position $(r,c)$ in an $H$-row by $W$-column patch grid.

\begin{table}[htbp]
  \centering
  \caption{\neomme{} processor format. Colors mark structural roles rather than model inputs.}
  \label{tab:processor-format}
  \scriptsize
  \setlength{\tabcolsep}{2pt}
  \begin{tabular}{p{0.28\textwidth}p{0.66\textwidth}}
    \toprule
    \textbf{Context} & \textbf{Formatting} \\
    \midrule
    Query prefix & \queryfmt{[QUERY]} \\
    Document prefix & \docfmt{[DOC]} \\
    Image stream prefix & \imagefmt{[IMG]} \\
    Patch-row boundary & \rowfmt{[ROW]} \\
    Query expansion & \maskfmt{[MASK]} \\
    Image patch at grid position $(r,c)$ & \patchfmt{$p_{r,c}$} \\
    \midrule
    \multicolumn{2}{c}{\textbf{Resulting model input sequence}} \\
    \midrule
    \faQuestionCircle\ Query &
    \queryfmt{[QUERY]}\; \texttt{$q_1\ \cdots\ q_{L_q}$}\;
    \maskfmt{[MASK]$^{\times 10}$} \\
    \faFilePdf\ Text document &
    \docfmt{[DOC]}\;\texttt{$d_1\ \cdots\ d_{L_d}$} \\
    \faImage\ Page screenshot &
    \docfmt{[DOC]}\;\imagefmt{[IMG]}\;
    \patchfmt{$[(p_{r,0}\ \cdots\ p_{r,W-1})$}\;\rowfmt{[ROW]}\patchfmt{$]_{r=0}^{H-1}$} \\
    \bottomrule
  \end{tabular}
\end{table}

The processor inserts learned structural tokens into each model input. \queryfmt{[QUERY]} and \docfmt{[DOC]} distinguish the two retrieval sides, while image documents add \imagefmt{[IMG]} before the row-major patch grid and \rowfmt{[ROW]} after each patch row to preserve its two-dimensional structure. Queries also append 10 learned \maskfmt{[MASK]} tokens, following ColBERT's query augmentation \citep{khattab2020colbert}; their effect is not isolated in an ablation. \autoref{tab:processor-format} gives the full layouts.

\FloatBarrier
\subsection{Training mixture details}

\noindent\textbf{Text-only stream.} The text stream contains 1,581,039 queries across five source pools. The LightOn sources provide the English, mLateOn, code, and organic pools \citep{sourty2026denseon}, while an additional multilingual pool supplies the rest. The translated mLateOn pools add further language coverage. The weights in \autoref{tab:p2-text-data} define the sampling distribution within this stream.

\begin{table}[htbp]
  \centering
  \caption{Retrieval text-source weights.}
  \label{tab:p2-text-data}
  \small
  \begin{tabular}{p{0.34\textwidth}rp{0.35\textwidth}}
    \toprule
    Source pool & Weight & Coverage \\
    \midrule
    \href{https://huggingface.co/datasets/lightonai/embeddings-fine-tuning-filtered-en}{\hfemoji\ Filtered English retrieval} & 0.09 & English, 10 candidates per query \\
    \href{https://huggingface.co/collections/lightonai/mdenseon-and-mlateon}{\hfemoji\ mLateOn multilingual} & 0.73 & Multilingual retrieval splits \\
    \href{https://huggingface.co/datasets/lightonai/embeddings-fine-tuning-filtered-code}{\hfemoji\ Code retrieval} & 0.13 & Code-search splits \\
    Additional multilingual retrieval & 0.03 & 13 languages, eight candidates per query \\
    \href{https://huggingface.co/datasets/lightonai/embeddings-fine-tuning-filtered-en}{\hfemoji\ Organic retrieval (en)}\newline \href{https://huggingface.co/datasets/lightonai/embeddings-fine-tuning-filtered-fr}{\hfemoji\ Organic retrieval (fr)} & 0.02 & English and French \\
    \midrule
    Total & 1.00 & \\
    \bottomrule
  \end{tabular}
\end{table}

\noindent\textbf{Multimodal stream.} The multimodal stream combines the ColPali train set, a multilingual document-image dataset, and the VisRAG in-domain and synthetic sets \citep{faysse2025colpali,yu2024visrag}. Each source has a configured weight of 0.25. \autoref{tab:p2-image-data} gives the final mixture sizes, and multiple queries can point to the same document. The original pairs cover English, French, German, Italian, and Spanish.

\begin{table}[htbp]
  \centering
  \caption{Image retrieval mixture.}
  \label{tab:p2-image-data}
  \small
  \begin{tabular}{lrrrl}
    \toprule
    Source & Weight & \# Queries & \# Documents & Languages \\
    \midrule
    \hflink{https://huggingface.co/datasets/vidore/colpali_train_set}{ColPali} & 0.25 & 148,123 & 118,195 & en \\
    Multilingual document images & 0.25 & 351,651 & 280,679 & en, it, fr, de, es \\
    \hflink{https://huggingface.co/datasets/openbmb/VisRAG-Ret-Train-In-domain-data}{VisRAG} & 0.25 & 154,187 & 122,752 & en \\
    \href{https://huggingface.co/datasets/openbmb/VisRAG-Ret-Train-Synthetic-data}{\hfemoji\ VisRAG synthetic} & 0.25 & 298,780 & 239,200 & en \\
    \midrule
    Total & 1.00 & 952,741 & 760,826 & 6 \\
    \bottomrule
  \end{tabular}
  \par\vspace{2pt}
  \small
  Query augmentation adds Portuguese, bringing the total to six languages.
\end{table}

\FloatBarrier
\noindent\textbf{Query augmentation.} Before hard-negative mining, we sampled 100,000 of the 760,826 pages and asked Qwen3.5-9B\textsuperscript{\ref{fn:qwen35-9b-hf}} to generate two queries per page. For each query, we uniformly sampled one of six types: multi-hop, compare-contrast, open-ended, boolean, enumerative, or numerical. We sampled question, instruction, and keyword formats with probabilities 0.60, 0.25, and 0.15. We sampled English with probability 0.35, French with probability 0.25, and Spanish, German, Italian, and Portuguese with probability 0.10 each. The prompt required a precise grounded answer and wording that differed from the page. After filtering for parsable, self-contained queries, 191,915 queries remained, increasing the multimodal query count from 760,826 to 952,741 without adding pages. During hard-negative mining, Qwen3-VL-Reranker-8B\textsuperscript{\ref{fn:qwen3-vl-reranker-hf}} kept generated pairs whose positive page scored at least 0.1.

\FloatBarrier
\subsection{Contrastive training details}
\label{app:contrastive-training}

\noindent\textbf{Hard-negative sampling.} Text examples use the hard negatives provided with the mLateOn multilingual dataset referenced in \autoref{tab:retrieval-mixture}. Visual hard negatives are self-mined in two stages. The first stage trains \neomme{}-Retriever with positive pages and in-batch negatives but no explicit hard negatives. The resulting model checkpoint then mines a fixed 32-candidate window for each visual query. The second stage repeats the same retrieval fine-tuning protocol with one positive and seven negatives sampled from that window. Cross-rank gathering adds positives and negatives from the other ranks to the candidate pool. Negatives with stored mining scores at least $0.98\times$ the positive, and nonfinite candidates, are removed. The gate does not recompute the score with the current model during training.

\noindent\textbf{Memory-efficient training.} GradCache separates representation computation from the contrastive loss, so the encoder can process the gathered candidate pool in smaller activation chunks \citep{gao2021gradcache}. \lik{} fuses token dot products with an online maximum reduction, which avoids writing the full token-similarity tensor to high-bandwidth memory and reduces the MaxSim memory peak \citep{lac2026lik}. The \neomme-260M throughput control averaged 1.6\% higher throughput with \lik{}, although run-to-run variation overlaps. \autoref{tab:lik-throughput} and~\autoref{tab:lik-training-resources} report the measurements.

\FloatBarrier

\section{Supporting pretraining results}
\label{app:pretraining-support}

\subsection{Tokenizer evaluation}
\label{app:tokenizer}

All tokenizers encode identical texts without added special tokens. We compare \neomme{} with ModernBERT \citep{warner2024modernbert}, LFM2.5-Encoder-230M \citep{liquidai2026lfm25encoders}, mmBERT-base \citep{marone2025mmbert}, and EuroBERT-210m \citep{boizard2025eurobert}. The language evaluation uses all 1,012 aligned sentences per language in FLORES-200 devtest \citep{nllb2022}, which keeps the compared content fixed across languages. The aggregate reductions reported in \autoref{sec:architecture} are ratios of token counts summed over the 14 languages, not averages of the per-language rates below. The domain evaluation uses 2,000 documents per source, each capped at 4,000 characters. Flag icons come from \href{https://github.com/twitter/twemoji}{Twemoji} under CC-BY 4.0.

\begin{table}[htbp]
  \centering
  \caption{Tokenizer compression on the 14 target languages in FLORES-200 devtest. All tokenizers encode identical texts without added special tokens. Values are tokens per UTF-8 byte, and lower values are better.}
  \label{tab:tokenizer-metrics}
  \scriptsize
  \setlength{\tabcolsep}{3.5pt}
  \begin{tabular}{lrrrrr}
    \toprule
    Source &
    \neomme{}\textsuperscript{\ref{fn:neomme-260m-hf}} &
    ModernBERT\tablefootnote{\label{fn:tokenizer-modernbert}\hflink{https://huggingface.co/answerdotai/ModernBERT-base}{answerdotai/ModernBERT-base}.} &
    LFM2.5\tablefootnote{\label{fn:lfm25-encoder-hf}\hflink{https://huggingface.co/LiquidAI/LFM2.5-Encoder-230M}{LiquidAI/LFM2.5-Encoder-230M}.} &
    mmBERT\tablefootnote{\label{fn:tokenizer-mmbert}\hflink{https://huggingface.co/jhu-clsp/mmBERT-base}{jhu-clsp/mmBERT-base}.} &
    EuroBERT\tablefootnote{\label{fn:tokenizer-eurobert}\hflink{https://huggingface.co/EuroBERT/EuroBERT-210m}{EuroBERT/EuroBERT-210m}.} \\
    Vocabulary & 131,072 & 50,368 & 65,536 & 256,000 & 128,256 \\
    \midrule
    \flagemoji{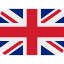} English & \textbf{0.1930} & 0.2075 & 0.2097 & 0.2044 & 0.2057 \\
    \flagemoji{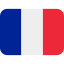} French & \textbf{0.2122} & 0.2926 & 0.2434 & 0.2243 & 0.2652 \\
    \flagemoji{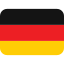} German & 0.2306 & 0.3194 & 0.2452 & \textbf{0.2165} & 0.2748 \\
    \flagemoji{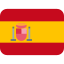} Spanish & 0.2154 & 0.2942 & 0.2478 & \textbf{0.2136} & 0.2617 \\
    \flagemoji{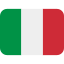} Italian & \textbf{0.2265} & 0.3107 & 0.2685 & 0.2313 & 0.2830 \\
    \flagemoji{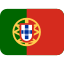} Portuguese & \textbf{0.2222} & 0.3064 & 0.2725 & 0.2250 & 0.2713 \\
    \flagemoji{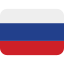} Russian & 0.1533 & 0.2779 & 0.1812 & \textbf{0.1432} & 0.1687 \\
    \flagemoji{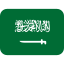} Arabic & \textbf{0.1661} & 0.3865 & 0.2482 & 0.1904 & 0.2126 \\
    \flagemoji{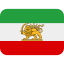} Persian & \textbf{0.1622} & 0.4165 & 0.3895 & 0.1742 & 0.1851 \\
    \flagemoji{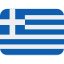} Greek & \textbf{0.1576} & 0.3209 & 0.5066 & 0.2154 & 0.2112 \\
    \flagemoji{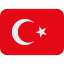} Turkish & \textbf{0.2344} & 0.4079 & 0.4209 & 0.2558 & 0.2576 \\
    \flagemoji{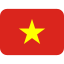} Vietnamese & \textbf{0.1790} & 0.4798 & 0.5565 & 0.2019 & 0.2059 \\
    \flagemoji{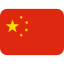} Chinese & \textbf{0.2419} & 0.4671 & 0.2991 & 0.2456 & 0.2881 \\
    \flagemoji{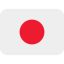} Japanese & \textbf{0.1896} & 0.3572 & 0.2293 & 0.1941 & 0.2457 \\
    \bottomrule
  \end{tabular}
\end{table}

\begin{table}[htbp]
  \centering
  \caption{Tokenizer compression on document, math, and code sources. All tokenizers encode the same 2,000 documents per source without added special tokens. Values are tokens per UTF-8 byte, and lower values are better.}
  \label{tab:tokenizer-metrics-b}
  \scriptsize
  \setlength{\tabcolsep}{3.5pt}
  \begin{tabular}{lrrrrr}
    \toprule
    Source &
    \neomme{}\textsuperscript{\ref{fn:neomme-260m-hf}} &
    ModernBERT\textsuperscript{\ref{fn:tokenizer-modernbert}} &
    LFM2.5\textsuperscript{\ref{fn:lfm25-encoder-hf}} &
    mmBERT\textsuperscript{\ref{fn:tokenizer-mmbert}} &
    EuroBERT\textsuperscript{\ref{fn:tokenizer-eurobert}} \\
    Vocabulary & 131,072 & 50,368 & 65,536 & 256,000 & 128,256 \\
    \midrule
    \faFilePdf\ English PDF & 0.2439 & 0.2535 & 0.2552 & 0.2616 & \textbf{0.2390} \\
    \faCalculator\ Math & \textbf{0.2510} & 0.2865 & 0.2889 & 0.2989 & 0.2792 \\
    \faPython\ Python & \textbf{0.2148} & 0.3092 & 0.2869 & 0.4059 & 0.2428 \\
    \faJsSquare\ JavaScript & \textbf{0.2092} & 0.3086 & 0.2850 & 0.3964 & 0.2462 \\
    \faDatabase\ SQL & \textbf{0.2665} & 0.3605 & 0.3533 & 0.3852 & 0.2924 \\
    \bottomrule
  \end{tabular}
\end{table}

Across all 204 FLORES-200 language-script pairs, \neomme{} emits 65.29\% more tokens than mmBERT and 15.28\% more than EuroBERT in aggregate, with the largest deficits in Tibetan, South Asian, and Southeast Asian scripts.

\FloatBarrier

\subsection{Image-conditioned generation examples}
\label{app:generation-examples}
The selected pretraining generations below use \neomme-260M\textsuperscript{\ref{fn:neomme-260m-hf}} and are qualitative examples rather than task-level captioning or optical character recognition measurements.

\begin{table}[!htbp]
  \centering
  \caption{Selected image inputs and text generated by the \neomme-260M model. \code{[...]} marks truncation.}
  \label{tab:generation-examples}
  \small
  \renewcommand{\arraystretch}{1.25}
  \begin{tabular}{
    >{\centering\arraybackslash}m{0.34\textwidth} m{0.59\textwidth}
  }
    \toprule
    Image & Predicted text \\
    \midrule
    \includegraphics[
      width=0.27\textwidth,
      height=0.18\textheight,
      keepaspectratio
    ]{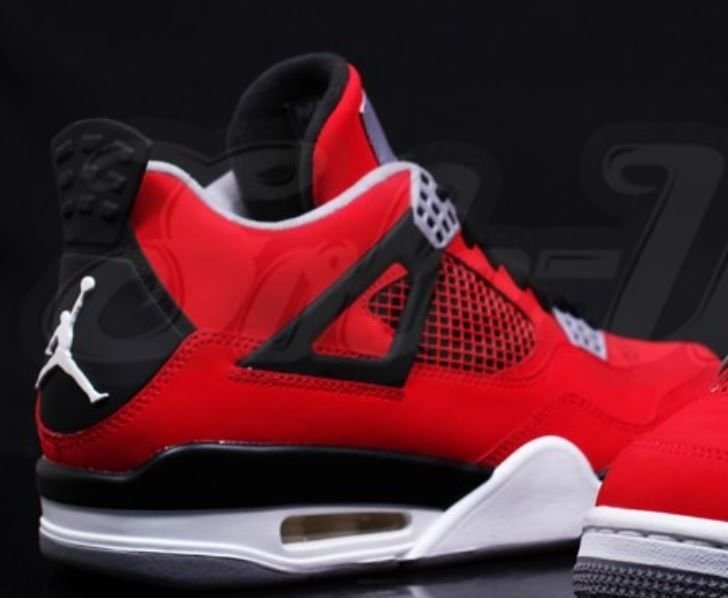}
    &
    \code{This image displays: A pair of red Nike sneakers with black laces and white Nike logo accents. The sneakers are displayed on a black background.}
    \\
    \midrule
    \includegraphics[
      width=0.27\textwidth,
      height=0.18\textheight,
      keepaspectratio
    ]{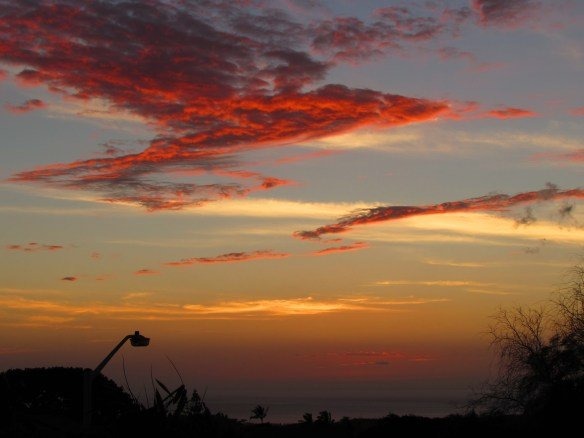}
    &
    \code{This image displays: A sunset over a river. The sky is orange and yellow, with a few clouds. The river is blue and still. There are no people in the image. The image is a photograph.}
    \\
    \midrule
    \includegraphics[
      width=0.30\textwidth,
      height=0.24\textheight,
      keepaspectratio
    ]{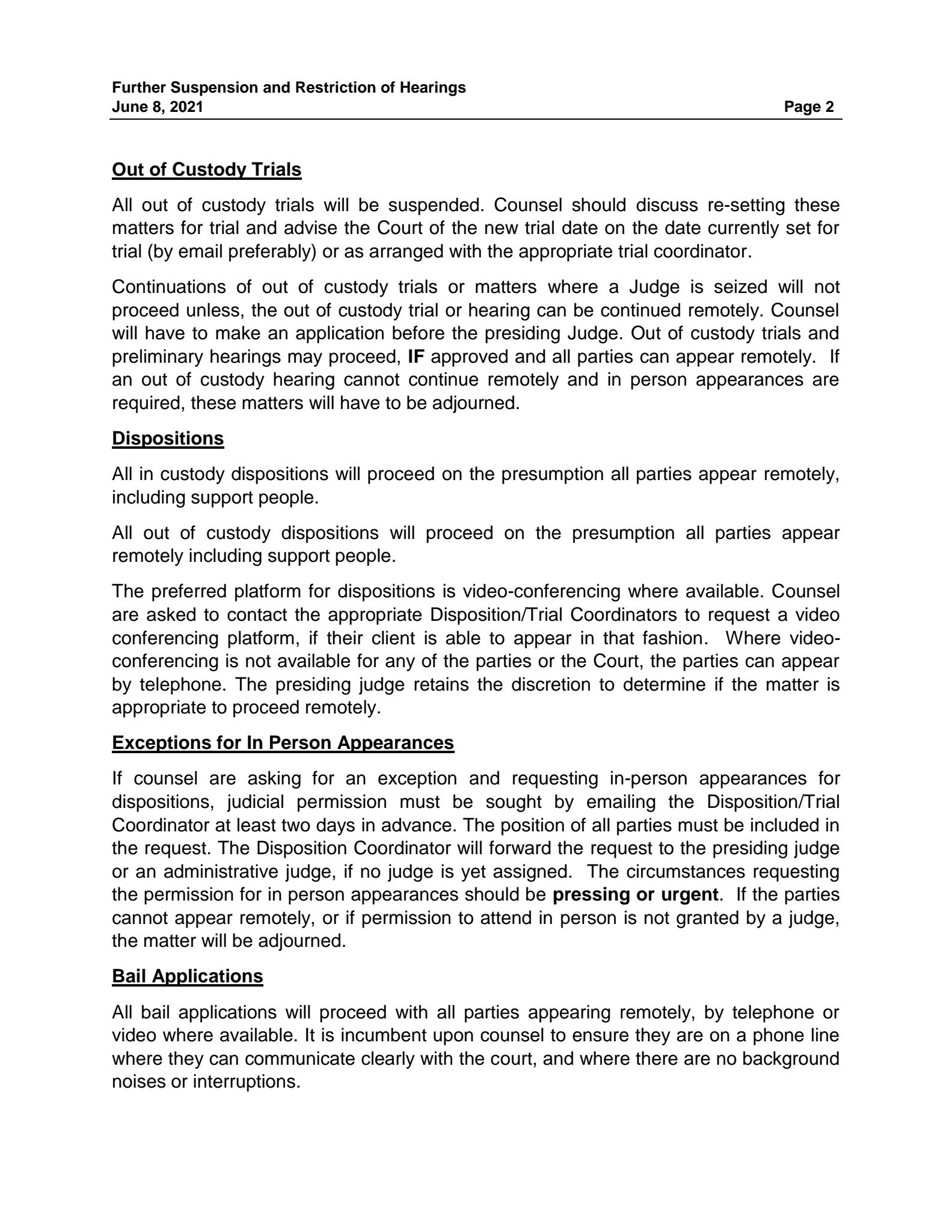}
    &
    \code{Further Suspension and Restriction of Hearings}\newline
    \code{June 8, 2021}\newline
    \code{Page 2}\newline
    \code{\#\# Out of Custody Trials}\newline
    \code{All out of custody trials will be suspended. Counsel should discuss re-setting these matters for trial and advise the Court of the new trial date on the date currently set for trial}\newline
    \code{[...]}
    \\
    \midrule
    \includegraphics[
      width=0.32\textwidth,
      height=0.21\textheight,
      keepaspectratio
    ]{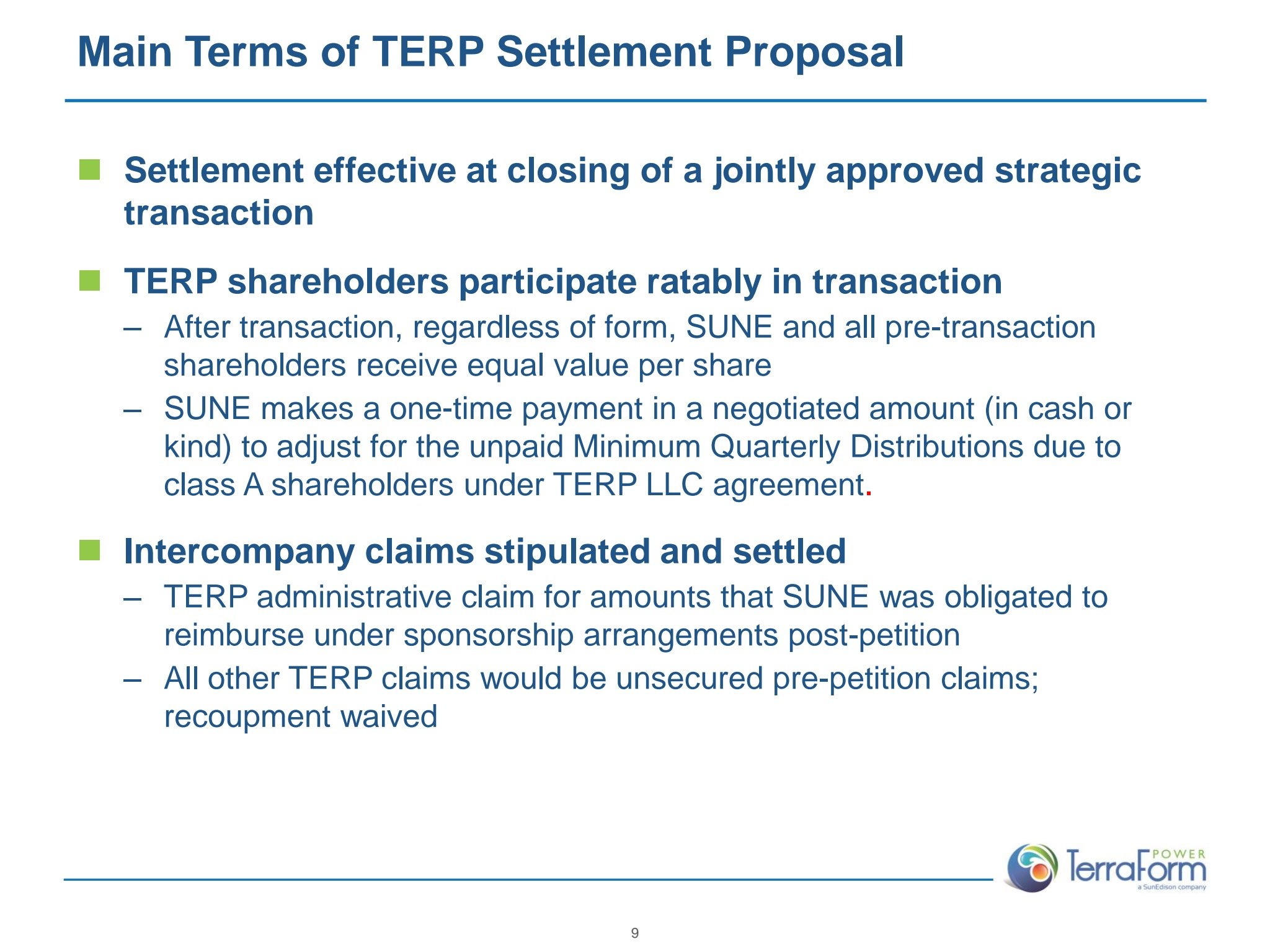}
    &
    \code{\# Main Terms of TERP Settlement Proposal}\newline
    \code{\#\# Settlement effective at closing of a jointly approved strategic transaction}\newline
    \code{\#\#\# TERP shareholders participate ratably in transaction}\newline
    \code{- After transaction, regardless of form, SUNE and all pre-transaction shareholders receive equal value per share}\newline
    \code{---}\newline
    \code{9}\newline
    \code{[...]}
    \\
    \bottomrule
  \end{tabular}
\end{table}

\FloatBarrier

\section{Retrieval evaluation and complete results}
\label{app:retrieval-evidence}
This section describes the retrieval evaluation protocol, complete metric and domain breakdowns, all trained Matryoshka widths, and supporting ablations \citep{kusupati2022matryoshka}.

\subsection{Evaluation protocol details}
\label{app:retrieval-evaluation-details}
\noindent\textbf{Benchmarks.} ViDoRe v1, v2, and v3 test document retrieval \citep{mace2025vidorev2,loison2026vidorev3}. BEIR-15 follows the MTEB 2.18.7 benchmark definition and matches the mLateOn evaluation task list \citep{thakur2021beir,muennighoff2023mteb}. It contains 14 of the original 18 zero-shot BEIR datasets and adds the MS MARCO development set. BioASQ, Signal-1M, TREC-NEWS, and Robust04 are omitted because their corpora are not distributed as downloadable BEIR datasets and require separate reconstruction. CQADupstack's 12 child corpora are averaged through the official aggregate task and counted once.

\noindent\textbf{Scoring.} The visual evaluation uses downscale-only resizing with a maximum image side length of 2,048 pixels and bfloat16 execution. MeanMaxSim uses exact all-pairs scoring for ViDoRe. BEIR-15 uses the official MTEB scorer over each full corpus, with 8,192-token query and document limits, title followed by body, a compressed multi-vector index, and exact rescoring of top candidates. The dense head uses the same pipeline with one normalized vector per input.

\noindent\textbf{Metric aggregation and evaluation.} For each task, we average the metric over judged queries and then give every task equal weight in the reported aggregate. ViDoRe v3 contains eight reported tasks, and BEIR-15 contains 15 tasks. The reported results use the 260M and 800M models at step 20,000.

\FloatBarrier

\subsection{Complete retrieval results}
\label{app:retrieval-results}
\noindent\textbf{ViDoRe benchmark comparison.} \autoref{tab:vidore-v3-pareto-models} lists every model and retrieval head included in \autoref{fig:vidore-v3-model-size}. The ViDoRe v3 column is the task-macro mean over the benchmark tasks. We reran ColModernVBERT on v1 to match the exact protocol used for \neomme{}, and on v2 because MTEB does not provide a full-benchmark result \citep{teiletche2025modernvbert}. We reran ColSmol-500M on v3 because MTEB does not provide a full-benchmark result there either.

\begin{table*}[htbp]
  \centering
  \caption{Visual document retrieval performance on the ViDoRe benchmarks for all compared models. A dash marks a result that was not reported. Emb. dim. is the per-token vector width for late-interaction models and the single-vector width for dense models.}
  \label{tab:vidore-v3-pareto-models}
  \scriptsize
  \setlength{\tabcolsep}{2.6pt}
  \begin{tabular}{@{}lcrrrrr@{}}
  \toprule
  \multicolumn{4}{c}{Model details} & \multicolumn{3}{c}{ViDoRe} \\
  \cmidrule(lr){1-4}\cmidrule(lr){5-7}
  Model & Late-interaction & Params. (B) & Emb. dim. & v1 nDCG@5 & v2 nDCG@5 & v3 nDCG@10 \\
  \midrule
  \multicolumn{7}{l}{\paramsmall} \\
  BiModernVBERT\tablefootnote{\label{fn:results-bimodernvbert}\hflink{https://huggingface.co/ModernVBERT/bimodernvbert}{ModernVBERT/bimodernvbert} \citep{teiletche2025modernvbert}.}\textsuperscript{\textdagger} & & 0.25 & 768 & -- & -- & 0.1541 \\
  ColModernVBERT\textsuperscript{\ref{fn:results-colmodernvbert}} & \checkmark & 0.25 & 128 & 0.8058\textsuperscript{\ensuremath{\ddagger}} & 0.4068\textsuperscript{\ensuremath{\ddagger}} & 0.2612\textsuperscript{\textdagger} \\
  ColSmol-256M\textsuperscript{\ref{fn:results-colsmol}}\textsuperscript{\textdagger} & \checkmark & 0.26 & 128 & 0.7974 & 0.3482 & 0.2073 \\
  \neomme-260M, dense\textsuperscript{\ref{fn:results-neomme-260m}}\textsuperscript{\ensuremath{\ddagger}} & & 0.26 & 1,024 & 0.7552 & 0.4075 & 0.3907 \\
  \neomme-260M, late-interaction\textsuperscript{\ref{fn:results-neomme-260m}}\textsuperscript{\ensuremath{\ddagger}} & \checkmark & 0.26 & 128 & 0.8598 & 0.5218 & 0.5226 \\
  \midrule
  \multicolumn{7}{l}{\parammedium} \\
  ColSmol-500M\textsuperscript{\ref{fn:results-colsmol-500m}} & \checkmark & 0.50 & 128 & 0.8249\textsuperscript{\textdagger} & 0.4550\textsuperscript{\textdagger} & 0.3397\textsuperscript{\ensuremath{\ddagger}} \\
  SigLIP-So400M\tablefootnote{\label{fn:results-siglip}\hflink{https://huggingface.co/google/siglip-so400m-patch14-384}{google/siglip-so400m-patch14-384} \citep{zhai2023siglip}.}\textsuperscript{\textdagger} & & 0.88 & 1,152 & 0.5638 & 0.3300 & 0.1723 \\
  \neomme-800M, dense\textsuperscript{\ref{fn:results-neomme-800m}}\textsuperscript{\ensuremath{\ddagger}} & & 0.80 & 1,792 & 0.7993 & 0.4475 & 0.4391 \\
  \neomme-800M, late-interaction\textsuperscript{\ref{fn:results-neomme-800m}}\textsuperscript{\ensuremath{\ddagger}} & \checkmark & 0.80 & 128 & 0.8744 & 0.5591 & 0.5560 \\
  Vultron Flash (0.8B)\textsuperscript{\ref{fn:results-vultron}}\textsuperscript{\textdagger} & \checkmark & 0.85 & 320 & 0.8815 & 0.6036 & 0.5649 \\
  \midrule
  \multicolumn{7}{l}{\paramlarge} \\
  Qwen3-VL (2B)\tablefootnote{\label{fn:results-qwen3vl-2b}\hflink{https://huggingface.co/Qwen/Qwen3-VL-Embedding-2B}{Qwen/Qwen3-VL-Embedding-2B} \citep{li2026qwen3vlembedding}.}\textsuperscript{\textdagger} & & 2.13 & 2,048 & -- & -- & 0.5289 \\
  ColQwen2-v1.0\tablefootnote{\label{fn:results-colqwen2}\hflink{https://huggingface.co/vidore/colqwen2-v1.0}{vidore/colqwen2-v1.0} \citep{faysse2025colpali}.}\textsuperscript{\textdagger} & \checkmark & 2.21 & 128 & 0.8923 & 0.5604 & 0.4418 \\
  ColPali-v1.3\textsuperscript{\ref{fn:results-colpali}}\textsuperscript{\textdagger} & \checkmark & 2.92 & 128 & 0.8475 & 0.5472 & 0.4295 \\
  NemoRetriever (1B)\tablefootnote{\label{fn:results-nemoretriever-1b}\hflink{https://huggingface.co/nvidia/llama-nemoretriever-colembed-1b-v1}{nvidia/llama-nemoretriever-colembed-1b-v1} \citep{xu2025nemoretriever}.}\textsuperscript{\textdagger} & \checkmark & 2.42 & 2,048 & 0.9050 & 0.6296 & 0.5548 \\
  ColNomic (3B)\tablefootnote{\label{fn:results-colnomic-3b}\hflink{https://huggingface.co/nomic-ai/colnomic-embed-multimodal-3b}{nomic-ai/colnomic-embed-multimodal-3b} \citep{nomic2025multimodal}.}\textsuperscript{\textdagger} & \checkmark & 3.00 & 128 & 0.8986 & 0.5568 & 0.5640 \\
  ColQwen2.5-v0.2\textsuperscript{\ref{fn:results-colqwen25}}\textsuperscript{\textdagger} & \checkmark & 3.75 & 128 & 0.8954 & 0.6006 & 0.5244 \\
  Tomoro (4B)\tablefootnote{\label{fn:results-tomoro-4b}\hflink{https://huggingface.co/TomoroAI/tomoro-colqwen3-embed-4b}{TomoroAI/tomoro-colqwen3-embed-4b} \citep{huang2025tomoro}.}\textsuperscript{\textdagger} & \checkmark & 4.00 & 320 & 0.9057 & 0.6469 & 0.6016 \\
  Jina v4\tablefootnote{\label{fn:results-jina-v4}\hflink{https://huggingface.co/jinaai/jina-embeddings-v4}{jinaai/jina-embeddings-v4} \citep{gunther2025jina}.}\textsuperscript{\textdagger} & & 3.93 & 2,048 & 0.9035 & 0.5823 & 0.4961 \\
  NemoRetriever (3B v1)\tablefootnote{\label{fn:results-nemoretriever-3b}\hflink{https://huggingface.co/nvidia/llama-nemoretriever-colembed-3b-v1}{nvidia/llama-nemoretriever-colembed-3b-v1} \citep{xu2025nemoretriever}.}\textsuperscript{\textdagger} & \checkmark & 4.41 & 3,072 & 0.9100 & 0.6332 & 0.5707 \\
  Nemo ColEmbed (3B v2)\tablefootnote{\label{fn:results-nemo-colembed-3b-v2}\hflink{https://huggingface.co/nvidia/llama-nemotron-colembed-vl-3b-v2}{nvidia/llama-nemotron-colembed-vl-3b-v2} \citep{moreira2026nemotron}.}\textsuperscript{\textdagger} & \checkmark & 4.41 & 3,072 & 0.9174 & 0.6338 & 0.5970 \\
  Vultron Core (4.5B)\tablefootnote{\label{fn:results-vultron-core}\hflink{https://huggingface.co/vultr/VultronRetrieverCore-Qwen3.5-4.5B}{vultr/VultronRetrieverCore-Qwen3.5-4.5B} \citep{georgiou2026vultroncore}.}\textsuperscript{\textdagger} & \checkmark & 4.54 & 320 & 0.9221 & 0.6612 & 0.6372 \\
  Ops ColQwen3 (4B)\tablefootnote{\label{fn:results-ops-colqwen3}\hflink{https://huggingface.co/OpenSearch-AI/Ops-Colqwen3-4B}{OpenSearch-AI/Ops-Colqwen3-4B} \citep{opensearch2026opscolqwen}.}\textsuperscript{\textdagger} & \checkmark & 4.80 & 2,560 & 0.9136 & 0.6866 & 0.6127 \\
  Nemotron (4B)\tablefootnote{\label{fn:results-nemotron-4b}\hflink{https://huggingface.co/nvidia/nemotron-colembed-vl-4b-v2}{nvidia/nemotron-colembed-vl-4b-v2} \citep{moreira2026nemotron}.}\textsuperscript{\textdagger} & \checkmark & 4.80 & 2,560 & 0.9162 & 0.6449 & 0.6142 \\
  ColNomic (7B)\tablefootnote{\label{fn:results-colnomic-7b}\hflink{https://huggingface.co/nomic-ai/colnomic-embed-multimodal-7b}{nomic-ai/colnomic-embed-multimodal-7b} \citep{nomic2025multimodal}.}\textsuperscript{\textdagger} & \checkmark & 7.00 & 128 & 0.8972 & 0.6025 & 0.5764 \\
  Qwen3-VL (8B)\tablefootnote{\label{fn:results-qwen3vl-8b}\hflink{https://huggingface.co/Qwen/Qwen3-VL-Embedding-8B}{Qwen/Qwen3-VL-Embedding-8B} \citep{li2026qwen3vlembedding}.}\textsuperscript{\textdagger} & & 8.14 & 4,096 & -- & -- & 0.5829 \\
  Tomoro (8B)\tablefootnote{\label{fn:results-tomoro-8b}\hflink{https://huggingface.co/TomoroAI/tomoro-colqwen3-embed-8b}{TomoroAI/tomoro-colqwen3-embed-8b} \citep{huang2025tomoro}.}\textsuperscript{\textdagger} & \checkmark & 8.00 & 320 & 0.9076 & 0.6540 & 0.6160 \\
  Vultron Prime (8B)\tablefootnote{\label{fn:results-vultron-prime}\hflink{https://huggingface.co/vultr/VultronRetrieverPrime-Qwen3.5-8B}{vultr/VultronRetrieverPrime-Qwen3.5-8B} \citep{georgiou2026vultronprime}.}\textsuperscript{\textdagger} & \checkmark & 8.39 & 320 & 0.9208 & 0.6818 & 0.6472 \\
  Nemotron (8B)\tablefootnote{\label{fn:results-nemotron-8b}\hflink{https://huggingface.co/nvidia/nemotron-colembed-vl-8b-v2}{nvidia/nemotron-colembed-vl-8b-v2} \citep{moreira2026nemotron}.}\textsuperscript{\textdagger} & \checkmark & 8.70 & 4,096 & 0.9265 & 0.6516 & 0.6354 \\
  \bottomrule
  \end{tabular}
  \par\vspace{2pt}
  \scriptsize
  \textsuperscript{\textdagger} Scores from MTEB. \textsuperscript{\ensuremath{\ddagger}} Results from our evaluation.
\end{table*}

\noindent\textbf{Aggregate metrics.} \autoref{tab:visual-all-metrics} reports all aggregate visual-retrieval metrics at step 20,000. Visual retrieval uses a maximum image side length of 2,048 pixels. Late-interaction uses the full 128-dimensional token representation, and dense uses the full model width.

\begin{table}[htbp]
  \centering
  \caption{Visual retrieval aggregates. A dash marks a metric that was not reported. All results use the evaluation protocol in \autoref{sec:evaluation-protocol}.}
  \label{tab:visual-all-metrics}
  \scriptsize
  \setlength{\tabcolsep}{3.6pt}

  \textbf{ViDoRe v3}\\[2pt]
  \begin{tabular}{llrrrrrr}
    \toprule
    Model & Head & nDCG@5 & nDCG@10 & Recall@5 & Recall@10 & Recall@100 & MAP \\
    \midrule
    \multirow{2}{*}{\neomme-260M\textsuperscript{\ref{fn:results-neomme-260m}}} & Late-interaction & 0.4991 & 0.5226 & 0.4637 & 0.5664 & 0.8326 & 0.4555 \\
    & Dense & 0.3680 & 0.3907 & 0.3484 & 0.4419 & 0.7478 & 0.3274 \\
    \cmidrule(lr){1-8}
    \multirow{2}{*}{\neomme-800M\textsuperscript{\ref{fn:results-neomme-800m}}} & Late-interaction & \textbf{0.5329} & \textbf{0.5560} & \textbf{0.4906} & \textbf{0.5970} & \textbf{0.8607} & \textbf{0.4878} \\
    & Dense & 0.4145 & 0.4391 & 0.3877 & 0.4904 & 0.7937 & 0.3719 \\
    \bottomrule
  \end{tabular}

  \vspace{6pt}
  \textbf{ViDoRe v2}\\[2pt]
  \begin{tabular}{llrrrrrr}
    \toprule
    Model & Head & nDCG@5 & nDCG@10 & Recall@5 & Recall@10 & Recall@100 & MAP \\
    \midrule
    ColModernVBERT\textsuperscript{\ref{fn:results-colmodernvbert}} & Late-interaction & 0.4068 & 0.4364 & -- & -- & 0.8475 & -- \\
    \cmidrule(lr){1-8}
    \multirow{2}{*}{\neomme-260M\textsuperscript{\ref{fn:results-neomme-260m}}} & Late-interaction & 0.5218 & 0.5505 & -- & -- & 0.9232 & -- \\
    & Dense & 0.4075 & 0.4334 & 0.3938 & 0.4979 & 0.8091 & 0.3662 \\
    \cmidrule(lr){1-8}
    \multirow{2}{*}{\neomme-800M\textsuperscript{\ref{fn:results-neomme-800m}}} & Late-interaction & \textbf{0.5591} & \textbf{0.5814} & -- & -- & \textbf{0.9259} & -- \\
    & Dense & 0.4475 & 0.4697 & \textbf{0.4347} & \textbf{0.5319} & 0.8434 & \textbf{0.4075} \\
    \bottomrule
  \end{tabular}

  \vspace{6pt}
  \textbf{ViDoRe v1}\\[2pt]
  \begin{tabular}{llrrrrrr}
    \toprule
    Model & Head & nDCG@5 & nDCG@10 & Recall@5 & Recall@10 & Recall@100 & MAP \\
    \midrule
    ColModernVBERT\textsuperscript{\ref{fn:results-colmodernvbert}} & Late-interaction & 0.8058 & 0.8177 & -- & -- & \textbf{0.9805} & -- \\
    \cmidrule(lr){1-8}
    \multirow{2}{*}{\neomme-260M\textsuperscript{\ref{fn:results-neomme-260m}}} & Late-interaction & 0.8598 & 0.8666 & 0.9011 & 0.9222 & 0.9752 & 0.8511 \\
    & Dense & 0.7552 & 0.7691 & 0.8352 & 0.8776 & 0.9603 & 0.7378 \\
    \cmidrule(lr){1-8}
    \multirow{2}{*}{\neomme-800M\textsuperscript{\ref{fn:results-neomme-800m}}} & Late-interaction & \textbf{0.8744} & \textbf{0.8801} & \textbf{0.9175} & \textbf{0.9353} & 0.9786 & \textbf{0.8641} \\
    & Dense & 0.7993 & 0.8098 & 0.8685 & 0.9006 & 0.9681 & 0.7836 \\
    \bottomrule
  \end{tabular}
\end{table}

\noindent\textbf{Resolution impact.} \autoref{tab:resolution-impact} reports MeanMaxSim nDCG@10 under a downscale-only longest-side cap. The cost columns show the maximum for a square page at each cap; pages with smaller native dimensions or nonsquare aspect ratios use fewer vectors. Vector counts include image patches and structural positions. Raw storage assumes a 128-dimensional float32 late-interaction representation and excludes compression and index overhead.

\begin{table*}[htbp]
  \centering
  \caption{Retrieval quality and uncompressed square-page representation size across image resolutions. All results use the evaluation protocol in \autoref{sec:evaluation-protocol}.}
  \label{tab:resolution-impact}
  \scriptsize
  \setlength{\tabcolsep}{3.5pt}
  \begin{tabular}{rrr@{\hspace{8pt}}rrrrrr}
    \toprule
    \multirow{2}{*}{\shortstack{Longest side\\(pixels)}} &
    \multicolumn{2}{c}{Square-page representation} &
    \multicolumn{2}{c}{ViDoRe v1} &
    \multicolumn{2}{c}{ViDoRe v2} &
    \multicolumn{2}{c}{ViDoRe v3} \\
    \cmidrule(lr){2-3}\cmidrule(lr){4-5}\cmidrule(lr){6-7}\cmidrule(lr){8-9}
    & Vectors & \shortstack{Raw float32\\(MB)} &
    260M & 800M & 260M & 800M & 260M & 800M \\
    \midrule
    768   & 602   & 0.31 & 0.7231 & 0.7771 & 0.3897 & 0.4454 & 0.3340 & 0.4049 \\
    1,024 & 1,058 & 0.54 & 0.8278 & 0.8533 & 0.4843 & 0.5251 & 0.4562 & 0.5029 \\
    1,536 & 2,354 & 1.21 & 0.8633 & 0.8775 & 0.5419 & 0.5728 & 0.5134 & 0.5496 \\
    2,048 & 4,162 & 2.13 & \textbf{0.8666} & \textbf{0.8801} & \textbf{0.5505} & \textbf{0.5814} & \textbf{0.5226} & \textbf{0.5560} \\
    \bottomrule
  \end{tabular}
\end{table*}

\noindent\textbf{ViDoRe v3 domains.} \autoref{tab:vidore-v3-domains} reports nDCG@10 for every task and a subset of the compared models. The \neomme-800M row has a mean score of 0.5560.

\begin{table*}[htbp]
  \centering
  \caption{ViDoRe v3 nDCG@10 results by task.}
  \label{tab:vidore-v3-domains}
  \scriptsize
  \resizebox{\textwidth}{!}{%
  \begin{tabular}{lrrrrrrrrr}
    \toprule
    \multirow{2}{*}{Model} &
    \multirow{2}{*}{Params.} &
    \multicolumn{8}{c}{ViDoRe v3 domains} \\
    \cmidrule(lr){3-10}
    & & HR & Fin.-EN & Industrial & Pharma. & CS & Energy & Physics & Fin.-FR \\
    \midrule
    \multicolumn{10}{l}{\paramsmall} \\
    ColModernVBERT\textsuperscript{\ref{fn:results-colmodernvbert}}\textsuperscript{\textdagger} & 250M & 0.1982 & 0.2876 & 0.1567 & 0.3219 & 0.3761 & 0.3251 & 0.2262 & 0.1977 \\
    ColSmol-256M\textsuperscript{\ref{fn:results-colsmol}}\textsuperscript{\textdagger} & 256M & 0.1646 & 0.2323 & 0.1287 & 0.2785 & 0.2880 & 0.2483 & 0.1614 & 0.1568 \\
  \neomme-260M\textsuperscript{\ref{fn:results-neomme-260m}}\textsuperscript{\ensuremath{\ddagger}} & 260M & \textbf{0.5520} & \textbf{0.5639} & \textbf{0.3989} & \textbf{0.5963} & \textbf{0.6743} & \textbf{0.5922} & \textbf{0.4246} & \textbf{0.3782} \\
    \midrule
    \multicolumn{10}{l}{\parammedium} \\
    ColSmol-500M\textsuperscript{\ref{fn:results-colsmol-500m}} & 500M & 0.2552 & 0.3697 & 0.1985 & 0.4126 & 0.5054 & 0.4052 & 0.3053 & 0.2656 \\
    Vultron Flash\textsuperscript{\ref{fn:results-vultron}}\textsuperscript{\textdagger} & 850M & 0.5837 & \textbf{0.6060} & \textbf{0.4619} & \textbf{0.6309} & \textbf{0.7382} & 0.6114 & \textbf{0.4795} & 0.4079 \\
  \neomme-800M\textsuperscript{\ref{fn:results-neomme-800m}}\textsuperscript{\ensuremath{\ddagger}} & 800M & \textbf{0.5946} & 0.6054 & 0.4455 & 0.6187 & 0.7026 & \textbf{0.6180} & 0.4482 & \textbf{0.4151} \\
    \midrule
    \multicolumn{10}{l}{\paramlarge} \\
    ColQwen2 v1.0\textsuperscript{\ref{fn:results-colqwen2}}\textsuperscript{\textdagger} & 2.21B & \textbf{0.4511} & \textbf{0.3903} & \textbf{0.3834} & 0.5221 & \textbf{0.6861} & \textbf{0.4856} & 0.4162 & 0.1997 \\
    ColPali v1.3\textsuperscript{\ref{fn:results-colpali}}\textsuperscript{\textdagger} & 2.92B & 0.4480 & 0.3444 & 0.3557 & \textbf{0.5311} & 0.6528 & 0.4692 & \textbf{0.4174} & \textbf{0.2177} \\
    \bottomrule
  \end{tabular}%
  }
  \par\vspace{2pt}
  \scriptsize
  \textsuperscript{\textdagger} Scores from MTEB. \textsuperscript{\ensuremath{\ddagger}} Results from our evaluation.
\end{table*}

\noindent\textbf{Task-level results.} \autoref{tab:selected-retrieval-results} reports task-level results for the \neomme-260M late-interaction head on ViDoRe v3 and BEIR-15.

\begin{table}[htbp]
  \centering
  \caption{Selected task-level \neomme-260M\textsuperscript{\ref{fn:results-neomme-260m}} late-interaction results. All results use the evaluation protocol in \autoref{sec:evaluation-protocol}.}
  \label{tab:selected-retrieval-results}
  \scriptsize
  \setlength{\tabcolsep}{3pt}
  \begin{minipage}[t]{0.57\textwidth}
    \centering
    \textbf{ViDoRe v3}\\[2pt]
    \begin{tabular}{lrrr}
      \toprule
      Task & nDCG@5 & nDCG@10 & Recall@100 \\
      \midrule
      Human resources & 0.5255 & 0.5520 & 0.8884 \\
      Finance, English & 0.5389 & 0.5639 & 0.8665 \\
      Industrial & 0.3883 & 0.3989 & 0.6530 \\
      Pharmaceuticals & 0.5822 & 0.5963 & 0.8657 \\
      Computer science & 0.6464 & 0.6743 & 0.9409 \\
      Energy & 0.5651 & 0.5922 & 0.8938 \\
      Physics & 0.3977 & 0.4246 & 0.7903 \\
      Finance, French & 0.3485 & 0.3782 & 0.7621 \\
      \midrule
      Task macro mean & 0.4991 & 0.5226 & 0.8326 \\
      \bottomrule
    \end{tabular}
  \end{minipage}
  \hfill
  \begin{minipage}[t]{0.38\textwidth}
    \centering
    \textbf{BEIR-15}\\[2pt]
    \begin{tabular}{lr}
      \toprule
      Task & nDCG@10 \\
      \midrule
      ArguAna & 0.4188 \\
      ClimateFEVER & 0.2400 \\
      CQADupstack & 0.3529 \\
      DBPedia & 0.3848 \\
      FEVER & 0.9176 \\
      FiQA-2018 & 0.3657 \\
      HotpotQA & 0.7222 \\
      MS MARCO & 0.3852 \\
      NFCorpus & 0.3173 \\
      NQ & 0.5336 \\
      Quora & 0.6945 \\
      SCIDOCS & 0.1514 \\
      SciFact & 0.7181 \\
      TREC-COVID & 0.7635 \\
      Touch\'e-2020 & 0.3556 \\
      \midrule
      Task macro mean & 0.4881 \\
      \bottomrule
    \end{tabular}
  \end{minipage}
\end{table}

\FloatBarrier

\subsection{Matryoshka results}
\label{app:matryoshka-results}
Matryoshka training trains each model to produce dense representations at several widths. \autoref{tab:matryoshka-all} reports the retrieval quality at every trained width for \neomme-260M and \neomme-800M on ViDoRe v3, v2, and v1, using task-macro nDCG@10.

\begin{table}[htbp]
  \centering
  \caption{nDCG@10 results on ViDoRe v3, v2, and v1 at every trained Matryoshka representation size. All results use the evaluation protocol in \autoref{sec:evaluation-protocol}.}
  \label{tab:matryoshka-all}
  \scriptsize
  \setlength{\tabcolsep}{3pt}
  \begin{tabular*}{0.61\textwidth}{@{\extracolsep{\fill}}lrrrr@{}}
    \toprule
    \multicolumn{2}{c}{Model details} &
    \multicolumn{3}{c}{ViDoRe (nDCG@10)} \\
    \cmidrule(lr){1-2}\cmidrule(lr){3-5}
    Model & Dense width & v3 & v2 & v1 \\
    \midrule
    \multirow{4}{*}{\neomme-260M\textsuperscript{\ref{fn:results-neomme-260m}}}
      & 128 & 0.3607 & 0.4063 & 0.7461 \\
      & 256 & 0.3761 & 0.4133 & 0.7568 \\
      & 512 & 0.3864 & 0.4228 & 0.7668 \\
      & 1,024 & \textbf{0.3907} & \textbf{0.4334} & \textbf{0.7691} \\
    \midrule
    \multirow{5}{*}{\neomme-800M\textsuperscript{\ref{fn:results-neomme-800m}}}
      & 128 & 0.3974 & 0.4332 & 0.7786 \\
      & 256 & 0.4217 & 0.4519 & 0.7999 \\
      & 512 & 0.4318 & 0.4627 & 0.8074 \\
      & 1,024 & 0.4352 & 0.4571 & 0.8089 \\
      & 1,792 & \textbf{0.4391} & \textbf{0.4697} & \textbf{0.8098} \\
    \bottomrule
  \end{tabular*}
\end{table}

\FloatBarrier

\subsection{Dual-head training ablation}
\label{app:ablations}
\neomme{}-Retriever jointly trains a late-interaction head and a dense head on one shared backbone. We test whether optimizing both objectives improves either representation compared with optimizing its corresponding objective alone. The \neomme-260M experiment compares late-interaction-only, dense-only, and joint training from the same initial checkpoint. All three runs use the same data seed, query order, negative draws, data mixture, schedule, and evaluation protocols. They differ only in the active losses. \autoref{tab:head-ablations} reports the task-macro results, and \autoref{fig:head-ablation-curves} shows nDCG@10 over training.

\begin{center}
  \begin{minipage}{\textwidth}
  \begin{minipage}[t]{0.50\textwidth}
    \vspace{0pt}
    \centering
    \captionsetup{justification=raggedright,singlelinecheck=false}
    \captionof{table}{Retrieval-head objective ablation for \neomme-260M. Arrows show nDCG@10-point changes from the corresponding single-objective run. All results use the same evaluation protocol.}
    \label{tab:head-ablations}
    \small
    \renewcommand{\arraystretch}{1.15}
    \setlength{\tabcolsep}{2.5pt}
    \begin{tabular*}{\linewidth}{@{\extracolsep{\fill}}>{\raggedright\arraybackslash}p{0.21\textwidth}cccc@{}}
      \toprule
      & \multicolumn{2}{c}{ViDoRe v3} & \multicolumn{2}{c}{BEIR-15} \\
      \cmidrule(lr){2-3}\cmidrule(lr){4-5}
      Architecture & Late & Dense & Late & Dense \\
      \midrule
      LI head & 0.5088 & -- & 0.4774 & -- \\
      Dense~head & -- & 0.3906 & -- & 0.3240 \\
      Dual-head & 0.5226 & 0.3907 & 0.4881 & 0.3055 \\
      & \scriptsize\textcolor{green!50!black}{\ensuremath{\uparrow}\,$+1.38$} & \scriptsize\textcolor{green!50!black}{\ensuremath{\uparrow}\,$+0.01$} & \scriptsize\textcolor{green!50!black}{\ensuremath{\uparrow}\,$+1.07$} & \scriptsize\textcolor{red!70!black}{\ensuremath{\downarrow}\,$-1.85$} \\
      \bottomrule
    \end{tabular*}
  \end{minipage}
  \hfill
  \begin{minipage}[t]{0.47\textwidth}
    \vspace{0pt}
    \centering
    \captionsetup{justification=raggedright,singlelinecheck=false}
    \includegraphics[width=\linewidth]{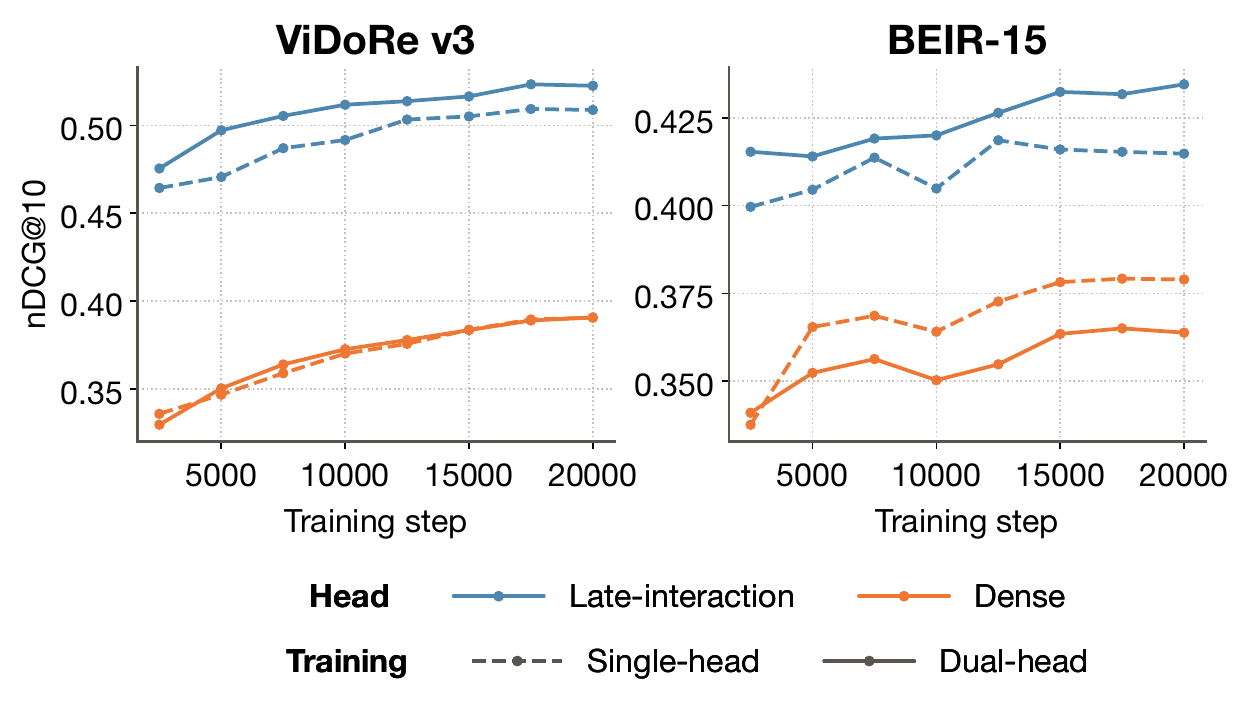}
    \captionof{figure}{nDCG@10 during joint and single-objective \neomme-260M retrieval training on ViDoRe v3 and BEIR-15.}
    \label{fig:head-ablation-curves}
  \end{minipage}
  \end{minipage}
\end{center}

Joint training raises late-interaction task-macro nDCG@10 by 1.38 points on ViDoRe v3, while dense retrieval changes by 0.01 point. A paired analysis across all 14,514 judged queries estimates a 1.39-point late-interaction gain with a 95\% confidence interval of 1.11 to 1.67 points ($p=10^{-4}$). The paired dense change is 0.03 point with a confidence interval of $-0.26$ to 0.32 points ($p=0.81$), so the experiment does not show a dense improvement. On BEIR-15, joint training raises late-interaction by 1.07 points and reduces dense retrieval by 1.85 points. These results are consistent with one-directional transfer from the dense objective to late-interaction on ViDoRe v3, not a general improvement to both heads. Each run uses one training seed, so the paired intervals do not measure run-to-run variation.

\FloatBarrier

\section{Retrieval efficiency}
\label{app:retrieval-efficiency}

\subsection{Efficiency measurement details}
\label{app:indexing-details}
\begin{figure*}[htbp]
  \centering
  \begin{subfigure}[t]{0.49\textwidth}
    \centering
    \includegraphics[width=\linewidth]{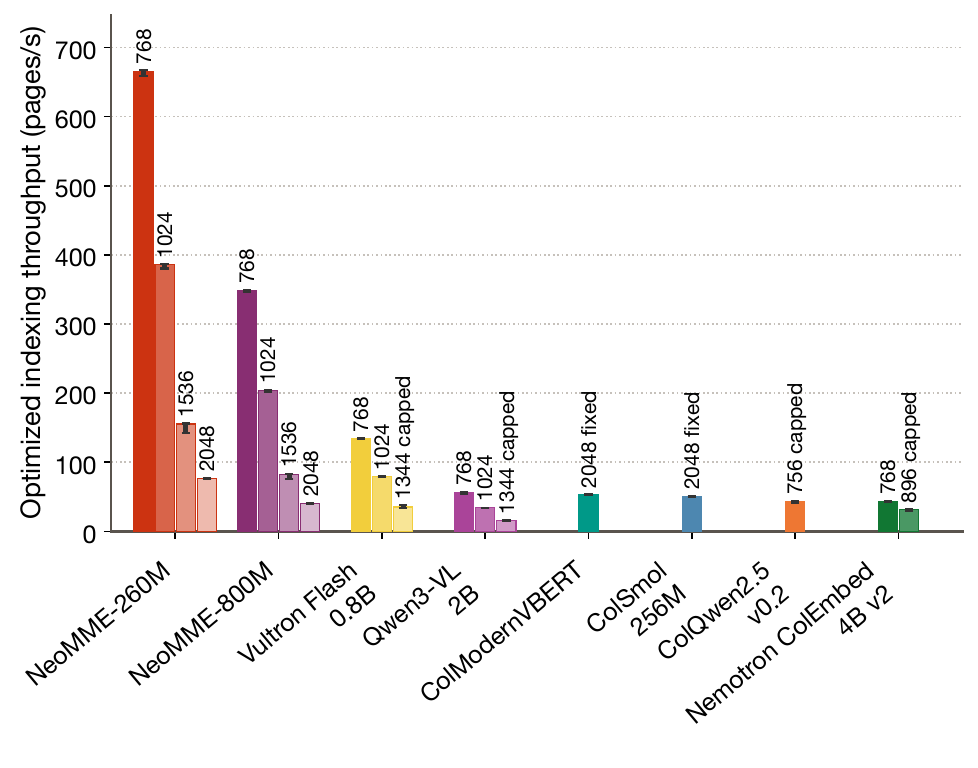}
    \caption{NVIDIA H100 80GB HBM3.}
    \label{fig:indexing-h100}
  \end{subfigure}
  \hfill
  \begin{subfigure}[t]{0.49\textwidth}
    \centering
    \includegraphics[width=\linewidth]{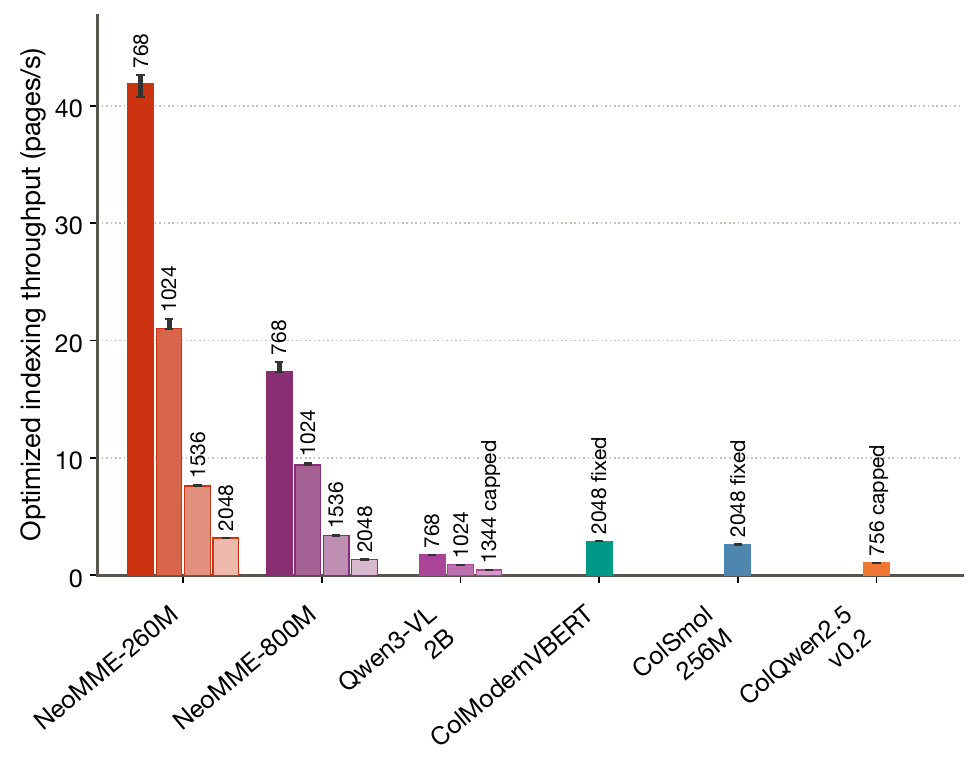}
    \caption{Apple M5 Pro.}
    \label{fig:indexing-m5-pro}
  \end{subfigure}
  \caption{Image indexing throughput over the resolution grid on NVIDIA H100 and Apple M5 Pro.}
  \label{fig:indexing-hardware}
\end{figure*}

\noindent\textbf{Timed path and benchmark inputs.} We generate deterministic RGB squares and run each model's official image processor before timing. This separates processor runtime from model and accelerator execution, while the processor still determines each model's input shape and token count. We time device transfer, document encoding, output unpadding, conversion to contiguous float32 embeddings, and in-memory serialization. We do not time image processing, image decoding, storage input and output, compression, or index construction. On H100 and L40S, we use 128 images and 10 timed observations. On M5 Pro, we use 16 images and three observations, so those intervals are less precise. Each model and image size has a separately calibrated batch size. The tested square sizes are 768, 1,024, 1,536, and 2,048 pixels, plus a model's native square when it is not already in the grid. We report each processor's effective image sizes in \autoref{tab:indexing-setup}. \neomme{} encodes every requested square, while the external processors rescale each input to a fixed pixel or token budget. ColSmol and ColModernVBERT use fixed 2,048-pixel tiling at every supplied size. We omit repeated inputs that map to the same model input and do not measure complete page ingestion. The processor settings for ColModernVBERT and Nemotron ColEmbed VL 4B v2 follow their respective model reports \citep{teiletche2025modernvbert,moreira2026nemotron}.

\noindent\textbf{Hardware and software setup.} For CUDA measurements, we use bfloat16 and \code{torch.compile}. The L40S measurements use four OpenMP threads. We benchmark \neomme{} with the Transformers implementation at revision \code{c0f8594}\footnote{\url{https://github.com/huggingface/transformers/commit/c0f8594234be10908e3e588f7a8e784d8cfbee33}} and use FlashAttention 2 on CUDA. On M5 Pro, \neomme{} and Qwen3-VL use compiled execution. The ColPali-engine models use eager execution because the PyTorch 2.13 MPS compiler cannot generate their graphs. For compiled runs, we set PyTorch Dynamo's recompile limit to 256\footnote{\url{https://docs.pytorch.org/docs/2.11/user_guide/torch_compiler/torch.compiler_troubleshooting.html}} and exclude any run whose log reports that the limit was reached. Vultron Flash and Nemotron ColEmbed do not support MPS. Within each device, every model uses the same machine settings and benchmark implementation. Because the execution environments differ across devices, we compare models only within the same device.

\begin{table*}[htbp]
  \centering
  \caption{Processor behavior and effective image settings in the indexing benchmark.}
  \label{tab:indexing-setup}
  \scriptsize
  \setlength{\tabcolsep}{3pt}
  \begin{tabular}{p{0.25\textwidth}p{0.30\textwidth}p{0.25\textwidth}r}
    \toprule
    Model & Processor behavior & Squares shown (pixels) & Vectors at cap \\
    \midrule
    \neomme-260M\textsuperscript{\ref{fn:results-neomme-260m}} & Encodes each requested square & 768 / 1,024 / 1,536 / 2,048 & 4,162 \\
    \neomme-800M\textsuperscript{\ref{fn:results-neomme-800m}} & Encodes each requested square & 768 / 1,024 / 1,536 / 2,048 & 4,162 \\
    ColModernVBERT\textsuperscript{\ref{fn:results-colmodernvbert}} & Fixed 2,048-pixel tiling & 2,048 & 1,149 \\
    ColSmol-256M\textsuperscript{\ref{fn:results-colsmol}} & Fixed 2,048-pixel tiling & 2,048 & 1,139 \\
    Vultron Flash\textsuperscript{\ref{fn:results-vultron}} & Caps at its visual-token budget & 768 / 1,024 / 1,344 & 1,775 \\
    Qwen3-VL-Embedding-2B\tablefootnote{\hflink{https://huggingface.co/Qwen/Qwen3-VL-Embedding-2B}{Qwen/Qwen3-VL-Embedding-2B}.} & Caps at its visual-token budget & 768 / 1,024 / 1,344 & 1 \\
    ColQwen2.5-v0.2\textsuperscript{\ref{fn:results-colqwen25}} & Caps at 602,112 pixels & 756 & 740 \\
    Nemotron ColEmbed VL 4B v2\tablefootnote{\hflink{https://huggingface.co/nvidia/nemotron-colembed-vl-4b-v2}{nvidia/nemotron-colembed-vl-4b-v2}.} & Caps at 802,816 pixels & 768 / 896 & 798 \\
    \bottomrule
  \end{tabular}
\end{table*}

\begin{table}[H]
  \centering
  \caption{Document-encoding throughput at each model's largest effective square. All values use the same measurement protocol.}
  \label{tab:indexing-results}
  \scriptsize
  \setlength{\tabcolsep}{4pt}
  \begin{tabular}{lrrrrr}
    \toprule
    Model & \shortstack{Square\\(pixels)} & \shortstack{Vectors\\per page} & H100 & L40S & M5 Pro \\
    \midrule
    \neomme-260M\textsuperscript{\ref{fn:results-neomme-260m}} & 2,048 & 4,162 & 76.8 & 51.3 & 3.2 \\
    \neomme-800M\textsuperscript{\ref{fn:results-neomme-800m}} & 2,048 & 4,162 & 40.4 & 21.2 & 1.4 \\
    Vultron Flash\textsuperscript{\ref{fn:results-vultron}} & 1,344 & 1,775 & 35.2 & 20.2 & -- \\
    Qwen3-VL-Embedding-2B & 1,344 & 1 & 15.5 & 8.1 & 0.4 \\
    ColModernVBERT\textsuperscript{\ref{fn:results-colmodernvbert}} & 2,048 & 1,149 & 53.3 & 26.0 & 2.9\textsuperscript{\textdagger} \\
    ColSmol-256M\textsuperscript{\ref{fn:results-colsmol}} & 2,048 & 1,139 & 50.7 & 24.8 & 2.6\textsuperscript{\textdagger} \\
    ColQwen2.5-v0.2\textsuperscript{\ref{fn:results-colqwen25}} & 756 & 740 & 42.7 & 16.1 & 1.1\textsuperscript{\textdagger} \\
    Nemotron ColEmbed VL 4B v2 & 896 & 798 & 31.3 & 16.8 & -- \\
    \bottomrule
  \end{tabular}
  \par\vspace{2pt}
  \scriptsize
  Values are median pages per second. A dash means that the model does not support the hardware target. \textsuperscript{\textdagger} means that the M5 Pro measurement uses eager execution because compilation could not generate the model graph.
\end{table}

\FloatBarrier

\subsection{Query-encoding latency}
\label{app:query-encoding-details}
We measure batch-one query-encoding latency using a fixed set of 649 NanoBEIR queries. For each reported row, we run one complete warmup pass and then three measured passes. We time tokenization or query processing, the model forward pass, and transfer of the output embedding to host memory. For the CPU measurements, we use a SkyPilot-provisioned CPU-only Kubernetes pod with 128 CPU cores and 1 TB RAM, matching ModernVBERT's query-latency setup. We set \code{OMP\_NUM\_THREADS=128} and use float32. The L40S and M5 Pro measurements use bfloat16. The indexing and query-encoding benchmarks use the same Transformers implementation for \neomme{}. We report each device separately because the execution environments differ. The measurements cover query encoding only and exclude candidate generation, approximate search, MeanMaxSim scoring, and end-to-end request latency.

\begin{figure}[htbp]
  \centering
  \includegraphics[width=\textwidth]{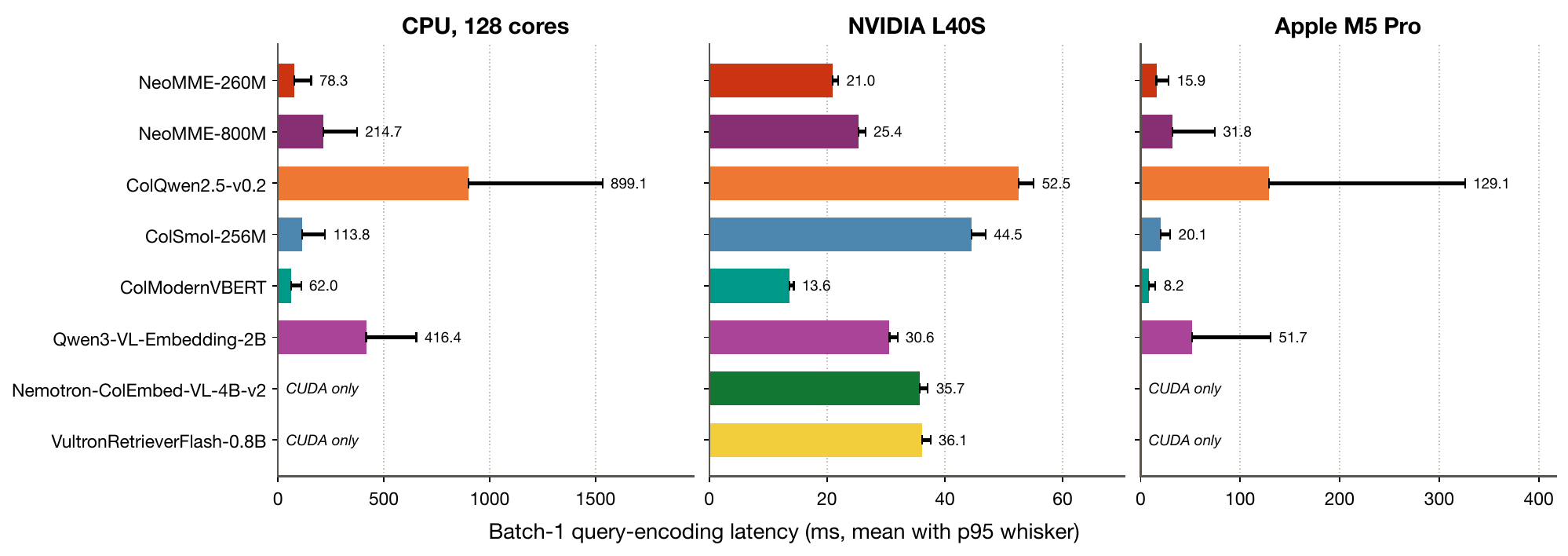}
  \caption{Batch-one text query-encoding latency for visual retrievers on a 128-core CPU host, one NVIDIA L40S, and an Apple M5 Pro. Bars show the mean, and error bars extend to the 95th percentile.}
  \label{fig:query-encoding-latency}
\end{figure}

\neomme-260M is second only to ColModernVBERT on every device in \autoref{fig:query-encoding-latency}. Its mean latency is 21.0 ms on the L40S, 78.3 ms on the CPU host, and 15.9 ms on the M5 Pro. The other measured visual language model retrievers range from 30.6 to 52.5 ms, 113.8 to 899.1 ms, and 20.1 to 129.1 ms, respectively. \neomme-800M is also faster than the larger visual language model retrievers on every device. Across the tested CUDA, MPS, and CPU execution paths, \neomme{} has low query-encoding latency.

\FloatBarrier

\subsection{Late-Interaction Kernels (\lik{})}
\label{app:lik}
\noindent\textbf{Method.} \lik{} computes exact MaxSim scores in tiles and keeps only running token maxima instead of writing the full query-token by document-token similarity tensor to high-bandwidth memory \citep{lac2026lik}. During training, it saves the winning document-token indices needed to route gradients. The \lik{} design guide\footnote{\url{https://hcompai.github.io/late-interaction-kernels/how-it-works.html}} gives the full derivation and kernel diagrams.

\noindent\textbf{Measurement scope.} \autoref{tab:lik-throughput} combines \neomme-260M throughput from matched release runs with selected long-document kernel measurements. \autoref{tab:lik-training-resources} selects end-to-end training results from the complete \lik{} benchmark report\footnote{\url{https://github.com/hcompai/late-interaction-kernels/blob/main/docs/benchmarks.md}}, which records the full sweeps, software versions, baselines, and timing protocol. All measurements use one H100 80 GB SXM and float32 accumulation. Training uses bfloat16 inputs. The \neomme-260M values are the mean and standard deviation from the 10 to 70 minute measurement window. ColQwen2 uses LoRA rank 32, gradient checkpointing, and ColPali training pages \citep{faysse2025colpali}. PyLate uses gradient checkpointing and MS MARCO triplets. Its batch-1,024 baseline uses \code{score\_mini\_batch\_size=64}, PyLate's own chunking mitigation. Long-document measurements use float16 inputs with 16 queries, 32 documents, and 32 query tokens. The results cover memory and short-run step time, not convergence or downstream retrieval quality.

\begin{table}[htbp]
  \centering
  \caption{Selected \neomme-260M throughput and long-document MaxSim results. \textsuperscript{\textdagger} results come from the \lik{} benchmark report, and \textsuperscript{\ensuremath{\ddagger}} results come from matched release runs.}
  \label{tab:lik-throughput}
  \scriptsize
  \begin{minipage}[t]{0.38\textwidth}
    \centering
    \textbf{Training throughput}\\[3pt]
    \setlength{\tabcolsep}{3pt}
    \begin{tabular}{lrr}
      \toprule
      \multicolumn{3}{l}{\neomme-260M\textsuperscript{\ref{fn:results-neomme-260m}}\textsuperscript{\ensuremath{\ddagger}}} \\
      \midrule
      Configuration & Tokens/s & Std. dev. \\
      \midrule
      With \lik{} & 383,614 & 21,404 \\
      Without \lik{} & 377,439 & 19,576 \\
      \bottomrule
    \end{tabular}
  \end{minipage}
  \hfill
  \begin{minipage}[t]{0.60\textwidth}
    \centering
    \textbf{Long-document MaxSim\textsuperscript{\textdagger}}\\[3pt]
    \setlength{\tabcolsep}{2pt}
    \begin{tabular}{l lrrr}
      \toprule
      Document length & Method & Forward & Backward & Peak memory \\
      \midrule
      \multirow{2}{*}{4,096} & Naive & 0.65 ms & 1.82 ms & 672 MB \\
      & \lik{} & \textbf{0.12 ms} & \textbf{0.54 ms} & \textbf{193 MB} \\
      \addlinespace
      \multirow{2}{*}{8,192} & Naive & OOM & OOM & OOM \\
      & \lik{} & \textbf{0.18 ms} & \textbf{0.44 ms} & \textbf{321 MB} \\
      \bottomrule
    \end{tabular}
  \end{minipage}
\end{table}

\begin{table}[htbp]
  \centering
  \caption{Selected \lik{} training results from the \lik{} benchmark report.}
  \label{tab:lik-training-resources}
  \small
  \setlength{\tabcolsep}{4.5pt}
  \begin{tabular}{lllrrr}
    \toprule
    System & Workload & Measure & Baseline & \lik{} & Improvement \\
    \midrule
    \multirow{2}{*}{ColQwen2\tablefootnote{\hflink{https://huggingface.co/vidore/colqwen2-base}{vidore/colqwen2-base}.}}
    & MaxSim, batch 128 & Peak memory & 7.81 GiB & \textbf{61 MiB} &
    \textbf{$130\times$ lower} \\
    & Whole training & Maximum batch & 64 & \textbf{128} &
    \textbf{$2\times$ higher} \\
    \addlinespace
    \multirow{2}{*}{PyLate\tablefootnote{\hflink{https://huggingface.co/lightonai/GTE-ModernColBERT-v1}{lightonai/GTE-ModernColBERT-v1}.}}
    & Whole training, batch 1,024 & Peak memory & 62.18 GiB &
    \textbf{57.67 GiB} &
    \textbf{7\% lower} \\
    & Whole training, batch 1,024 & Step time & 6.02 s &
    \textbf{4.81 s} &
    \textbf{$1.25\times$ faster} \\
    \bottomrule
  \end{tabular}
\end{table}

\FloatBarrier

\Needspace*{10\baselineskip}
\section{Retrieval case demonstrations}
\label{app:retrieval-case-demonstrations}
The three cases below show how \neomme{}-Retriever 260M ranks pages on three ViDoRe v3 tasks \citep{loison2026vidorev3}. Each case reports one query, the top-3 retrieved pages of the full task corpus, their MeanMaxSim scores, and each page's query relevance. ViDoRe v3 grades a page (1) fully relevant when it answers the query on its own, (2) critically relevant when it holds facts the answer needs but leaves the rest to other pages, or (3) unjudged when no annotator graded it for that query. Green frames mark relevant pages and red frames mark unjudged ones.

\begin{tcolorbox}[title=ViDoRe v3 retrieval case 1, retrievalshowcase]
  \begin{tcolorbox}[title=Query (French), retrievalquery]
    En quelle année les activités du réseau PES ont-elles été marquées par la célébration de son dixième anniversaire, et à travers quel segment la commémoration a-t-elle été diffusée ?
  \end{tcolorbox}
  \retrievalcasemeta{\href{https://huggingface.co/datasets/vidore/vidore_v3_hr}{\hfemoji\ \code{vidore/vidore\_v3\_hr}} (1,110 pages)}{\neomme{}-Retriever 260M (late-interaction)}
  \begin{tcbraster}[raster columns=3, raster column skip=0.2em,
    raster valign=top, raster force size=false]
    \retrievalresultcard{retrievalrelevant}{1}{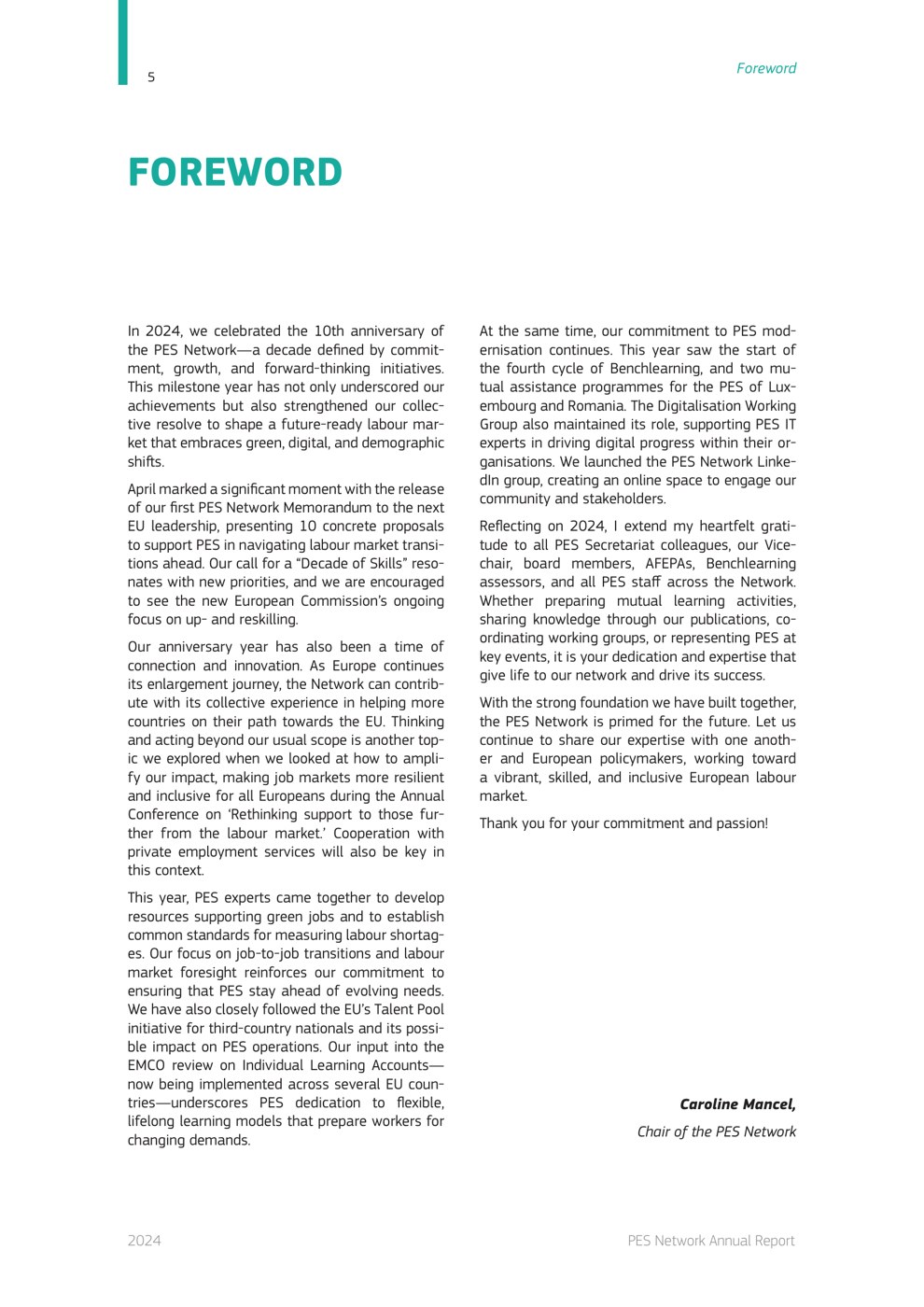}{0.7205}{Critically relevant}
    \retrievalresultcard{retrievalrelevant}{2}{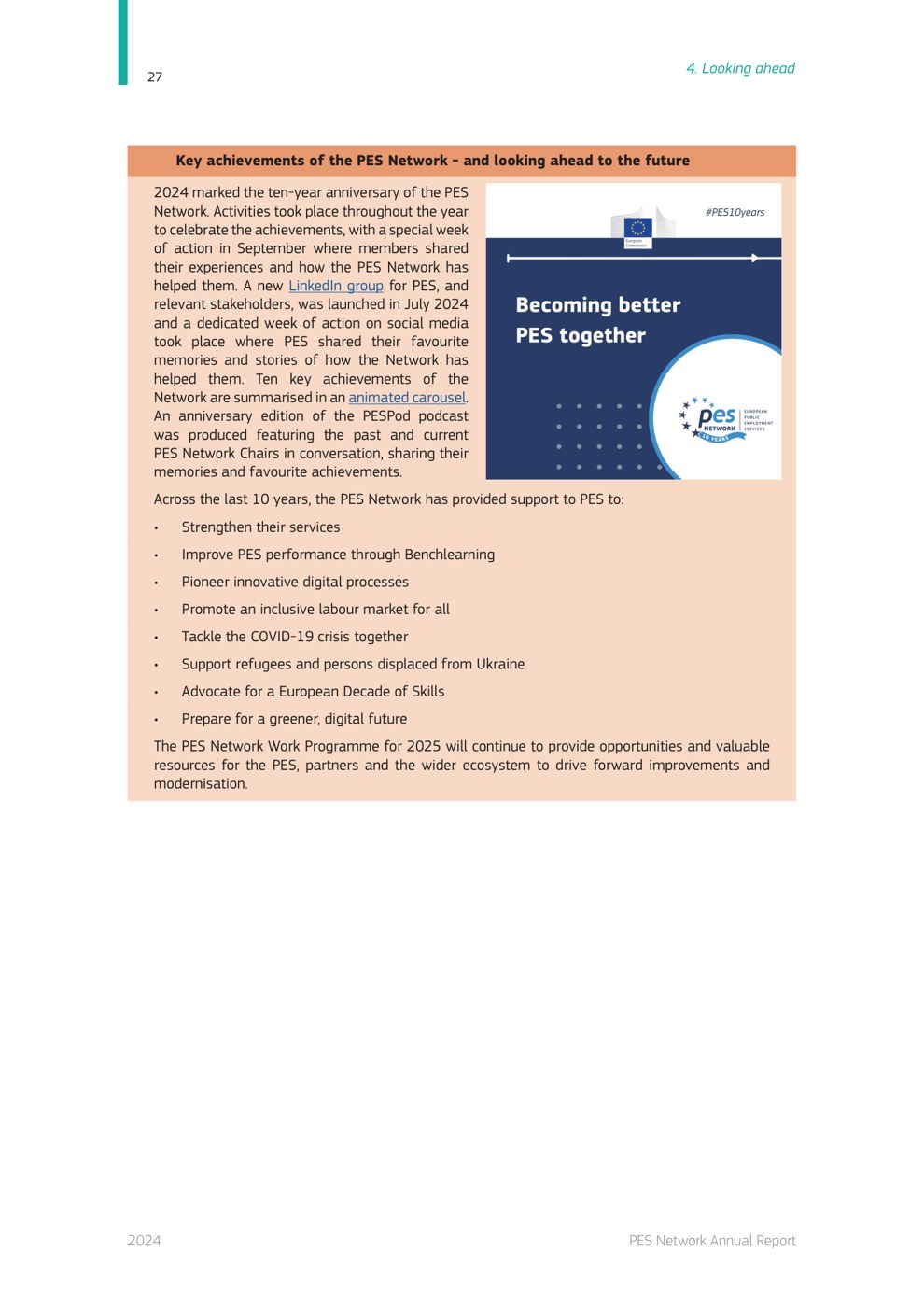}{0.7068}{Critically relevant}
    \retrievalresultcard{retrievalrelevant}{3}{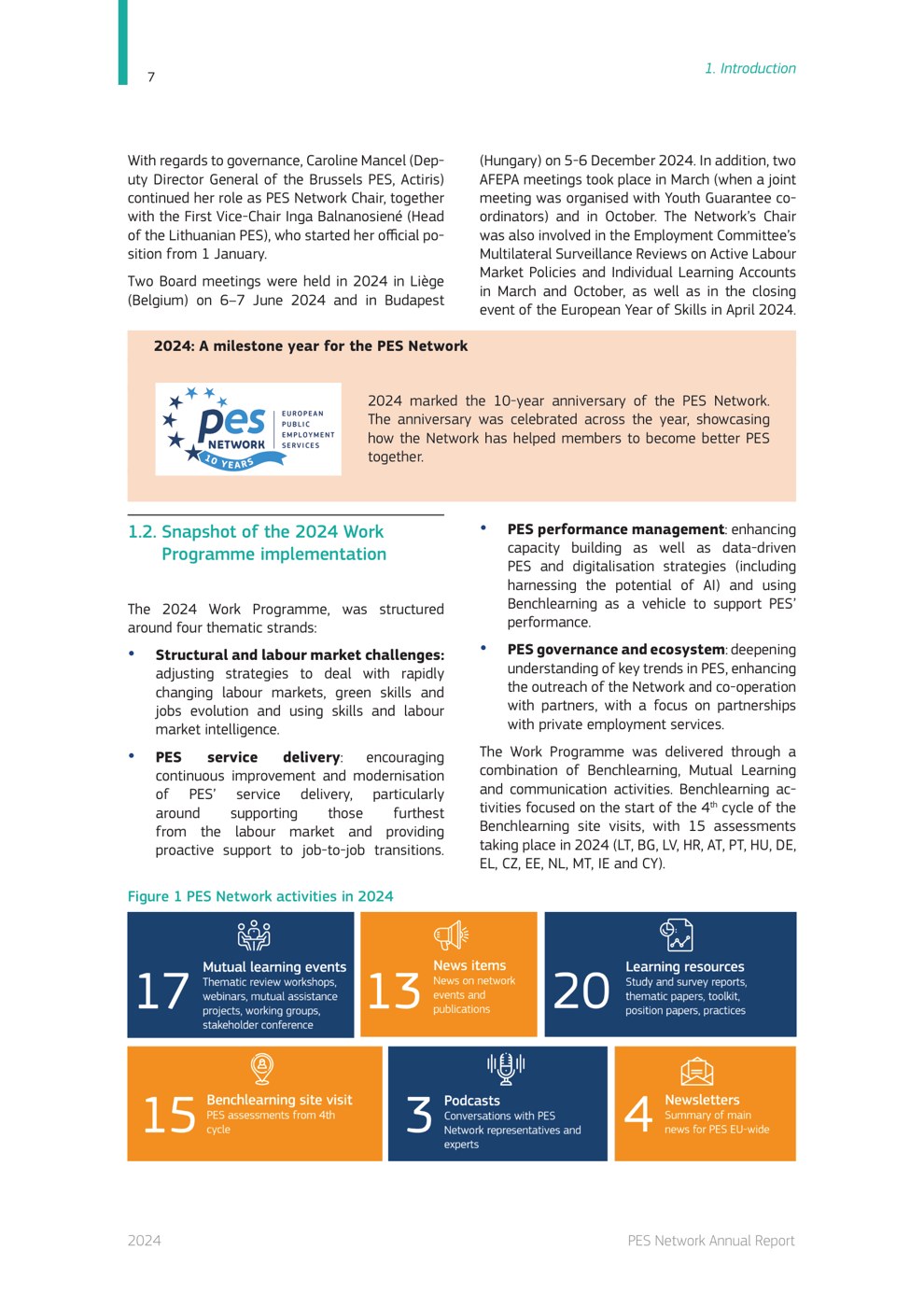}{0.6989}{Critically relevant}
  \end{tcbraster}
\end{tcolorbox}

\begin{tcolorbox}[title=ViDoRe v3 retrieval case 2, retrievalshowcase]
  \begin{tcolorbox}[title=Query (English), retrievalquery]
    Can crazing be caused by the paints used in a plastic panel paint restoration process?
  \end{tcolorbox}
  \retrievalcasemeta{\href{https://huggingface.co/datasets/vidore/vidore_v3_industrial}{\hfemoji\ \code{vidore/vidore\_v3\_industrial}} (5,244 pages)}{\neomme{}-Retriever 260M (late-interaction)}
  \begin{tcbraster}[raster columns=3, raster column skip=0.2em,
    raster valign=top, raster force size=false]
    \retrievalresultcard{retrievalrelevant}{1}{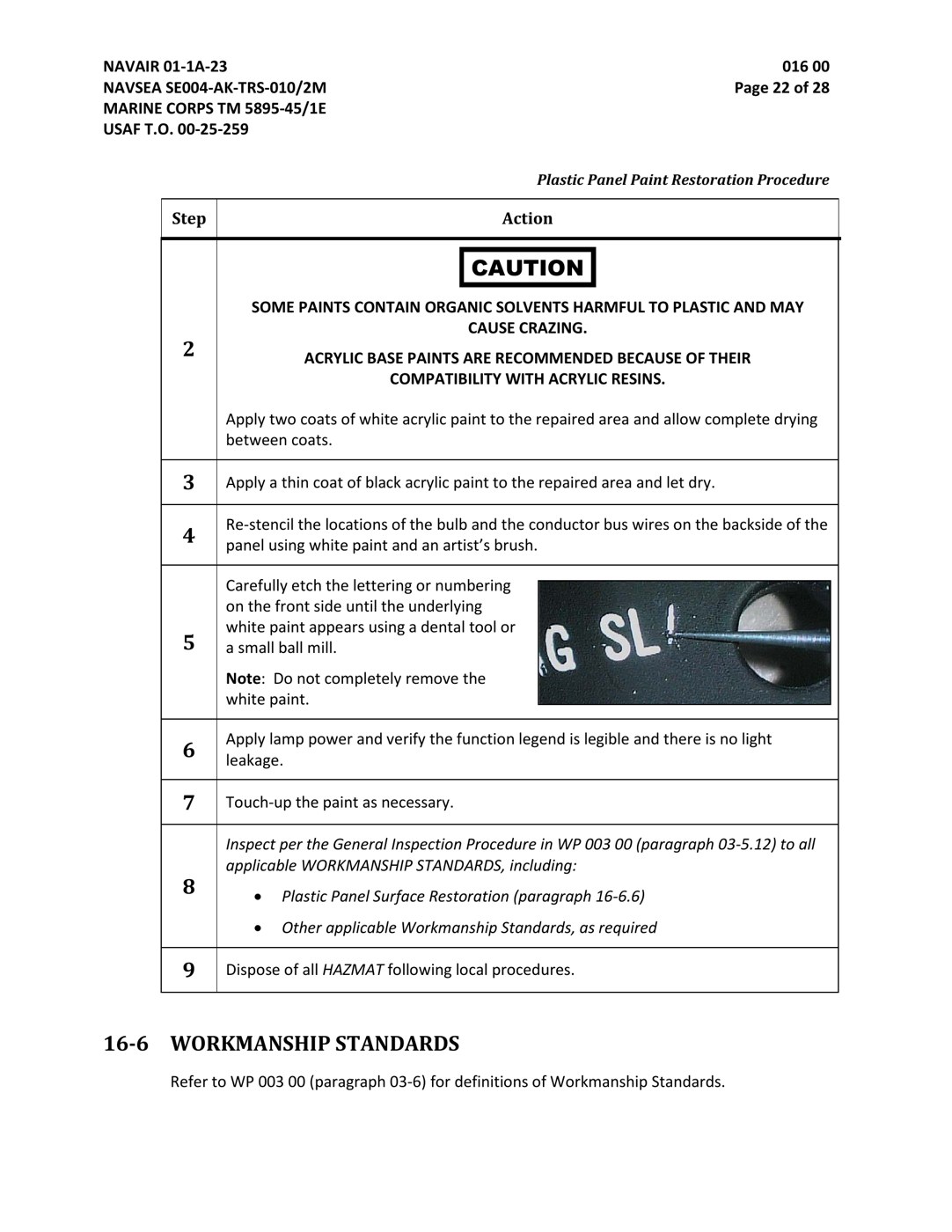}{0.8502}{Fully relevant}
    \retrievalresultcard{retrievalunjudged}{2}{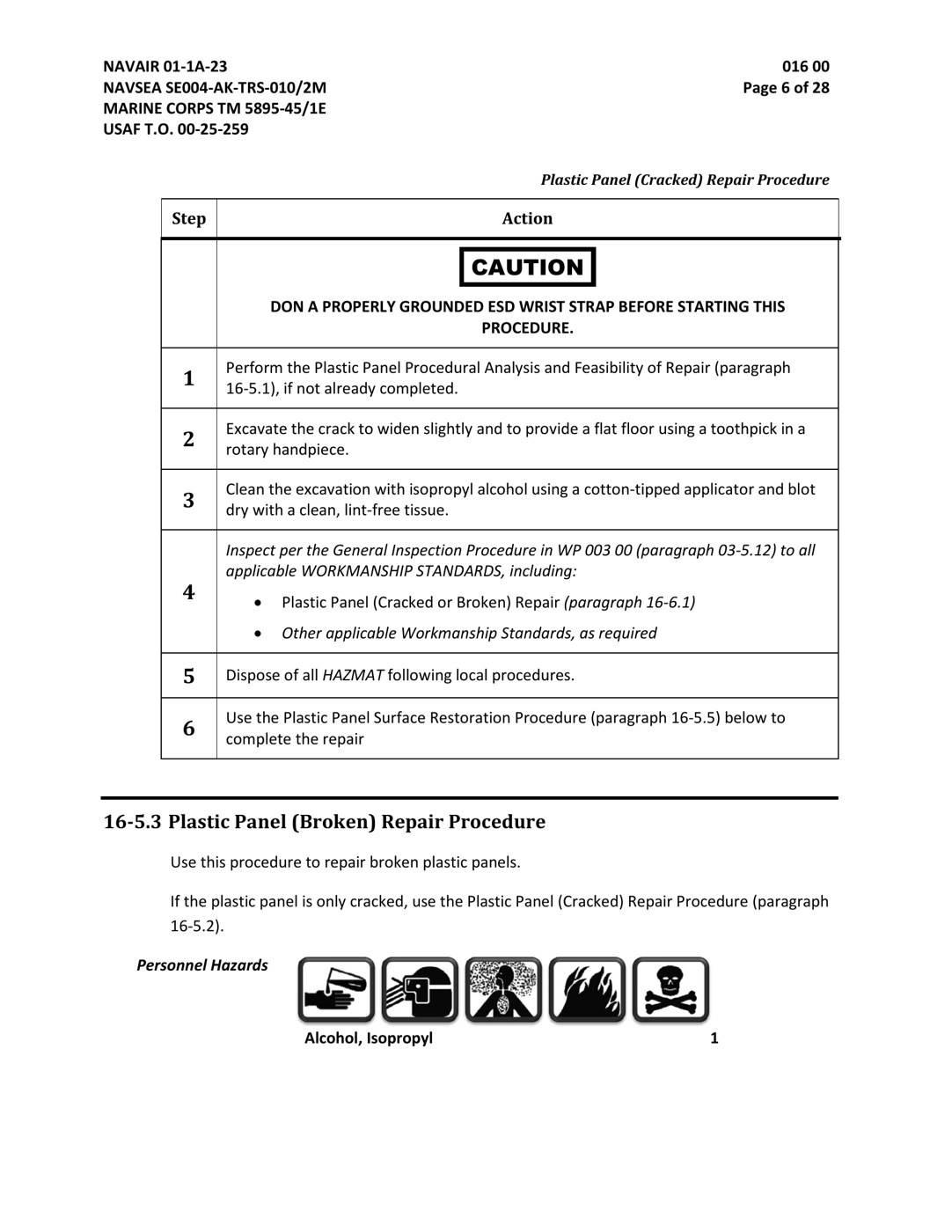}{0.7275}{Unjudged}
    \retrievalresultcard{retrievalunjudged}{3}{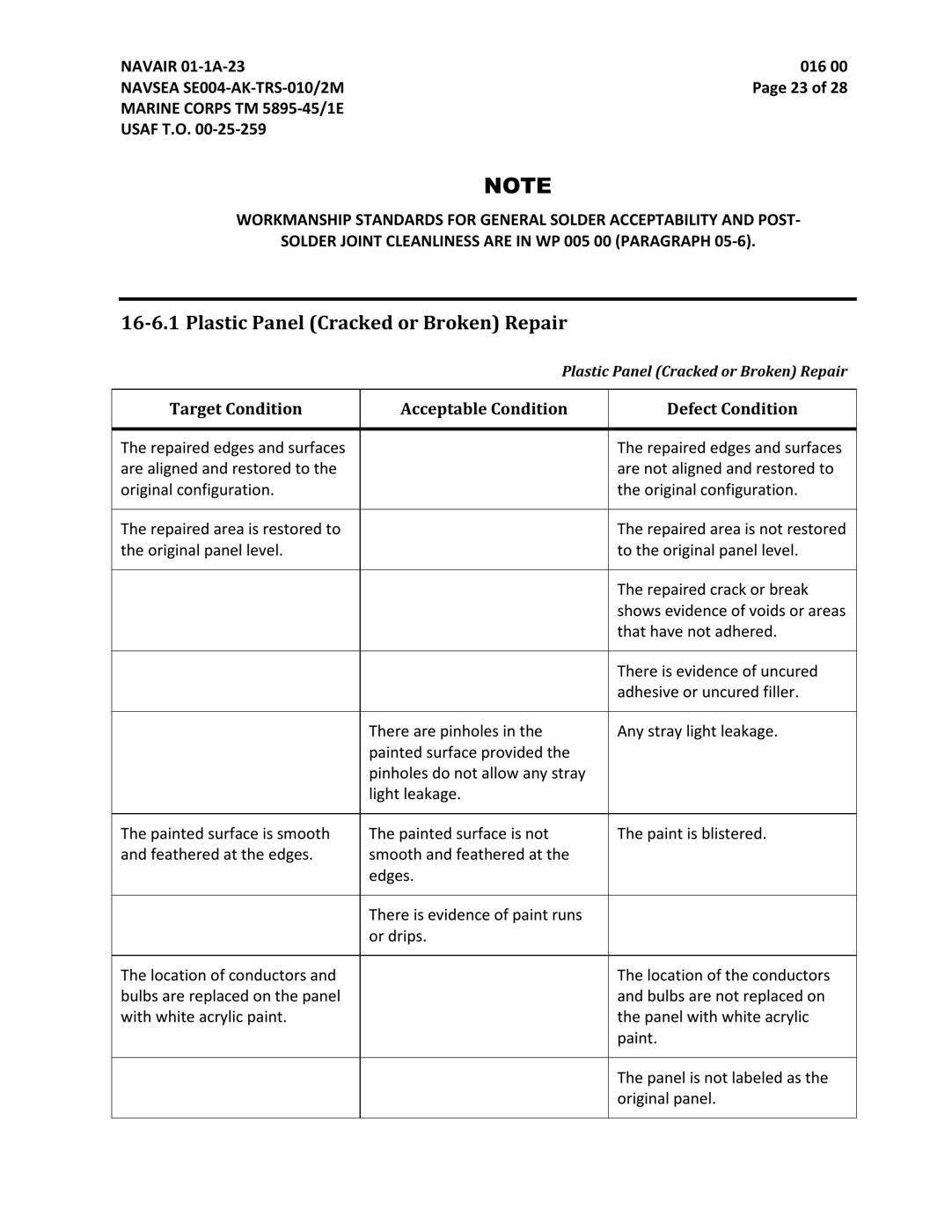}{0.7146}{Unjudged}
  \end{tcbraster}
\end{tcolorbox}

\begin{tcolorbox}[title=ViDoRe v3 retrieval case 3, retrievalshowcase]
  \begin{tcolorbox}[title=Query (German), retrievalquery]
    Wie unterscheidet sich der Ansatz von ISMP zur Fehlermeldung von dem der FDA?
  \end{tcolorbox}
  \retrievalcasemeta{\href{https://huggingface.co/datasets/vidore/vidore_v3_pharmaceuticals}{\hfemoji\ \code{vidore/vidore\_v3\_pharmaceuticals}} (2,313 pages)}{\neomme{}-Retriever 260M (late-interaction)}
  \begin{tcbraster}[raster columns=3, raster column skip=0.2em,
    raster valign=top, raster force size=false]
    \retrievalresultcard{retrievalunjudged}{1}{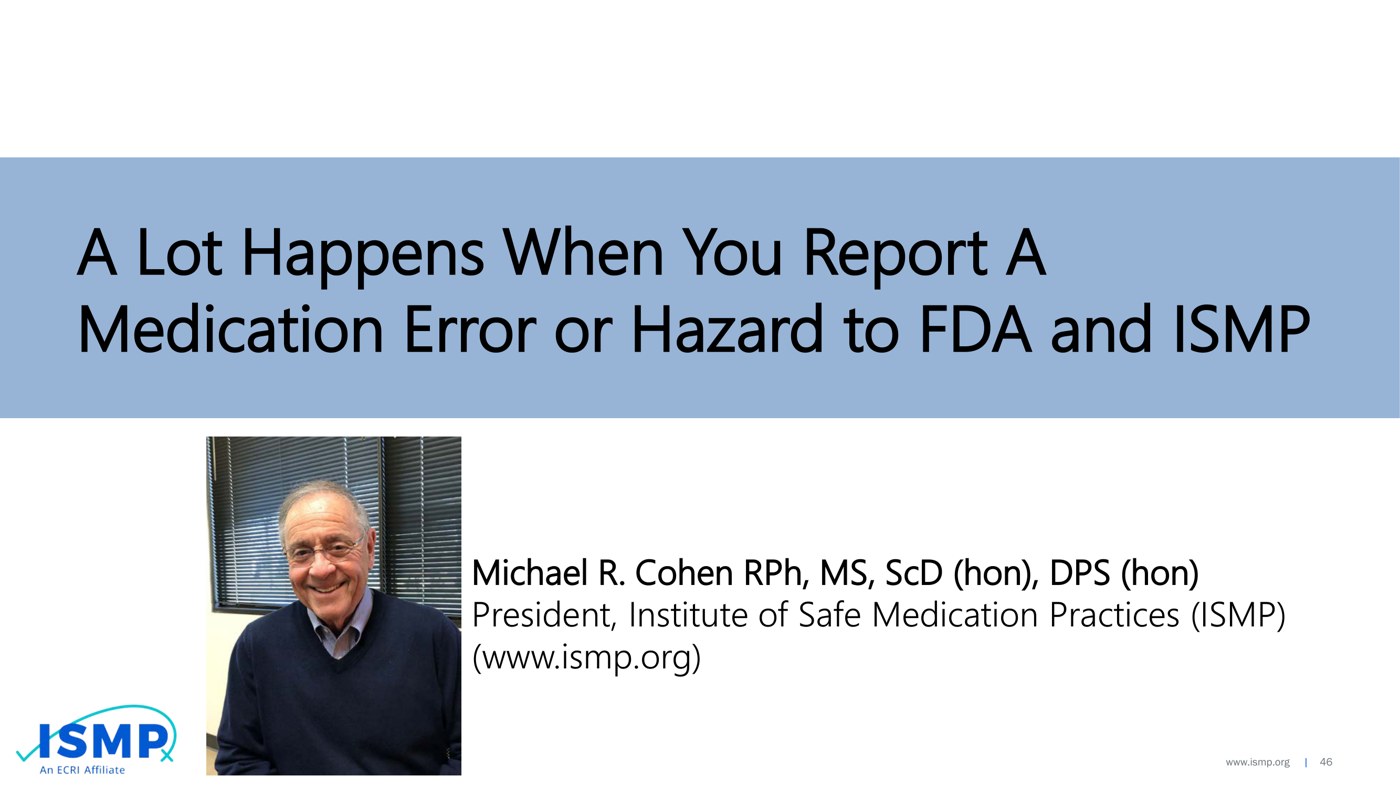}{0.7850}{Unjudged}
    \retrievalresultcard{retrievalunjudged}{2}{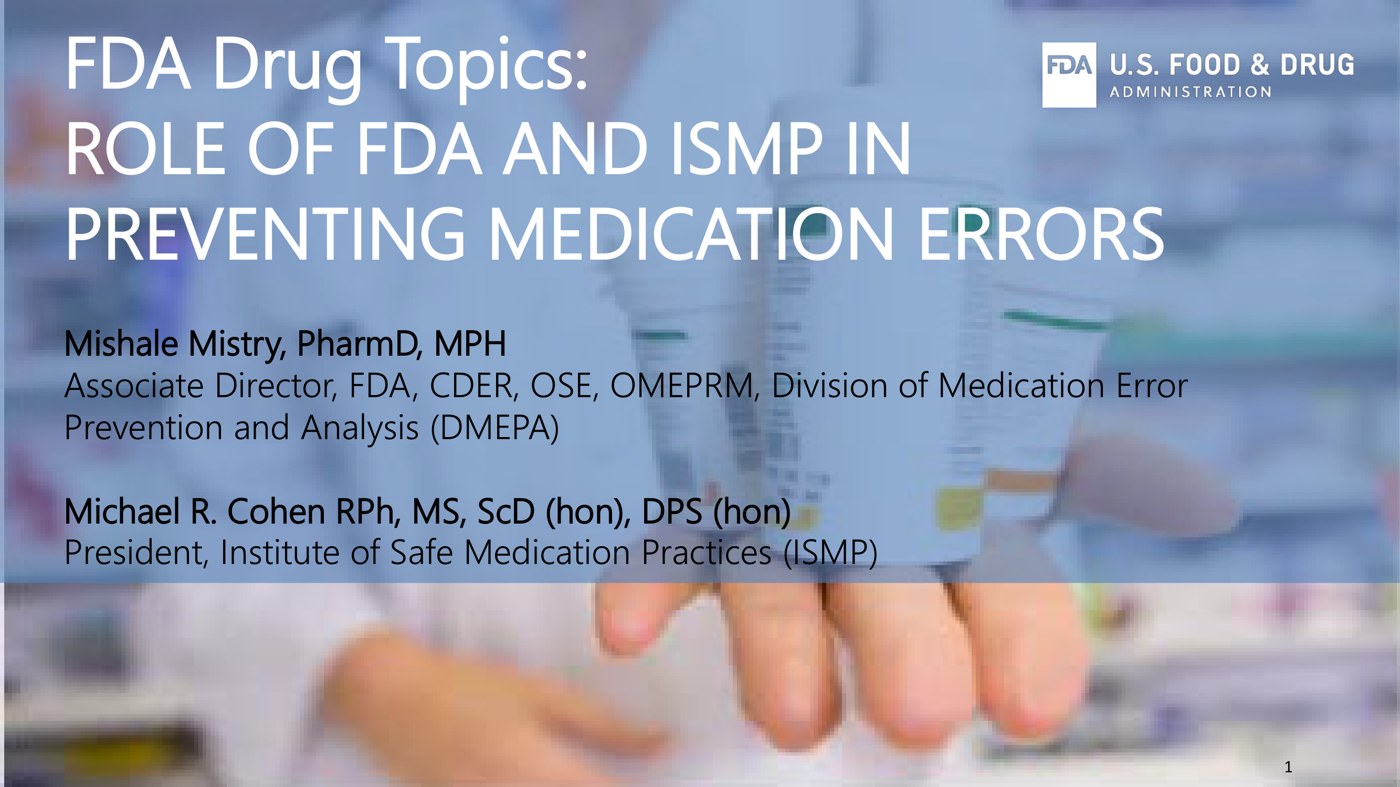}{0.7804}{Unjudged}
    \retrievalresultcard{retrievalrelevant}{3}{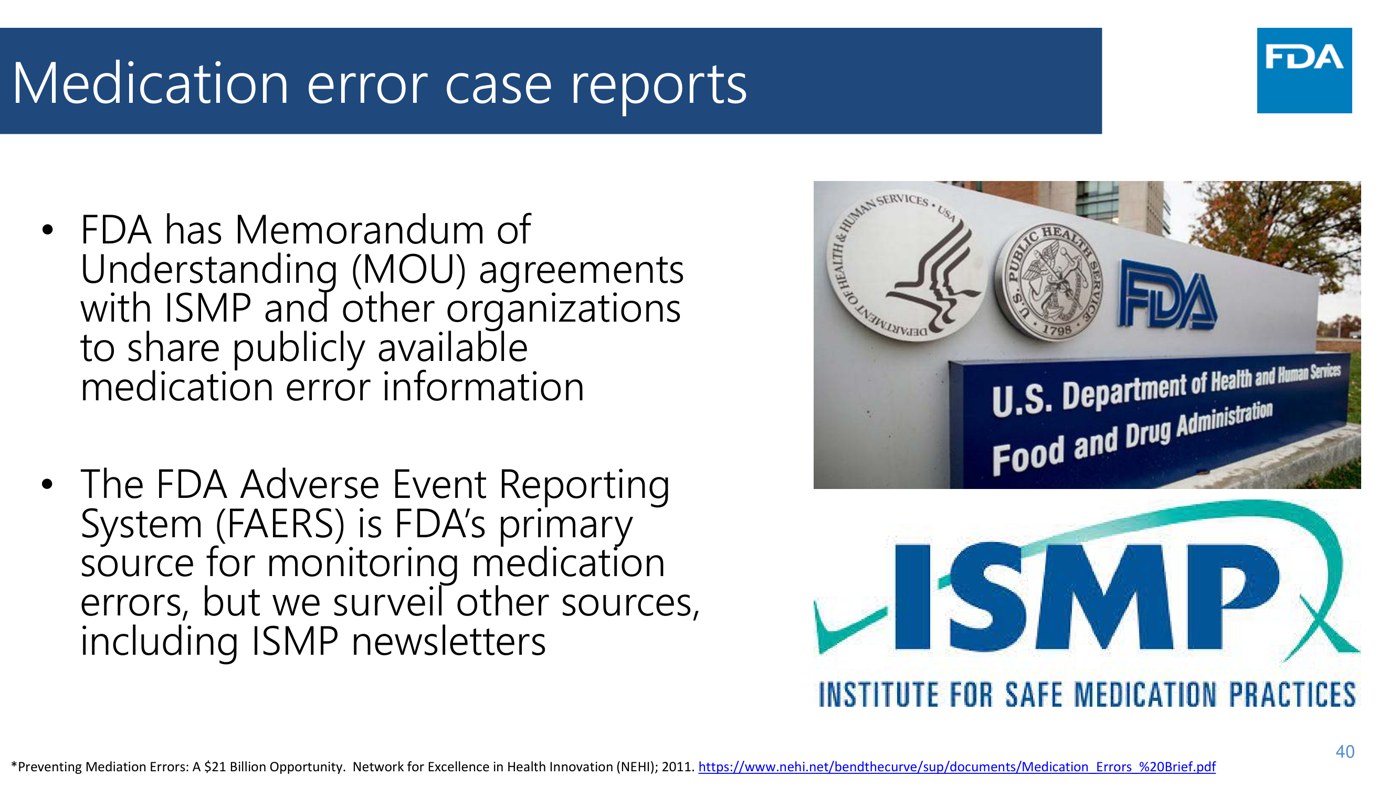}{0.7572}{Critically relevant}
  \end{tcbraster}
\end{tcolorbox}

\end{document}